\documentclass[aps,prx,superscriptaddress,twocolumn,10pt]{revtex4-2} 
\usepackage[margin=1in]{geometry}
\usepackage{graphicx,amsfonts}
\usepackage[caption=false,position=top,labelformat=empty]{subfig}
\usepackage{titletoc}
\usepackage{physics,xcolor}
\usepackage{orcidlink}
\usepackage{empheq}
\usepackage[most]{tcolorbox}
\definecolor{pink}{rgb}{1, 0.2, 0.5}
\usepackage{hyperref}[hypertexnames=false]
\hypersetup{
	colorlinks   = true, 
	urlcolor     = pink,
	linkcolor    = pink, 
	citecolor   = pink 
}
\usepackage{bm}

\newtcbox{\eqbox}[1][]{%
    nobeforeafter, math upper, tcbox raise base,
    enhanced, colframe=black,
    colback=black!5!white, boxrule=0.5pt, left=1mm, right=1mm, boxsep=2pt,
    #1}

\usepackage{tikz}
\newcommand*\beamsplitter[1]{\tikz[]{
    \draw [fill] (0,#1) circle [radius=0.06]; 
    \draw[thick] (0,#1) -- (0,0);
    \draw [fill] (0,0) circle [radius=0.06];
}}

\titlecontents{section}
  [0em]                                  
  {\addvspace{0.5\baselineskip}\bfseries} 
  {\contentslabel{2.0em}}                
  {}                                     
  {\hfill\contentspage}                  

\titlecontents{subsection}
  [2.0em]
  {\small}
  {\contentslabel{2.5em}}                
  {}
  {\hfill\contentspage}

\titlecontents{subsubsection}
  [2.0em]
  {\small}
  {\contentslabel{2.5em}}                
  {}
  {\hfill\contentspage}

\newcommand{\printAppendixTOC}{%
    \section*{Appendix Contents}
    \startcontents[appendix]%
    \printcontents[appendix]{}{1}{}
}

\newcommand{\ee}{\end{equation}}
\newcommand{\be}{\begin{equation}}
\newcommand{\p}[1]{\left( #1 \right)}
\newcommand{\br}[1]{\left[#1\right]}

\newcommand{\hc}[1]{#1^\dagger} 

\newcommand{\dir}[1]{#1^{(d)}} 
\newcommand{\gho}[1]{#1^{(g)}} 
 
\renewcommand\bra[1]{{\langle{#1}|}}
\makeatletter
\renewcommand\ket[1]{%
	\@ifnextchar\bra{\k@t{#1}\!}{\k@t{#1}}%
}
\newcommand\k@t[1]{{|{#1}\rangle}}
\makeatother

\newcommand{\subalign}[1]{%
	\begin{subequations}
		\begin{align}
			#1
		\end{align}
	\end{subequations}
}
\renewcommand{\eqref}[1]{Eq.~(\ref{#1})} 
\newcommand{\secref}[1]{Sec.~\ref{#1}} 
\newcommand{\figref}[1]{Fig.~\ref{#1}} 
\newcommand{\appref}[1]{App.~\ref{#1}} 
\newcommand{\refref}[1]{Ref.~\citenum{#1}} 

\newcommand{\ra}{\rightarrow}

\newcommand{\g}{\gamma}
\newcommand{\lamSq}{|\lambda|^2}
\newcommand{\nbar}{\overbar{n}}
\newcommand{\overbar}[1]{\mkern 1.5mu\overline{\mkern-1.5mu#1\mkern-1.5mu}\mkern 1.5mu}

\newcommand{\cC}{\mathcal{C}}
\newcommand{\cD}{\mathcal{D}}
\newcommand{\cE}{\mathcal{E}}
\newcommand{\cF}{\mathcal{F}}

\newcommand{\cK}{\mathcal{K}}
\newcommand{\cO}{\mathcal{O}}
\newcommand{\cM}{\mathcal{M}}
\newcommand{\cN}{\mathcal{N}}

\newcommand{\cP}{\mathcal{P}}
\newcommand{\cR}{\mathcal{R}}

\newcommand{\cT}{\mathcal{T}}

\newcommand{\tin}{\textrm{in}}
\newcommand{\tout}{\textrm{out}}
\newcommand{\nvec}{\vec{n}}

\newcommand{\lvec}{\vec{l}}

\newcommand{\sigvec}{\vec{\sigma}}
\newcommand{\ivec}{\vec{\iota\hspace{2pt}}}
\newcommand{\lamvec}{\vec{\lambda}}

\DeclareMathOperator*{\Motimes}{\text{\raisebox{0.25ex}{\scalebox{0.7}{$\bigotimes$}}}} 

\newcommand{\ketm}[1]{ \ket{#1}\!\rangle} 
\newcommand{\bram}[1]{ \langle\!\bra{#1}}
\newcommand{\kbm}[2]{\ketm{#1}\!\bram{#2}}

\newcommand{\vac}{\ketm{0}} 

\definecolor{turquoise}{rgb}{0.0, 0.85, 0.82}

\newcommand{\snlNM}{Quantum Algorithms and Applications Collaboratory, Department of Quantum Computer Science, Sandia National Laboratories, Albuquerque, NM 87185, USA} 
\newcommand{\snlCA}{Quantum Algorithms and Applications Collaboratory, Department of Quantum Information Science, Sandia National Laboratories, Livermore, CA 94550, USA}

\begin{document}

\title{Ghost unitaries for loss-resilient photonic resource-state generation} 

\author{Kevin J. Randles\orcidlink{0000-0002-4739-4313}}
\email{kjrandl@sandia.gov}
\affiliation{\snlNM}
\author{Manuel H. Mu\~noz-Arias}
\affiliation{\snlCA}
\author{Kevin Thompson\orcidlink{0000-0001-5669-2200}}
\affiliation{\snlNM} 

\begin{abstract}
Entangled few-photon states, such as Bell and Greenberger--Horne--Zeilinger (GHZ) states, are central resources for discrete-variable photonic quantum information processing. In many architectures, these states are generated probabilistically by first heralding single photons from biphoton sources and then interfering them in linear-optical circuits. In this work, we show that the entanglement structure of said biphoton sources can be directly exploited for noise-engineering to deliver higher quality resource states. Using the transpose identity of multi-mode squeezed vacuum states, we introduce ``ghost-unitary'' implementations in which the target resource-generation interferometer is applied to the heralding modes rather than to the delivered resource modes. When the dominant imperfection is interferometric loss, this relocation converts direct amplitude damping of the output into false-positive heralding events that require both higher-pair emission and photon loss. In the low-squeezing regime, these events are parametrically suppressed, yielding higher-fidelity conditional resource states.
We analyze this mechanism for single-qubit states, two-photon NOON states, Bell states, and a family of $n$-qubit GHZ states, as well as for single-qubit teleportation and type-I-fusion-based resource growth. Across these examples, we delineate finite parameter regimes in which ghost-unitary implementations outperform direct implementations, with particular relevance to frequency-bin photonics where active linear-optical transformations can be relatively lossy. We also prove that, in the setting considered here, the direct implementation of the $n$-GHZ family delivers identically damped outputs: equivalent to the ideal state followed by uniform amplitude damping. This extends the known $n=2,3$ cases and provides a structural result of independent interest for photonic quantum information processing.
\end{abstract}

\maketitle

\section{Introduction}
The generation of high-quality entanglement is central to quantum information science. In discrete-variable photonic quantum information processing (QIP), entangled photonic states serve as a fundamental resource for applications
such as fusion-based quantum computing, blind quantum computing, all-photonic quantum repeaters, and quantum teleportation, while also playing a key role in entanglement-based quantum cryptography and entanglement-enhanced metrology \cite{bartolucci2023fusion,barz2012demonstration,azuma2015all,hu2023progress,pan2012multiphoton,liu2021distributed}.

One widely studied use of photonic entanglement is quantum ghost imaging, where a target's transmittance pattern is reconstructed from correlations between two optical beams, even though only one beam interacts with the target~\cite{erkmen2010ghost}. 
In quantum implementations, these beams are prepared in a highly entangled biphoton state produced by spontaneous parametric down conversion (SPDC) or spontaneous four-wave mixing (SFWM). 
Such entanglement has also enabled related demonstrations of nonlocal transfer of high-dimensional unitary operations~\cite{paneru2025nonlocal}. These examples illustrate that biphoton entanglement can make transformations applied to one subsystem observable through another.

In this paper, we exploit the same biphoton entanglement structure for loss-resilient heralded photonic resource-state generation (RSG). Specifically, we use the transpose identity of bipartite multi-mode squeezed vacuum states to implement a target linear-optical transformation on the heralding subsystem rather than on the delivered resource subsystem. When the dominant imperfection is component loss inside the transformation, this relocation converts direct output amplitude damping into false-positive heralding events that require both higher-pair emission and photon loss. In the low-squeezing regime, these events are parametrically suppressed. Consequently, output infidelity scales with combined powers of the squeezing strength and loss, rather than with transformation loss alone.

This approach is especially attractive for frequency-bin implementations, where active frequency-domain linear-optical components can be substantially lossier than passive spatial or polarization optics \cite{lu2023frequency,munoz2026modeling,congia2026fully}. In such settings, shifting the lossy transformation from the delivered modes to the heralding modes can directly improve the conditional quality of the generated resource. In homage to quantum ghost imaging, we call these ghost-unitary implementations: the delivered modes host the resource state but do not directly undergo the target linear-optical transformation.

The paper is organized as follows. Section~\ref{sec:background} introduces 
heralded photonic RSG, multi-mode squeezed vacuum sources, and the transpose identity enabling ghost-unitary implementations. It also situates our approach relative to related works.
Sections~\ref{sec:HOM}--\ref{sec:scaling} compare the direct and ghost implementations of increasingly complex RSG schemes. 
Section~\ref{sec:HOM} develops the central analytic ideas in the setting of two-photon-NOON-state generation, where analytical expressions can be obtained that highlight the tradeoffs between the implementations. 
Section~\ref{sec:HBSG} applies the same framework to heralded Bell-state generation. 
Section~\ref{sec:scaling} studies scaling with resource size, including single-photon examples and a family of $n$-qubit GHZ-state generation circuits. 
Section~\ref{sec:qip_applications} then examines how resulting Bell-state errors cascade in downstream photonic primitives, namely single-qubit teleportation and type-I fusion. 
Section~\ref{sec:conclusion} concludes with a summary and outlook.
Lengthy derivations are relegated to the appendices.

\section{Background}\label{sec:background}
\subsection{Resource-state generation}
Within discrete-variable implementations, small entangled resource states can be probabilistically prepared by passing single photons through tailored linear-optical interferometers and detecting a subset of the modes. In such schemes, certain detection patterns herald the generation of the resource state on the unmeasured modes \cite{forbes2025heralded,randles2026limits}. 
Especially within integrated optics platforms, these input photons are often heralded by detecting one half of an entangled biphoton state.
Entanglement is a valuable resource, yet such single-photon generation consumes the biphoton entanglement solely as a heralding mechanism.

This strongly motivates investigating whether this biphoton entanglement structure can be leveraged for improved resource-state generation (RSG). Here we pursue this direction by analyzing ghost-unitary implementations, in which the transpose of the RSG interferometer is applied to the heralding idler modes rather than applying the interferometer directly to the signal modes, a subset of which are to host the itinerant resource state. 
In the absence of loss and for equal squeezing, these implementations are equivalent via the transpose trick. However, the ghost implementation can be more robust to interferometer loss in certain operating regimes. As concrete resource-state examples, we focus on the heralded generation of two-photon NOON (or ``HOM'') states, Bell states, and $n$-qubit Greenberger--Horne--Zeilinger (GHZ) states.

\subsection{Structure of multiple biphoton sources}
The ideal state produced by a biphoton source, e.g., based on SPDC or SFWM, is the two-mode-squeezed vacuum state
\be\label{eq:TMSW_pure}
    \ket{\textrm{TMSV}(\lambda)} = \sqrt{1-|\lambda|^2} \sum_{n=0}^\infty  \lambda^n \ketm{n, n}_{s,i}. 
\ee
Here $\ketm{\cdot}$ is used to refer to Fock basis states with a fixed number of photons in each mode.
The multi-mode squeezed vacuum (MMSV) state comprised of $M$ such states in disjoint modes with identical squeezing strengths, $\lambda \in \mathbb{C}$,
$\ket{\zeta} = \ket{\textrm{TMSV}(\lambda)}^{\otimes M}$,
can be decomposed into subspaces with fixed total photon number $S$ on both the signal and idler modes. Namely, defining $\nvec = (n_0, n_1, \cdots, n_{M-1})$, $S_{\nvec} = \sum_{j=0}^{M-1} n_j$, and the set of corresponding occupations $\nvec_S = \{\nvec: S_{\nvec} = S\}$ one finds
\subalign{
    \ket{\zeta} &= \sum_{S=0}^\infty z_S \ket{\Phi_S}, \label{eq:zetaMMSV} \\
    z_S &= \p{1-|\lambda|^2}^{M/2} \lambda^{S} \sqrt{|\nvec_S|}, \\
    \ket{\Phi_S} &= \frac{1}{\sqrt{|\nvec_S|}}
     \sum_{\nvec \in \nvec_S} \ketm{\nvec, \nvec}_{s,i}.
}
Here $\ket{\Phi_S}$ is a maximally entangled state of the signal and idler modes, and $|\nvec_S|= \binom{S+M-1}{S}$ 
is the number of ways $S$ photons can occupy $M$ modes.

\subsection{Transpose trick}
Maximally entangled bipartite systems, with canonical state $\ket{\Phi} = \frac{1}{\sqrt{d}} \sum_{i=0}^{d-1} \ket{ii}$ in the Schmidt basis, satisfy the \emph{transpose property}, where for any linear transformation on half of the bipartition:
\be\label{eq:transpose_trick} 
    \p{U \otimes I} \ket{\Phi} = \p{I \otimes U^T} \ket{\Phi}.
\ee
This can easily be seen by expanding the transformation in the Schmidt basis as $U = \sum_{ij} U_{ij} \ketbra{i}{j}$ for each subsystem \footnote{This identity follows directly from vectorization and the Choi–Jamio{\l}kowski isomorphism, e.g., see Section 2.2 of 
\textit{The theory of quantum information} by Watrous, Cambridge University Press, 2018.}. 

As the MMSV state $\ket{\zeta}$ is maximally entangled in each fixed-photon-number sector, by linearity one can use the transpose trick to shift a RSG unitary, $U$, from the signal modes onto the idler modes
\be
    \p{U_s \otimes I_i} \ket{\zeta} 
    = \p{I_s \otimes U^T_i} \ket{\zeta}.
\ee
Moreover, if the unitary is decomposable into two subunitaries, $U = V W$, they can be distributed across the signal and idler modes as
\be\label{eq:subunitaryTransposeTrick}
    \p{V_s W_s \otimes I_i} \ket{\zeta} 
    = \p{V_s \otimes W^T_i} \ket{\zeta}.
\ee
Graphically, this MMSV transpose trick can be expressed as 
\be
    \raisebox{-0.45\totalheight}{ \includegraphics[width=0.4\linewidth]{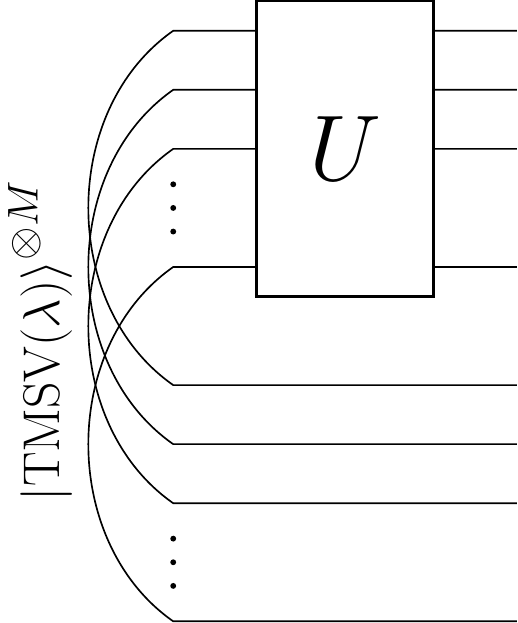} } 
    ~\scalebox{1.5}{=}~
    \raisebox{-0.45\totalheight}{ \includegraphics[width=0.4\linewidth]{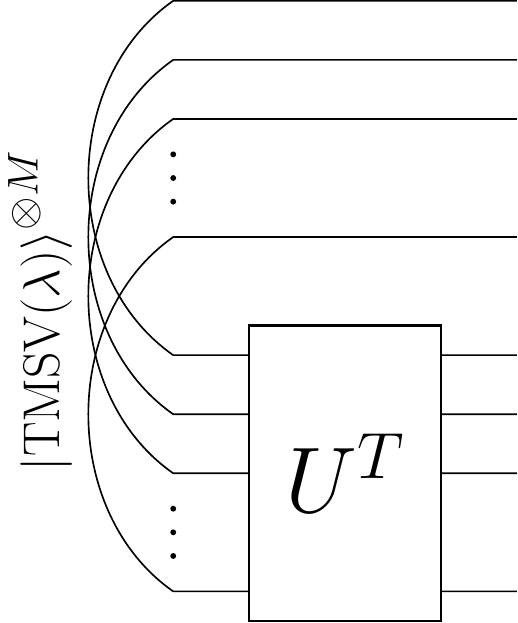} } 
    \nonumber
\ee
with the upper and lower lines corresponding to the signal and idler modes, respectively, which are in entangled biphoton states as indicated via the linking arcs.
Crucially for RSG, these equalities are equally valid if heralding measurements are performed on a subset of the modes.

The transpose trick concerns where the physical linear-optical transformation is implemented. Thus, loss associated with the components realizing $U$ is relocated when $U^T$ is implemented on the heralding subsystem. This is distinct from, though often aided by, commuting loss through a circuit. For the examples treated analytically below, uniform loss and detector inefficiency can be collected into lumped efficiencies, giving compact formulae. 
For more general RSG circuits with residual non-commuting loss, the ghost implementation is still well defined, but its effective heralding positive operator-valued measure (POVM) and output state are less transparent.

\subsection{Related works}
The transpose identity underlying \eqref{eq:transpose_trick} is central to gate teleportation and Choi-state resource constructions, beginning with Gottesman and Chuang \cite{gottesman1999demonstrating} and appearing operationally in photonic quantum computing through offline resource-state preparation and fusion-based architectures~\cite{browne2005resource,bartolucci2023fusion}. The present work is also structurally related to that of \refref{paneru2025nonlocal}, where biphoton entanglement is used to nonlocally transfer a unitary operation between subsystems.

Gimeno-Segovia also observed a related avenue for using the entanglement structure of SPDC sources more directly in photonic RSG (see Sec.~4.5.3 of \refref{gimeno2016towards}). In particular, SPDC sources can be engineered to produce a superposition of vacuum, a Bell state in the two-photon sector, and higher-order multi-pair terms. Such states can be injected directly into fusion-based resource-generation circuits, where successful fusion measurements both herald pair generation and project the remaining modes into the desired resource state. Our approach is complementary to these directions. Rather than using nonlocal transfer as an end in itself, or engineering the emitted two-photon sector directly, we exploit the entanglement structure of MMSV states to reshape loss by implementing a desired resource-generation unitary on the heralding subsystem.

\section{HOM-state generation}\label{sec:HOM}
To demonstrate the utility of the transpose trick for RSG, we begin by analyzing the generation of a two-photon NOON state. This example is rich enough to capture the interplay of various error mechanisms, while being simple enough to admit a complete analytical treatment. Thus, it is also useful in setting up our analytical methodology and notation for the more complicated Bell- and GHZ-state generation schemes we consider.

\begin{figure*}[ht!]
    \subfloat[(a)]{\includegraphics[height=4cm]{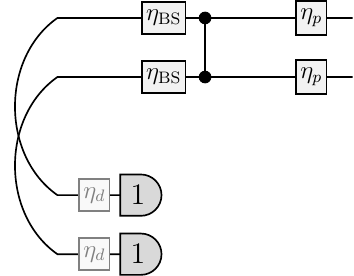}}
    \hfill
    \subfloat[(b)]{\includegraphics[height=4cm]{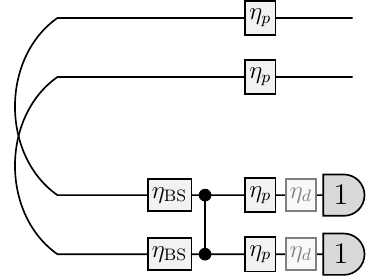}}
    \hfill
    \subfloat[(c)]{\includegraphics[height=4cm]{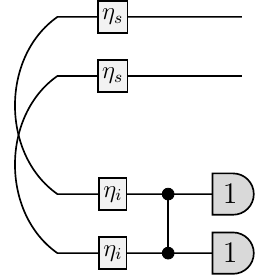}}
    \caption{
        HOM-state generation in the (a) direct and (b) ghost implementations. In (a), the lossy beamsplitter acts on the delivered signal modes (top); in (b), the transposed beamsplitter acts on the heralding idler modes (bottom). For equal squeezing, both cases can be mapped onto the unified model in (c), with the implementation dependence captured by the lumped efficiencies $\eta_s$ and $\eta_i$ of \eqref{eq:HOM_etas}.
    }
    \label{fig:HOM_circuits}
\end{figure*}

In their eponymous experiment \cite{hong1987measurement}, Hong, Ou, and Mandel demonstrated that if you inject two indistinguishable photons into two ports of a 50:50 beamsplitter, they bunch, yielding a two-photon NOON state that we deem the ``HOM state,''
$\ket{\psi_\textrm{HOM}} = \frac{1}{\sqrt{2}} \p{ \ketm{20} + \ketm{02} }$.
Such production of HOM states is commonly used to benchmark the quality of single photon sources \cite{ollivier2021hong}. Moreover, they can be used to boost entangling fusion measurements, increasing the probability that they succeed \cite{ewert20143,melkozerov2026single}.
The direct and ghost implementations of HOM-state generation are shown in \figref{fig:HOM_circuits}(a) and (b), respectively. Therein, we account for three types of loss: beamsplitter, propagation, and detector with respective efficiencies $\eta_{\rm BS}$, $\eta_p$, and $\eta_d$ and corresponding loss parameters $\g_j \equiv 1-\eta_j$.
Throughout this work we neglect global phases and use the beamsplitter convention 
$B=\raisebox{-0.3\totalheight}{\beamsplitter{0.4}} = \tfrac{1}{\sqrt{2}} 
\left( \begin{smallmatrix} 1 & i \\ i & 1 \end{smallmatrix} \right)$, which 
notably equals its own transpose \footnote{The real-entried 50:50 beamsplitter $\tfrac{1}{\sqrt{2}} 
\left( \begin{smallmatrix} 1 & 1 \\ 1 & -1 \end{smallmatrix} \right)$ could equally be used throughout this analysis without changing any results beyond imparting known phases in the heralded output states.}.

\subsection{Analytic methodology}
For both implementations we begin with the following two steps.
First, start with the initial MMSV state (here with $M=2$)
\be\label{eq:zeta_lambda}
    \ket{\zeta_{\vec{\lambda}}} = \Motimes_{j=0}^{M-1} \ket{\textrm{TMSV}(\lambda_j)},
\ee
where the subscript $\vec{\lambda} = (\lambda_0, \cdots, \lambda_{M-1})$ indicates that the squeezing parameters may undesirably vary.
Second, using the properties of photon loss (amplitude damping) channels---that single-mode loss is multiplicative and uniform loss commutes with linear optics as reviewed in \appref{app:gen_transpose_trick}---we can effectively shift all the loss onto the initial state
\be\label{eq:rho0_HOM}
    \rho_\tin = \Motimes_{j=0}^{M-1} \rho_\textrm{TMSV}(\lambda_j;\eta_{s,j}, \eta_{i,j}).
\ee
Here we defined the amplitude-damped TMSV state
\be\label{eq:rhoADdTMSV}
    \rho_\textrm{TMSV}(\lambda; \eta_s, \eta_i) = \cE_{\eta_s}\otimes\cE_{\eta_i}(\ketbra{\textrm{TMSV}(\lambda)}{\textrm{TMSV}(\lambda)})
\ee
(see \appref{app:LossyTMSVs} for details) with implementation-specific signal and idler mode efficiencies
\be\label{eq:HOM_etas}
    (\eta_s, \eta_i) = \begin{cases}
        (\eta_{\rm BS} \eta_p, \eta_d), & \textrm{direct} \\
        (\eta_p, \eta_{\rm BS} \eta_p \eta_d), & \textrm{ghost}
    \end{cases}.
\ee
On a single mode, such amplitude damping is characterized by the channel 
\be
    \cE_\eta(\rho) := \sum_{l=0}^\infty K_l \rho \hc{K}_l
\ee
with Kraus operators
\be\label{eq:adChannelKrausOp}
    K_l 
    =  \sqrt{\frac{\g^l}{l!}} \eta^{\hc{a} a/2} a^l.
\ee
We use the standard lossy photon-number-resolving detector (PNRD) model: lossy detection is equivalent to passing the input through an $\eta_d$ efficient amplitude-damping channel before ideal detection. 

With all of the losses subsumed into $\rho_\tin$, we need simply propagate it through the ideal linear optics, $U_\textrm{HOM}$, and perform ideal detection. 
Direct HOM-state generation requires a single photon on each signal mode, so we are concerned with a $(1,1)$ click pattern on the heralding idler modes with POVM element $\hat{\Pi}^{(11)} = \kbm{11}{11}$. 
This detection event occurs with probability $p_\textrm{HOM}$, heralding the output state $\rho_\textrm{HOM}$, which together are specified by the equation
\be
    \tilde\rho_\textrm{HOM} \equiv 
    p_\textrm{HOM} \rho_\textrm{HOM} 
    = \Tr_i\!\br{ U_\textrm{HOM} \rho_\tin \hc{U}_\textrm{HOM} \hat{\Pi}_i^{(11)}}.
\ee
To compare the implementations, we need to perform two such calculations (as is done in \appref{app:HOMStateFull}) one with the linear optics---the beamsplitter here---on the itinerant signal modes (direct) and another with it on the heralding idler modes (ghost):
\be
    U_\textrm{HOM} = \begin{cases}
        B_s \otimes I_i, & \textrm{direct} \\
        I_s \otimes B^T_i, & \textrm{ghost}
    \end{cases}.
\ee
Henceforth direct and ghost quantities will be indicated via superscript $(d)$ and $(g)$, respectively. Additionally, we will often omit the explicit $\otimes$ and leave such identity operators, $I$, implicit. 

\subsection{Direct implementation}
The probability of heralding is the probability of two independent heralded single-photon generations from the two lossy TMSV states 
\be\label{eq:dirp_HOM}
    \dir{p}_\textrm{HOM} = P_1(|\lambda_0|, \g_i) P_1(|\lambda_1|, \g_i)
\ee
with $\g_i = 1-\eta_i$ and
\be 
    P_1(|\lambda|, \g_i) = \frac{\eta_i (1 - |\lambda|^2) |\lambda|^2 }{(1- \g_i |\lambda|^2)^2}.
\ee
Each of two lossy single photon output states before the beamsplitter and any signal mode loss are of the simple form 
\be
    \rho^{(j)}_1 = (1-q_j) \hc{a}_j T_j(q_j) a_j 
\ee
with $q_j \equiv \g_i |\lambda_j|^2$ and
\be
    T(q) = (1-q) \sum_{k=0}^\infty q^k \kbm{k}{k}
    = (1-q) q^{\hc{a} a}
\ee
a normalized single-mode thermal state with average occupation $\nbar = q/(1-q)$. 

Thus, conditioned on the idler heralds, the pre-beamsplitter signal state is a product of photon-added thermal states,
$
    \rho_\tin^{(d)} \propto  
    \hc{a}_0 \hc{a}_1  T_0(q_0) T_1(q_1) a_0 a_1.
$
Passing through the symmetric beamsplitter maps the two-photon creation part as
\be\label{eq:HOM_twoPhoton_mapping}
    a_0^\dagger a_1^\dagger
    \mapsto
    \frac{i}{2} \p{ a_0^{\dagger 2}+a_1^{\dagger 2} }.
\ee
For equal squeezing, $q_0=q_1$, the thermal product is beamsplitter invariant.
For $q_0 \neq q_1$, the thermal background is transformed into a correlated two-mode Gaussian; the full state expression is given in \appref{app:HOMStateFull}. 
The fidelity of the output state, after passing it through the signal mode loss channel $\cE^{\otimes 2}_{\eta_s}$ is 
\be\label{eq:dirF_HOM}
    \dir{\cF}_\textrm{HOM} = \bra{\psi_\textrm{HOM}} \dir{\rho}_\textrm{HOM} \ket{\psi_\textrm{HOM}} 
    = \cF_1(\lambda_0) \cF_1(\lambda_1),
\ee
which is simply the product of the lossy single photon fidelities
\be
    \cF_1(\lambda) = \bram{1} \cE_{\eta_s}(\rho_1) \ketm{1} 
    = \eta_s \frac{(1- q)^2 \p{1 + g}}{h^{3}}.
\ee 
with $g \equiv \g_s q = \g_s \g_i |\lambda|^2$ and $h \equiv 1 - g$.

\subsection{Ghost implementation}
With the beamsplitter on the idler modes, using the cyclic property of the trace, we can rewrite the unnormalized state as
\subalign{
    \tilde\rho_\textrm{HOM} &= \Tr_i\!\p{ \rho_\tin \hat{\Pi}_i^{(\psi)} },
    \\
    \hat{\Pi}_i^{(\psi)} &= B^*_i \hat{\Pi}_i^{(11)} B^T_i
    = \ketbra{\psi_\textrm{HOM}}{\psi_\textrm{HOM}}.
}
That is, by backpropagating the target click-pattern POVM element through the idler-mode linear optics, we are projecting onto the amplitude-wise complex conjugated target state, though $\ket{\psi^*_\textrm{HOM}} = \ket{\psi_\textrm{HOM}}$ here.
The resulting heralded signal state, pre-signal-mode-loss, has the compact form
\be\label{eq:rhoG_HOM}
    \gho{\rho}_\textrm{HOM} \propto
    \p{ \lambda_0^{* 2} a_0^{\dagger 2} + \lambda_1^{* 2} a_1^{\dagger 2}} 
    q_0^{\hat{n}_0} q_1^{\hat{n}_1}
    \p{ \lambda_0^{2} a_0^{2} + \lambda_1^{2} a_1^{2}}, 
\ee
with $\hat{n}_j = \hc{a}_j a_j$. Under signal-mode-loss $\gho{\rho}_\textrm{HOM} \ra \cE^{\otimes 2}_{\eta_s}(\gho{\rho}_\textrm{HOM})$.
This expression makes the relevant physics transparent: idler loss produces thermal tails $q_j^{\hat{n}_j}$, while the heralding measurement coherently adds two photons to either output mode. The corresponding heralding probability and fidelity are given in \appref{app:HOMStateFull}. 

\subsection{Comparison for equal squeezing}
\begin{figure}[ht!]
    \subfloat[(a)]{ \includegraphics[height=6.5cm]{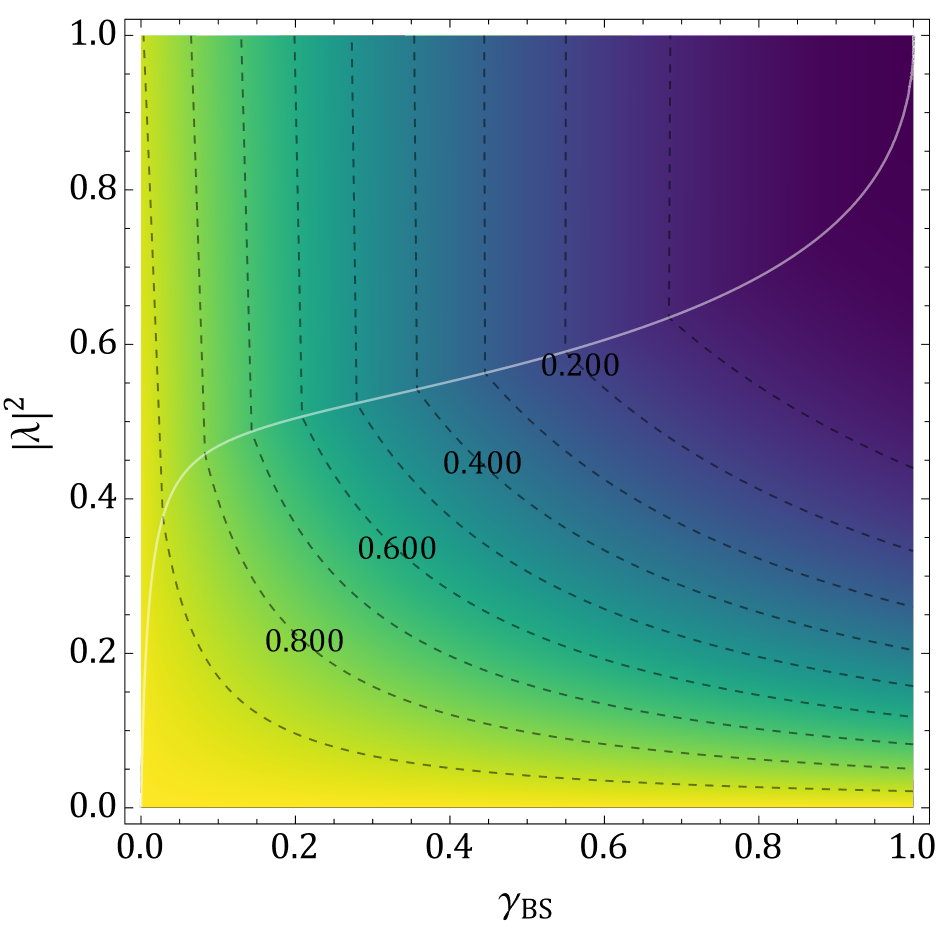}
    \raisebox{1.1cm}{\includegraphics[height=4.8cm]{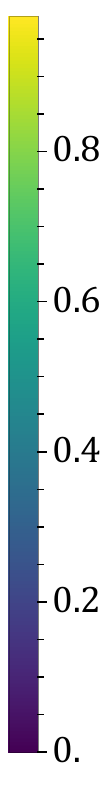}} }
    \\
    \subfloat[(b)]{ \includegraphics[height=6.5cm]{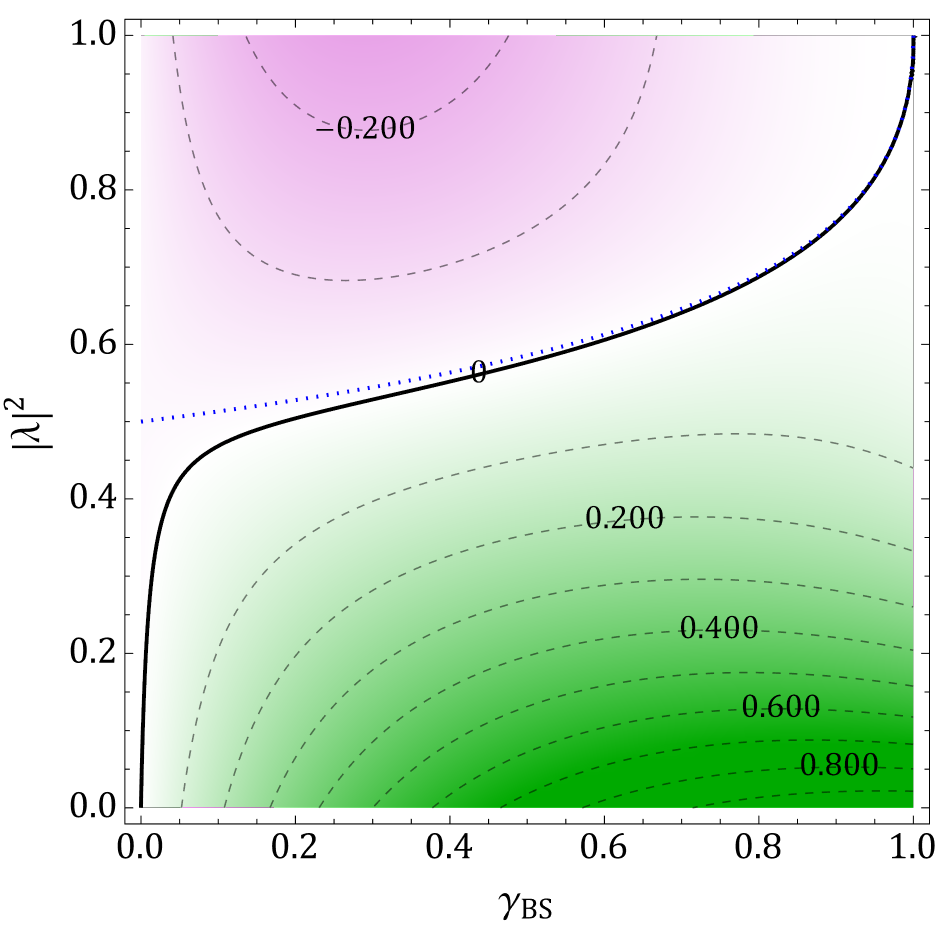}
    \raisebox{1cm}{\includegraphics[height=5.2cm]{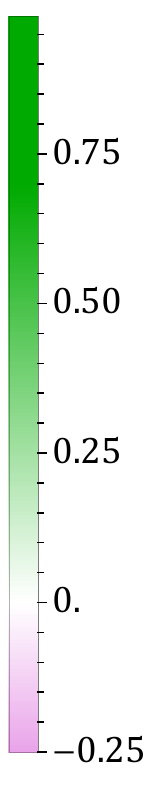}} } 
    \caption{
    HOM-state generation comparison for equal squeezing. 
    (a) Maximum fidelity $\max(\gho{\cF}_\textrm{HOM}, \dir{\cF}_\textrm{HOM})$ and (b) fidelity difference $\gho{\cF}_\textrm{HOM} - \dir{\cF}_\textrm{HOM}$ versus squeezing strength $\lamSq$ and beamsplitter loss rate $\g_{\rm BS} = 1 - \eta_{\rm BS}$. Here we fix $\eta_p = 0.99$ and $\eta_d = 0.98$.
    The solid line [white in (a), black in (b)]
    marks $\gho{\cF}_\textrm{HOM} = \dir{\cF}_\textrm{HOM}$; the dotted blue line in (b) is the only-beamsplitter-loss locus of \eqref{eq:HOM_bsLoss_curve}.
    }
    \label{fig:HOM_fidComp}
\end{figure}

These results simplify considerably in the desired limit with equal TMSV squeezing parameters, $\lambda_0 = \lambda_1 = \lambda$.
In this case, 
we can use the transpose trick to map the direct implementation onto an effective ghost implementation such that both cases are captured by the \emph{unified} model of \figref{fig:HOM_circuits}(c). 
In each case one can use the equal $\lambda$ limits of the formulae for $p_\textrm{HOM}$ and $\cF_\textrm{HOM}$ given in Eqs.~(\ref{eq:dirp_HOM}) and (\ref{eq:dirF_HOM}), respectively.
The distinction is whether beamsplitter loss is subsumed into $\eta_s$ or $\eta_i$. The fidelities are compared in \figref{fig:HOM_fidComp}.

As anticipated, the ghost implementation delivers significantly higher fidelity states for small $\lamSq$ and appreciable beamsplitter loss:
\subalign{
    \dir{\cF}_\textrm{HOM} &= (\eta_{\rm BS} \eta_p)^2 + \cO(\lamSq), \\
    \gho{\cF}_\textrm{HOM} &= \eta_p^2 + \cO(\lamSq).
}
As an instructive limiting case, if there is only beamsplitter loss ($\eta_p = \eta_d = 1$) the fidelities are
\subalign{
    \dir{\cF}_\textrm{HOM} &= \eta_{\rm BS}^2, \\
    \gho{\cF}_\textrm{HOM} &= \br{1-\lamSq (1- \eta_{\rm BS})}^4
}
from which one can easily identify the regime in which the ghost implementation ``wins,'' i.e., produces higher fidelity HOM states, as 
\be\label{eq:HOM_bsLoss_curve}
    \gho{\cF}_\textrm{HOM} \geq \dir{\cF}_\textrm{HOM}
    \Longleftrightarrow
    \lamSq \leq \frac{1}{1 + \sqrt{\eta_{\rm BS}}}. 
\ee
Thus, the ghost implementation wins for all beamsplitter efficiencies provided $\lamSq \leq 1/2$. The equal fidelity locus shifts when propagation and detector loss are present, but there is still a substantial region in which the ghost implementation wins. 

In this same limit, 
the product of the heralding probability and output fidelity, a quantity we deem the \emph{yield}, is equivalent in both implementations,
\be
    Y_\textrm{HOM} = p_\textrm{HOM} \times \cF_\textrm{HOM}
    = \br{ 
        \eta_{\rm BS} \p{1-\lamSq} \lamSq 
    }^2.
\ee
Thus, the advantage of the ghost implementation can be understood as an improved filtering of false-positive heralding events. It directly trades raw heralding probability for conditional-state fidelity, making it favorable in regimes where high-fidelity heralded states are the limiting resource; see \appref{app:homYield}.

We emphasize that this fidelity advantage is distinct from the usual HOM-dip visibility. The latter can remain high even for lossy direct generation, since it normalizes away the reduced two-photon rate. For applications such as boosted fusion, however, the full photon-number composition of the ancilla is relevant as loss and multiphoton components affect downstream success probabilities and errors.

\subsection{Impact of unequal squeezing}
Unequal squeezing $\lambda_0 \neq \lambda_1$ impacts the direct and ghost implementations in different ways.
In the direct implementation, the desired process is to locally herald two single-photon Fock states and then interfere them. The conditional fidelity of this process is not sensitive to imbalances in the TMSV squeezing strengths or phases. Rather, the desired HOM interference demands the photons be indistinguishable.
Meanwhile, having $\lambda_0 \approx \lambda_1$ is crucial in the ghost implementation 
as the interference is not simply of photons, but of source-emission histories. Namely, the successful $(1,1)$ idler heralding can occur via two paths: biphoton source $0$ emitted two pairs and source $1$ emitted none, or vice versa. The idler beamsplitter and detection serve to erase which-path information, yet unequal squeezing partially delineates the paths.

This distinction is clearest in the limit of no photon loss. In the direct implementation, local idler heralding prepares the signal input $\ketm{11}$, independent of the phases and relative magnitudes of $\lambda_0$ and $\lambda_1$. The signal beamsplitter transforms the creation operators as in \eqref{eq:HOM_twoPhoton_mapping}, mapping this input to the desired balanced HOM state, with its relative phase set by the beamsplitter convention. In the ghost implementation, by contrast, the idler measurement coherently combines the two source-emission histories, yielding the imbalanced HOM operator $\lambda_0^{* 2} a_0^{\dagger 2} + \lambda_1^{* 2} a_1^{\dagger 2}$ of \eqref{eq:rhoG_HOM} acting on the vacuum. The resulting output state is 
\begin{align}
    \ket{\gho{\psi}_\textrm{HOM}} 
    &\propto \lambda_0^{* 2} \ketm{20}  + \lambda_1^{* 2}  \ketm{02} \\
    &\propto 
    \p{\lambda_0^{* 2} + \lambda_1^{* 2}} \ket{\psi_+} 
    + \p{\lambda_0^{* 2} - \lambda_1^{* 2}} \ket{\psi_-}, \nonumber
\end{align}
where $\ket{\psi_+} = (\ketm{20} + \ketm{02})/\sqrt{2}$ is the target HOM state and  $\ket{\psi_-} = (\ketm{20} - \ketm{02})/\sqrt{2}$ is its orthogonal phase-flipped counterpart.
Thus, imbalanced squeezing provides which-path information leading to a coherent error. Moreover, we see that the ghost implementation requires phase stability between the biphoton sources.

\begin{figure}[ht!]
    \hfill \includegraphics[width=0.6\linewidth]{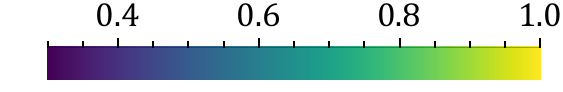} \hfill \\
    \includegraphics[width=\linewidth]{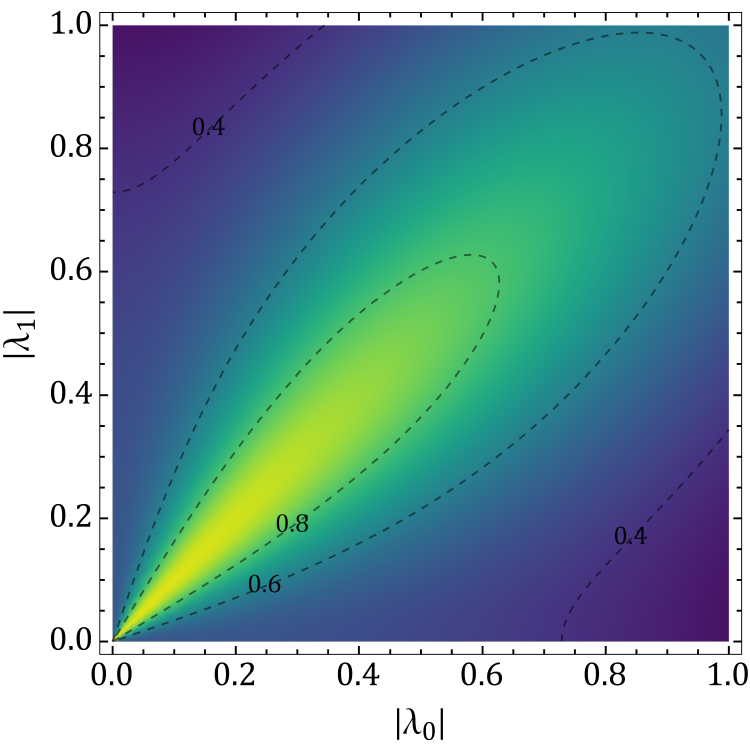}
    \begin{picture}(0,0)
        \put(30, 45){\includegraphics[width=0.16\textwidth]{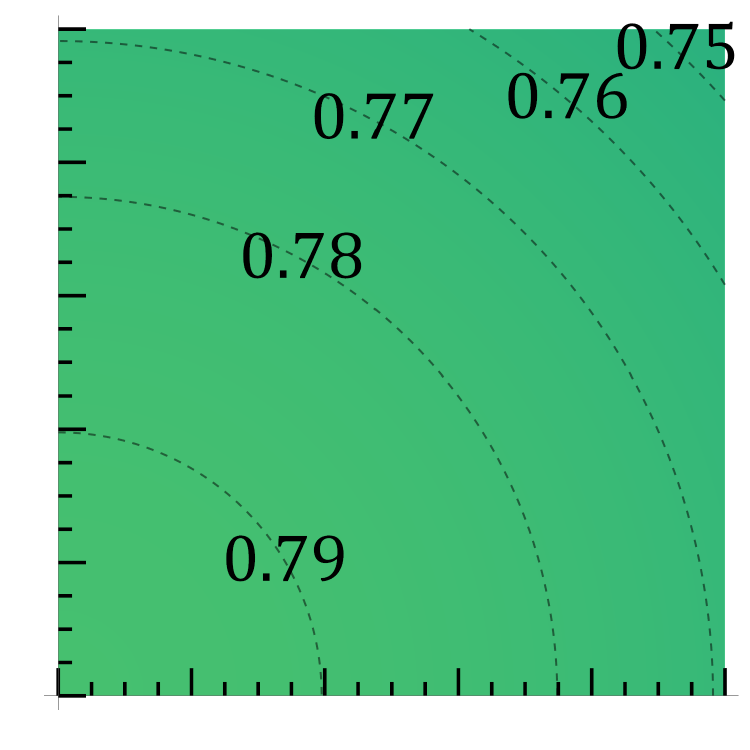}}
    \end{picture}
    \caption{
    HOM fidelity under unequal squeezing.
    Main panel: Ghost implementation fidelity as a function of the squeezing parameters $|\lambda_{0,1}|$ with $\eta_p = 0.99$, $\eta_d = 0.98$, 
    $\g_{\rm BS} = 0.1$, and 
    $\arg(\lambda_0/\lambda_1) = 0.1$.
    Inset: corresponding direct implementation fidelity using the same scale and axes. Both plots are symmetric under $\lambda_0 \leftrightarrow \lambda_1$.
    }
    \label{fig:HOM_lambdaComp}
\end{figure}

This distinction persists when photon loss is included. In the direct implementation, source imbalance primarily changes the probability of successful heralding; conditioned on success, the HOM fidelity is largely insensitive to the individual $\lambda_j$ and to their ratio. In the ghost implementation, source imbalance still provides partial which-path information and induces a coherent error, making the fidelity appreciably dependent on the relative magnitudes and phases of $\lambda_0$ and $\lambda_1$. As shown in \figref{fig:HOM_lambdaComp}, however, this sensitivity does not eliminate the ghost advantage. In the interferometric-loss-dominated regime plotted there, the direct fidelity is bounded above by $(\eta_p \eta_{\rm BS})^2 \approx 0.79$ due to the lossy signal beamsplitter, whereas the ghost implementation achieves substantially higher fidelity for small-to-modest squeezing strengths that are sufficiently well matched.

\section{Bell-state generation}\label{sec:HBSG}
\begin{figure*}[ht!]
    \subfloat[(a)]{\includegraphics[width=0.4\linewidth]{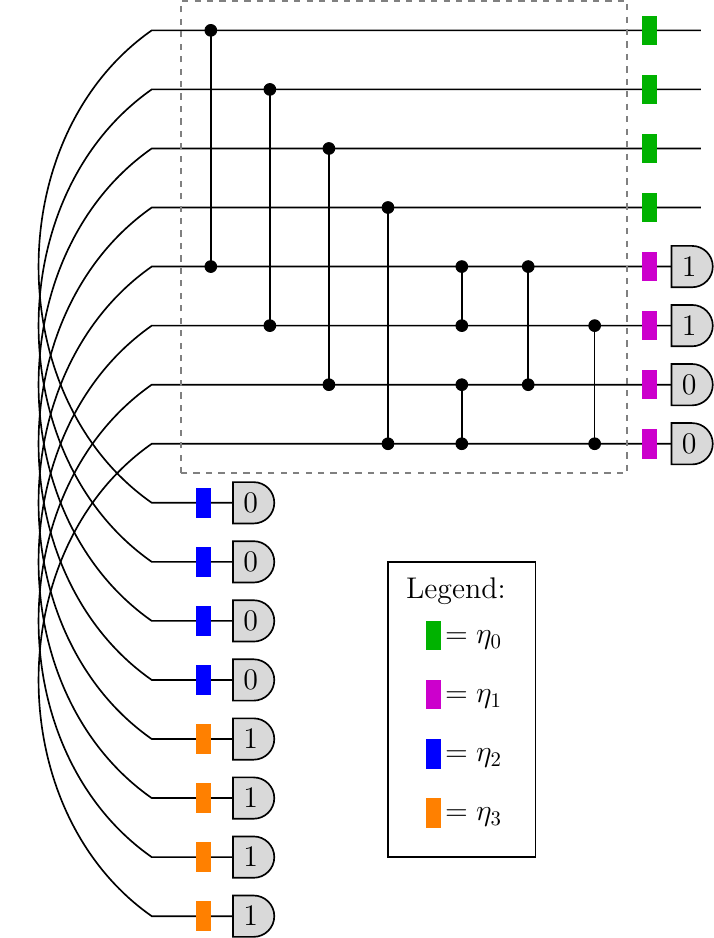}}
    \hfill
    \subfloat[(b)]{\includegraphics[width=0.4\linewidth]{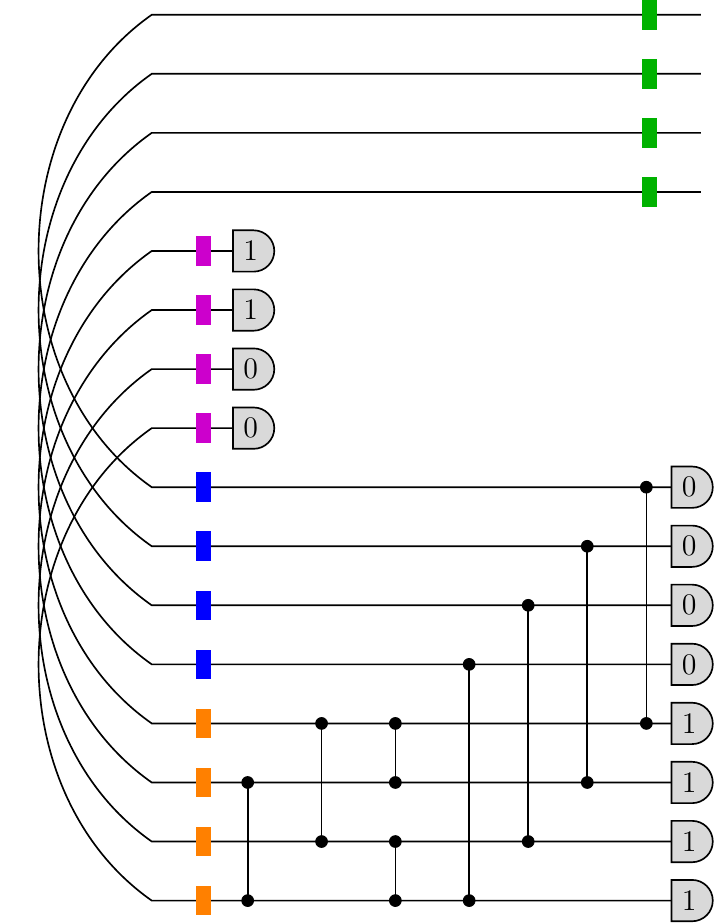}}
    \caption{
        Heralded Bell-state generation using the (a) direct and (b) ghost implementations. The boxed interferometer in (a) is the standard four-photon, eight-mode Bell-state-generation circuit; in the ghost implementation its transpose is applied to the idler modes. The lumped efficiencies $\eta_0, \cdots \eta_3$ (see legend) have implementation-dependent values; see \eqref{eq:BellEtas}. The representative click patterns $\sigvec = (1100)$ and $\ivec = (00001111)$ are shown.
    }
    \label{fig:Bell_circuits}
\end{figure*}

We now consider heralded Bell-state generation, for which many linear-optical schemes are known \cite{forbes2025heralded,randles2026limits}. 
For concreteness, we focus on the representative four-photon, eight-mode scheme shown in \figref{fig:Bell_circuits}(a), together with its ghost implementation in \figref{fig:Bell_circuits}(b). In the standard prescription, four single photons are injected into either the top or bottom four modes of the interferometer, $U$. For the direct implementation studied here, we instead use $M=8$ TMSV sources, which provides additional input-pattern flexibility and enables a unified comparison with the ghost implementation. The rationale for this choice, and a comparison with the more standard $M=4$ direct implementation, are given in \appref{app:FullHBSGResults}, where we show that the additional input-pattern flexibility compensates for the small fidelity cost in the regimes of interest.

The basic ideas and notation are analogous to HOM-state generation, but here success requires heralding on both the signal and idler modes. The valid idler-mode heralding patterns are the $N_\iota = 2^4$ bit-complement patterns $(b_0, b_1, b_2, b_3, \bar{b}_0, \bar{b}_1, \bar{b}_2, \bar{b}_3)$, with $b_j \in \{0, 1\}$ and $\bar{b}_j = 1 - b_j$. 
The valid signal-mode heralds are the $N_\sigma = 6$ two-photon detections on disjoint detectors: $(1100), (0011), (1001),(0110), (1010), (0101)$.
Taking $\ivec = (00001111)$ as a representative idler pattern, the first two, middle two, and last two signal patterns herald
$\ket{\Phi^+} = (\ketm{1010} + \ketm{0101})/\sqrt{2}$,
$\ket{\Psi^-} = (\ketm{1001} - \ketm{0110})/\sqrt{2}$,
and
$\ket{\chi^+} = (\ketm{1100} + \ketm{0011})/\sqrt{2}$ (a permuted dual-rail Bell state),
respectively. 
Other valid idler patterns give the same grouping up to known click-pattern-dependent phases. We therefore carry out the calculation for $\sigvec = (1100)$ and $\ivec = (00001111)$, with the remaining cases related by symmetry as described in \appref{app:GHZPermutations}.

\subsection{Unified model for equal squeezing}
For this circuit, assuming beamsplitters acting on a given mode separation have equal efficiencies and all signal modes suffer equal propagation loss, all direct signal mode loss can be commuted to the end. Thus, as was the case for HOM-state generation, for equal $\lambda$ values we can use the transpose trick to map the direct implementation onto the ghost one by changing how the lumped efficiencies $\eta_0, \cdots, \eta_3$ depend on underlying component efficiencies. Hence, assuming equal detector efficiencies, \figref{fig:Bell_circuits}(b) serves as a unified model here.
In particular, we take the implementation-specific lumped efficiencies to be
\be\label{eq:BellEtas}
    \begin{pmatrix}
        \eta_0\\ \eta_1\\ \eta_2\\ \eta_3
    \end{pmatrix}
    = \begin{cases}
    (\eta_T \eta_p, \eta_T \eta_{\rm BS}^2 \eta_p \eta_d, \eta_d, \eta_d), 
    & \textrm{direct} \\
    (\eta_p, \eta_d, \eta_T \eta_p \eta_d, \eta_T \eta_{\rm BS}^2 \eta_p \eta_d), 
    & \textrm{ghost}
    \end{cases},
\ee
where, analogous to~\eqref{eq:HOM_etas}, $\eta_p$ and $\eta_d$ are propagation and detector loss, respectively. 

Here we split beamsplitter loss into two kinds. Namely, referring to the direct implementation, we consider the transverse beamsplitters linking modes $(j,j+4)$ for $j=0,\cdots,3$ and the ``shorter'' beamsplitters on the heralding signal modes to have efficiencies $\eta_T$ and $\eta_{\rm BS}$, respectively. 
We do this specific separation because in the frequency domain, where the ghost implementation should be especially beneficial, the circuit can be implemented in a hybrid spatial-frequency manner: the transverse beamsplitters are implemented with a single (low loss) spatial beamsplitter and the others are implemented with lossier active frequency beamsplitters~\cite{munoz2026modeling,randles2026limits}. 

We thus proceed assuming equal squeezing strengths, $\lambda$, so that we can leverage this unified model and perform a single effective ghost implementation calculation that captures both implementations.
(We consider unequal squeezing in~\secref{sec:unequalSqueezing_GHZ}, where we find, as in the HOM case, that its dominant effect is to induce coherent errors in the ghost implementation.)
Backpropagating $\ivec$ through the transposed unitary on the idler modes, one finds the unnormalized output state
\be
    \tilde{\rho}_{\ivec} = p_{\ivec} \rho_{\ivec}
    = \Tr_i\!\p{ \rho_{\sigvec} \hat{\Pi}^{(i)}_{u^*} }, 
\ee
which is ideally proportional to the Bell-state $\ketbra{\Phi^+}{\Phi^+}$. Here 
\be 
    \hat{\Pi}^{(i)}_{u^*} = U_i^* \kbm{\ivec}{\ivec} U_i^T 
\ee
and 
$\rho_{\vec{\sigma}}$ is the full lossy input state post signal mode heralding, i.e., its unnormalized counterpart is
\be
    \tilde\rho_{\sigvec} = p_{\sigvec} \rho_{\sigvec}
    = {}_{s_h}\!\bra{\sigvec} \rho_\tin \ket{\sigvec}_{s_h}
\ee
with $\rho_\tin$ of \eqref{eq:rho0_HOM} with $M=8$ and the loss configuration shown in \figref{fig:Bell_circuits}.

For these click patterns the heralding probability, heralded output state, and fidelity with the target Bell state are 
\begin{subequations}\label{eq:Bell_system}
\begin{align}
    p_\textrm{Bell} &= p_{\sigvec} \times p_{\ivec},
    \\
    \rho_\textrm{Bell} &= \rho_{\ivec},
    \\
    \cF_\textrm{Bell} &= \bra{\Phi^+} \rho_\textrm{Bell} \ket{\Phi^+}.
\end{align}
\end{subequations}
Meanwhile, because of the various successful heralding click patterns, the overall probability of generating a Bell-like state increases to
\be
    p^{(\textrm{tot})}_\textrm{Bell} 
    = N_\sigma N_\iota \times  p_\textrm{Bell}
    = 96 \times p_\textrm{Bell}
\ee
(if the permuted Bell state, $\ket{\chi^+}$, is not accepted $N_\sigma= 6 \ra 4$ such that the prefactor $96 \ra 64$). 

\subsection{Comparison}
\begin{figure}[ht!]
    \subfloat[(a)]{%
        \begin{minipage}{\linewidth}
            \centering
            \hspace{5mm}\includegraphics[width=0.7\linewidth]{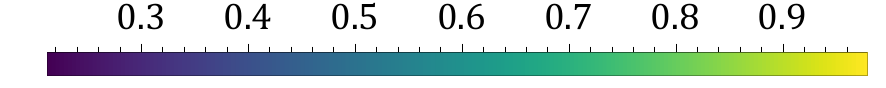}\\[-0.5ex]
            \includegraphics[width=\linewidth, trim={2pt 0 2pt 0}, clip]{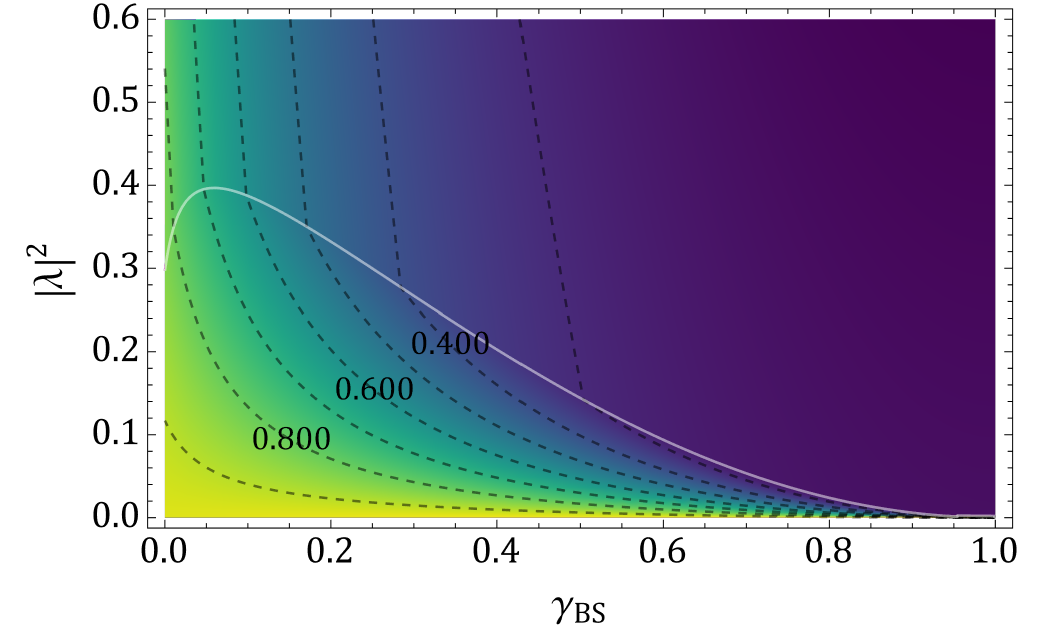}
        \end{minipage}%
    }
    \\
    \subfloat[(b)]{%
        \begin{minipage}{\linewidth}
            \centering
            \hspace{5mm}\includegraphics[width=0.7\linewidth]{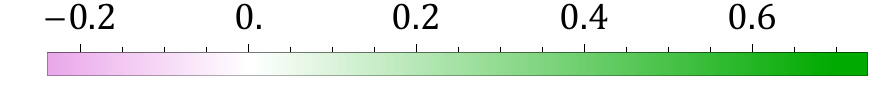}\\[-0.5ex]
            \includegraphics[width=\linewidth, trim={2pt 0 2pt 0}, clip]{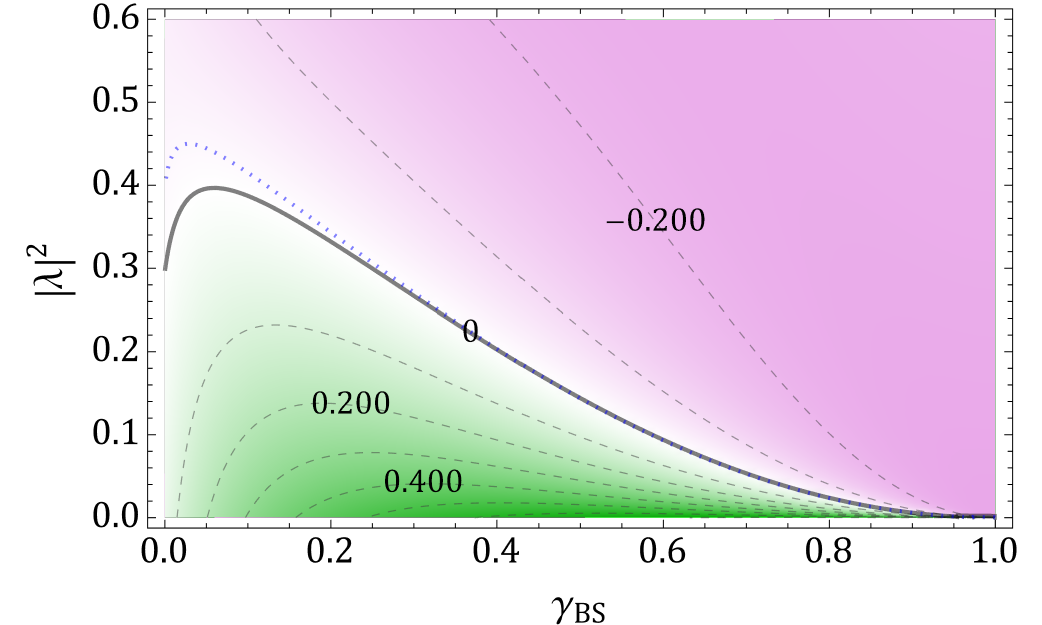}
        \end{minipage}%
    }
    \caption{
    Bell-state generation comparison for equal squeezing. (a) Maximum fidelity $\max(\gho{\cF}_\textrm{Bell}, \dir{\cF}_\textrm{Bell})$ and (b) fidelity difference $\gho{\cF}_\textrm{Bell} - \dir{\cF}_\textrm{Bell}$ versus $\lamSq$ and $\g_{\rm BS}$. We fix $\eta_p = 0.99$, $\eta_d = 0.98$, and use a hybrid spatial-frequency model with low-loss transverse spatial beamsplitter, $\eta_T = 0.99$. The results are qualitatively similar if we take $\eta_T = \eta_{\rm BS}$. The solid line marks $\gho{\cF}_\textrm{Bell} = \dir{\cF}_\textrm{Bell}$; the dotted blue line in (b) is the corresponding locus in the limit of only frequency-beamsplitter loss. 
    }
    \label{fig:BellFidelityCompMain}
\end{figure}

Calculating the quantities in \eqref{eq:Bell_system} is conceptually straightforward 
but lengthy, and is thus deferred to \appref{app:FullHBSGResults}. Here we focus on their behavior in representative parameter regimes, comparing the direct and ghost fidelities for equal squeezing in \figref{fig:BellFidelityCompMain}. 
As in the HOM example, the ghost implementation delivers substantially higher-fidelity resource states for small $\lamSq$ and appreciable interferometric loss, provided $\g_p, \g_d$ are minimal.

Expanding the fidelities for small $\lamSq$, one obtains 
\subalign{
    \dir{\cF}_\textrm{Bell} &= \p{ \frac{\eta_p \eta_T}{2- \eta_p \eta_d \eta_T \eta_{\rm BS}^2 } }^2 + \cO(\lamSq), \\
    \gho{\cF}_\textrm{Bell} &= \p{ \frac{\eta_p}{1- \g_d \eta_{\rm BS}^2 } }^2 + \cO(\lamSq).
}
These expressions are structurally quite different as is manifest in the limit of only interferometric loss ($\eta_p, \eta_d \ra 1$) for which, to leading order in $\lamSq$, $\dir{\cF}_\textrm{Bell} \sim [\eta_T/(2-\eta_T \eta_{\rm BS}^2)]^2$ and $\gho{\cF}_\textrm{Bell} \sim 1$. The direct Bell state suffers effective output loss from the lossy signal interferometer and false-positive signal heralds. In the ghost implementation, the delivered Bell modes avoid this direct loss, while false-positive idler heralds are suppressed at low $\lamSq$.

\section{Scaling across resource state size}\label{sec:scaling}
We now analyze how the utility of the ghost-unitary construction scales with resource-state size. In particular, we analyze how the ``winning fidelity region'' for the ghost implementation changes depending the size of the resource state generated and the number of TMSV states, $M$, used to generated it. 
This includes presenting the heralding probability and fidelity results for a whole class of $n$-qubit GHZ generation schemes in the limit of only interferometric loss. This class includes the Bell state generation scheme of~\secref{sec:HBSG} as a special case ($n=2$). 
We also consider two smaller examples involving the generation of single photon states.

\subsection[$n$-qubit GHZ state generation]{\texorpdfstring{$\bm{n}$}{$n$}-qubit GHZ state generation}\label{sec:nGHZMain}
Now we analyze the heralded generation of $n$-qubit GHZ states in both implementations using a class of RSG unitaries on $M=4n$ modes~\cite{gimeno2016towards}. 
Just as in the Bell-state generation scheme, many detection patterns are valid and herald successful generation. 
In particular, there are $N_\iota = 2^{2n}$ target idler mode heralding click patterns, $\ivec$, and $N_\sigma = 2^{n}$ target signal mode heralding click patterns, $\sigvec$. Proceeding under the assumption of equal $\lambda$, each of these patterns is equiprobable and thus the overall heralding probability is given by 
\be
     p_{G_n} = N_\iota N_\sigma p_{\ivec} p_{\sigvec}.
\ee
Ideally, each such click pattern heralds a target GHZ state of the form
\be\label{eq:nGHZ}
    \ket{G_n} = \frac{1}{\sqrt{2}} \p{\ket{0}^{\otimes n} + e^{i \phi_n} \ket{1}^{\otimes n} }
\ee
with the dual-rail qubit definitions $\ket{0} = \ketm{01}$,  $\ket{1} = \ketm{10}$,
and $\phi_n$ a click-pattern dependent phase that will be some integer multiple of $\pi/2$. 

Analogous to \figref{fig:Bell_circuits}(b), we model loss using four lumped efficiencies $\eta_0, \cdots, \eta_3$ (with loss rates $\g_j = 1 -\eta_j$) for the top signal, bottom signal, top idler, and bottom idler modes, respectively. 
Working in the Bargmann representation~\cite{bargmann1961hilbert} we are able to derive generic expressions for the unnormalized state $\tilde{\rho}_{G_n}$ in terms of a differential Bargmann operator acting on an analytic Bargmann function. However, to obtain closed-form expressions for arbitrary $n \geq 2$, we perform the implementation comparison in the presence of only interferometric loss,
\be\label{eq:ghz_etas}
    (\eta_0, \eta_1, \eta_2, \eta_3) 
    = \begin{cases}
    (\eta_\uparrow, \eta_\downarrow, 1, 1), 
    & \textrm{direct} \\
    (1, 1, \eta_\uparrow, \eta_\downarrow), 
    & \textrm{ghost}
    \end{cases}.
\ee
Here $\eta_\uparrow$ and $\eta_\downarrow$ denote the lumped efficiencies of the upper and lower modes of the RSG unitary, providing a common notation for the direct and ghost $n$-GHZ comparisons in this limit.
The corresponding derivations are relegated to~\appref{app:n_ghz}.

\subsubsection{Direct implementation: identically damped GHZ outputs}
In the direct implementation, \eqref{eq:ghz_etas} gives $\g_2 = \g_3 = 0$, corresponding to perfect idler-mode heralding, so a successful idler click pattern, $\ivec$, heralds ideal single photon inputs to the GHZ-generation interferometer. Each valid idler pattern involves one single photon and one vacuum herald per transverse pair, giving $p_{\ivec}=(P_0 P_1)^{2n}$ with $P_0 = 1 - \lamSq$ and $P_1 = (1 - \lamSq) \lamSq$ the respective probabilities of heralding zero and one photon(s) from a lossless TMSV state. Passing these ideal single photons through the lossy signal-mode interferometer, a valid signal-mode pattern occurs with probability $p_{\sigvec} = 2 \br{\eta_\downarrow (2-\eta_\downarrow)/8}^n$.
Accounting for all valid idler and signal patterns gives
\be\label{eq:direct_GHZ_prob}
    \dir{p}_{G_n} = 2 \br{
    \eta_\downarrow (2-\eta_\downarrow) (1-\lamSq)^4 |\lambda|^4
    }^n.
\ee

The more important point is the structure of the conditional output state. On a successful signal herald, the output is not merely close to the target GHZ state; under the lumped loss model considered here it is exactly equivalent to the ideal state followed by independent, uniform amplitude damping. Namely,
\begin{empheq}[box=\eqbox]{equation}\label{eq:direct_GHZ_identically_damped}
    \dir{\rho}_{G_n} = \cE^{\otimes 2 n}_{\eta_e}(\ketbra{G_n}{G_n}),
    \qquad
    \eta_e=\frac{\eta_\uparrow}{2-\eta_\downarrow},
\end{empheq}
so the fidelity with the target GHZ state is
\be
    \dir{\cF}_{G_n} = \eta_e^n.
\ee
We refer to \eqref{eq:direct_GHZ_identically_damped} as the identically damped property of this circuit family. The numerator of $\eta_e$ captures direct output mode loss, while the factor $2-\eta_\downarrow$ arises from false-positive lossy signal heralding events.
This property is not automatic. Loss on the heralding signal modes will necessarily lead to false-positive click patterns and hence to output states with fewer photons than desired. However, a generic heralded RSG circuit could convert such false positives into correlated loss, rail-dependent loss, or non-qubit leakage such as $\ketm{20}$ or $\ketm{11}$ within a dual-rail pair \cite{randles2026limits}. Here, all such effects combine exactly into uniform amplitude damping of the ideal GHZ state: surviving qubits remain in the dual-rail subspace, lost qubits are erased to $\ketm{00}$.

This observation is independent of the ghost-unitary comparison. The result is a property of the GHZ-generation scheme itself in the limit considered in this subsection, not of the TMSV sources. Thus, whenever the inputs can be modeled as ideal (or all identically damped) single photons, the same identically damped structure applies irrespective of the single-photon source platform. For instance, \eqref{eq:direct_GHZ_identically_damped} still holds for unequal squeezing conditioned on a perfect idler herald; only the heralding probability changes.
The Bell-state scheme considered in this work gives the (already known \cite{gimeno2016towards}) $n=2$ case. 
The $n=3$ version of this property was leveraged by \refref{varnava2008good} in their proposal for efficient linear-optical quantum computing. Reference \citenum{gimeno2016towards} extended the corresponding GHZ-generation circuits to arbitrary $n$, while noting that the identically damped property had not been proven for $n>3$. 
Equation~(\ref{eq:direct_GHZ_identically_damped}) resolves the question affirmatively for the full family. 

\begin{table*}[ht!]
\begin{tabular}{ccccccc}
Resource State & $M$ & $M_{s_h}$ & $N_\tout$ & Direct $\cF$ & Ghost $\cF$ & $Y^{(\textrm{tot})}$ \\ \hline
Single photon & 1 & 0 & 1 & $\eta$ & $(1-q)^2$ &
$\eta P_1$  \\
$\ket{+y}$ qubit & 2 & 0 & 1 & $\eta$ & $(1-q)^3$ & $\eta P_0 P_1$  \\
HOM & 2 & 0 & 2 & $\eta^2$ & $(1-q)^4$ & $\eta^2 P_1^2$ \\
Bell & 8 & 4 & 2 & $\eta_e^2$ & $(1-q_\uparrow)^6/(1+4\Upsilon+2\Upsilon^2)$ &
$3 (\eta_\uparrow \eta_\downarrow)^2 (P_0 P_1)^4$ \\
$n$-GHZ & $4n$ & $2n$ & $n$ & $\eta_e^n$ & $(1-q_\uparrow)^{3n}/\cM_n(\Upsilon)$ &
$2 (\eta_\uparrow \eta_\downarrow)^n (P_0 P_1)^{2n}$
\end{tabular}
\caption{
    Fidelity and yield comparison for the direct and ghost implementations across resource states of increasing size and generation complexity. Here $M$ is the number of modes acted on by the RSG unitary, 
    $M_{s_h}$ is the number of signal modes used for heralding, and $N_\tout$ is the number of output photons. The total yield $Y^{(\textrm{tot})} = \cF \times p^{(\textrm{tot})}$ is identical for the two implementations in the equal-squeezing, interferometric-loss-only setting considered here. 
    Remaining notation is defined in the main text.
}\label{tab:bsLossOnlyComparison}
\end{table*}

\subsubsection{Ghost implementation}
In the ghost implementation, \eqref{eq:ghz_etas} gives $\g_0 = \g_1 = 0$ such that the signal modes experience no loss. The heralded all-signal-mode state is given by coherently adding $2n$ photons to a tensor product of single-mode thermal states, $T(q)$.
Ideal signal mode heralding selects the desired GHZ branch along with erroneous branches induced by thermal-click false-positives. The resulting overall probability of success and fidelity are 
\be\label{eq:ghost_GHZ_prob}
    \gho{p}_{G_n} = 2 \cM_n(\Upsilon) \br{  \frac{ \eta_\uparrow \eta_\downarrow (1-\lamSq)^4 |\lambda|^4}{(1-q_\uparrow)^3} }^{n},
\ee
and 
\be\label{eq:ghost_GHZ_fid}
    \gho{\cF}_{G_n} = \frac{(1-q_\uparrow)^{3n}}{\cM_n(\Upsilon)},
\ee 
respectively with $q_j = \g_j \lamSq$. Here we have defined 
\be
    \cM_n(\Upsilon) = \frac{1}{2} \sum_{c=0}^n M(2n, c) (2 \Upsilon)^{n-c},
\ee
a polynomial in the quantity 
\be\label{eq:UpsilonDef}
    \Upsilon = \frac{\eta_\uparrow q_\downarrow}{\eta_\downarrow (1-q_\uparrow)},
\ee
where $M(2n, c) = \frac{2n}{2n-c} \binom{2n-c}{c}$ is the number of $c$ matchings on a $2n$-cycle graph. The matching polynomial counts the ways an accepted signal-heralding pattern can arise: matched edges correspond to desired branches and unmatched vertices correspond to heralding clicks supplied by thermal-background photons. Again, we see a reversed structure compared to the direct implementation; here false positives lead to additional photons in the output state rather than lost ones.

\subsubsection{Impact of unequal squeezing}\label{sec:unequalSqueezing_GHZ}
Similar to the HOM case, one can show that for the ghost implementation of the $n$-GHZ generation schemes considered here (and indeed many other heralded RSG schemes), the effect of unequal squeezing is to induce coherent errors. 
This can be understood in terms of the unequal squeezing strengths deforming the exact transpose identity (see \appref{app:nGHZ_unequal_lambda}).
Instead of the target GHZ state of \eqref{eq:nGHZ}, one gets an unbalanced state 
\be
    \ket{G_n(\lamvec)} 
    \propto  c_e(\lamvec) \ket{0}^{\otimes n} + e^{i \phi_n} c_o(\lamvec) \ket{1}^{\otimes n},
\ee
where 
\be
    c_e(\lamvec) = \prod_{k=0}^{n-1} \lambda_{2k}, \quad
    c_o(\lamvec) = \prod_{k=0}^{n-1} \lambda_{2k+1}
\ee
are aggregate amplitudes for the even and odd modes, respectively. How similar $c_e$ and $c_o$ are dictates the quality of the output state, namely, 
\be\label{eq:ghost_fidelity_unequal_squeezing}
    \gho{\cF}_{G_n}(\lamvec) = \frac{1}{2} + \frac{\Re(c_e c_o^*)}{ |c_e|^2 + |c_o|^2 }.
\ee
Meanwhile, the direct implementation fidelity is largely insensitive to source imbalance conditioned on successful local heralding. In particular, in the absence of idler-mode loss, unequal $\lambda$ only impacts the heralding probability. Thus, source imbalance only degrades the heralded state and its fidelity in combination with idler-mode loss as captured by the idler-side thermal factors $q$.

\subsection{Single-photon examples}
As much simpler examples, we consider the following two situations: 
(i) heralded single-photon generation, where we imagine using the transpose trick for a lossy single-mode phase shifter with efficiency $\eta$, 
and 
(ii)
the preparation of the non-trivial dual-rail qubit state, $\ket{+y} = (\ketm{01} + i \ketm{10})/\sqrt{2}$, as can be accomplished using the HOM setup but heralding on the detection pattern $(0,1)$.
Admittedly, the former example is of no use in isolation, 
nonetheless it serves as illustrative example.
The resulting heralding probabilities and fidelities are reported in the following subsection. They can be derived analogously to the HOM case (see \appref{app:SPSs}).

\subsection{Fidelity comparison and scaling}
The results for each RSG example considered are summarized in Table~\ref{tab:bsLossOnlyComparison}. We report the direct and ghost fidelities, along with the total yield $Y^{(\textrm{tot})} = \cF \times p^{(\textrm{tot})}$. In the equal-squeezing, interferometric-loss-only setting considered here, the yield is identical for the two implementations. Thus, it is reported once and can be used to obtain the corresponding overall heralding probabilities, $p^{(\textrm{tot})}$.
The notation follows the preceding sections. For the single-photon, qubit, and HOM rows, $\eta = 1-\g$ is the relevant component efficiency 
and $q = \g\lamSq$.
For the Bell and $n$-GHZ rows, $\eta_\uparrow$ and $\eta_\downarrow$ denote the lumped efficiencies of the upper and lower modes of the RSG unitary, 
$q_{\uparrow,\downarrow} = \g_{\uparrow,\downarrow} \lamSq$, 
$\eta_e = \eta_\uparrow/(2- \eta_\downarrow)$, 
and
$\Upsilon = \eta_\uparrow q_\downarrow/[\eta_\downarrow (1-q_\uparrow)]$. 
The Bell row is the $n=2$ member of the GHZ family up to additional accepted signal-heralding patterns when accepting permuted Bell states, which changes the GHZ yield prefactor from $2$ to $3$.

The table exposes a common structure. Here, the direct fidelity is the relevant effective efficiency raised to the number of output photons, $N_\tout$, reflecting the identically damped property of these schemes. 
The ghost fidelities contain a factor $(1-q)^{M-M_{s_h}+N_\tout}$, where $q$ is the relevant output-mode thermal parameter.
When signal-mode heralding is present, the additional denominator polynomial in $\Upsilon$ counts false-positive heralds in which thermal-background photons reproduce the accepted click pattern.
This highlights the central scaling contrast. In the direct case, the output fidelity is fixed by interferometric loss.
Meanwhile, in the ghost case, the fidelity can be increased by reducing $\lamSq$, at the cost of lowering the heralding probability. 

\begin{figure}[ht!]
    \subfloat[(a)]{%
        \begin{minipage}{\linewidth}
            \centering
            \hspace{5mm}\includegraphics[width=0.9\linewidth, trim={30pt 0 0 0}, clip]{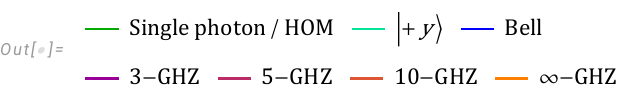}\\[-0.5ex]
            \includegraphics[width=\linewidth]{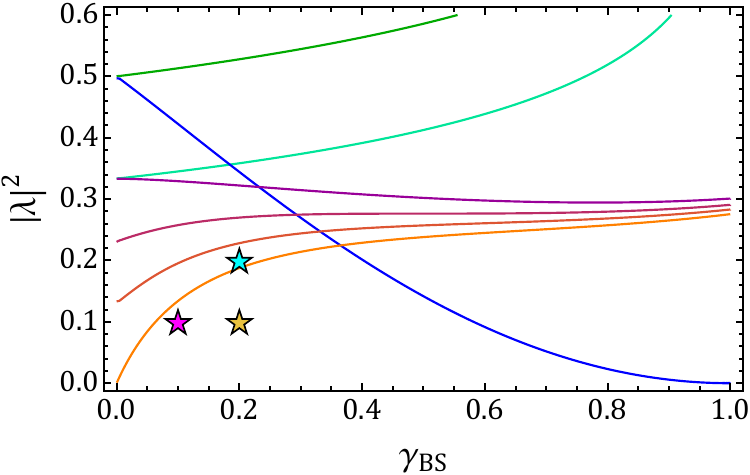}
        \end{minipage}%
    }
    \\
    \subfloat[(b)]{\includegraphics[width=\linewidth]{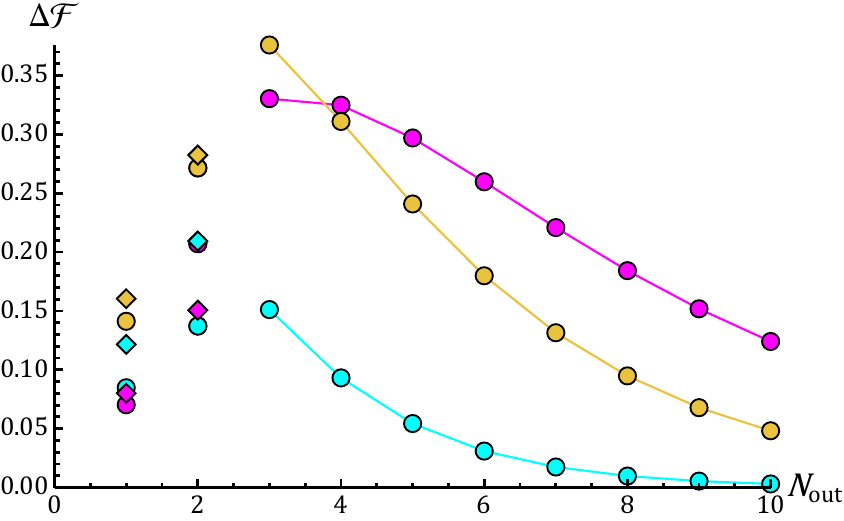} }
    \caption{
    Scaling of the ghost advantage with resource-state size.
    (a) Equal fidelity contours $\gho{\cF} = \dir{\cF}$ for the states in Table \ref{tab:bsLossOnlyComparison}. In each case, the ghost (direct) implementation yields higher fidelity resource states below (above) the contour. The $\infty$-GHZ curve denotes the $n\to\infty$ limit (see \appref{app:nGHZ_ghost}).
    (b) Fidelity difference $\Delta\cF = \gho{\cF} - \dir{\cF}$ versus output photon number $N_\tout$ for the three starred points in (a) that are chosen to be representative of frequency-domain implementations \cite{munoz2026modeling}. 
    Diamonds denote non-dual-rail-qubit resources; connected points with $N_\tout \geq 3$ correspond to GHZ states. 
    }
    \label{fig:trick_scaling_plots}
\end{figure}

Figure~\ref{fig:trick_scaling_plots} illustrates the scaling of the ghost advantage. In panel (a), the ghost-favorable region is substantial for small resources, but shrinks with increasing $N_\tout$. Panel (b) shows that the resulting fidelity gains can nevertheless be sizable at representative frequency-domain operating points. The decrease with $N_\tout$ is expected: both implementations have fidelities that scale approximately exponentially with output photon number, so even a fixed per-mode advantage translates into a shrinking absolute fidelity difference for larger states.
The plotted Bell and GHZ contours use the idealized hybrid-frequency limits $\eta_d = \eta_p = \eta_T = 1$. Their structural difference arises because the GHZ circuit requires an additional beamsplitter on each output qubit to get a $Z$-basis GHZ state, 
so $\eta_\uparrow = \eta_{\rm BS}$, whereas $\eta_\uparrow = 1$ for the Bell-state scheme. 

\section{Ghost implementation benefits for QIP applications}\label{sec:qip_applications}
\begin{figure}[ht!]
    \subfloat[(a)]{\includegraphics[width=0.95\linewidth]{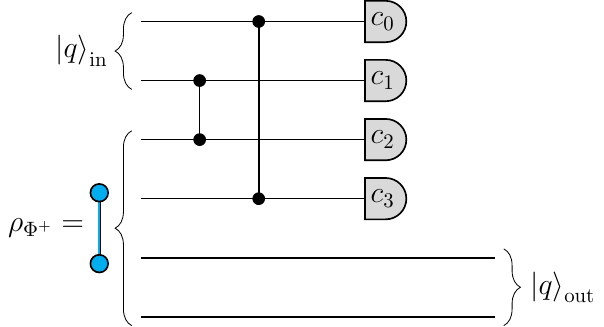}} \\
    \subfloat[(b)]{\includegraphics[width=0.75\linewidth]{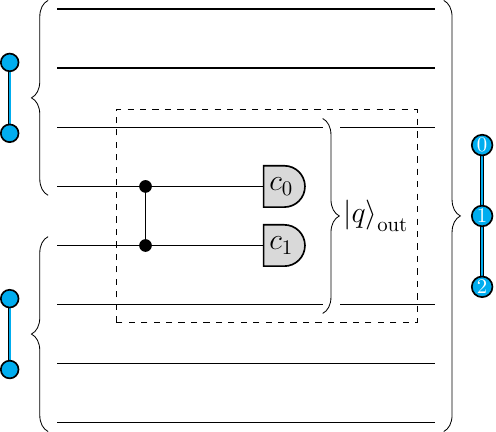}}
    \caption{Downstream primitives used to test how Bell-state errors propagate. (a) Single-qubit teleportation through a noisy Bell state using a type-II fusion. (b) Type-I fusion (dashed box) of two noisy Bell states to generate a 3-GHZ state. Successful click patterns herald the output up to known Pauli-frame corrections. The ideal Bell and 3-GHZ states are depicted as graph states up to local-Clifford equivalence.
    }
    \label{fig:teleportation_and_TI_fusion} 
\end{figure}

We have shown that, when compared with the direct implementation, the transpose trick leads to higher-fidelity resource states in parameter regimes of relevance for current photonic devices, especially within frequency-domain implementations. Now we show that these improved resource states can translate into improved performance in representative QIP primitives. In particular, we consider the tasks of teleporting a single-qubit state through a Bell state and of generating a 3-GHZ state via the fusion of two Bell states, see \figref{fig:teleportation_and_TI_fusion}. Such analysis is crucial for gauging the utility of the trick as part of larger protocols and tracking how multiphoton errors, the leading ghost-unitary output error, cascade in subsequent operations.
Here we show that for these representative primitives, the ghost-generated multiphoton errors can remain manageable in certain regimes and one retains the benefit of the ghost-generated resource state. Further analysis would be necessary to see whether other downstream protocols benefit.

\subsection{Single-qubit teleportation}\label{sec:qubit_tele}
The teleportation of a single photonic qubit $\ket{q} = \alpha \ket{0} + \beta \ket{1}$ through a Bell state can be probabilistically achieved using a type-II fusion \cite{browne2005resource} as illustrated in \figref{fig:teleportation_and_TI_fusion}(a). 
Therein, the detection click pattern $\cC = (c_0, c_1, c_2, c_3)$ is used to herald whether the fusion succeeded. 
For concreteness, we focus on the case where the Bell state used is a noisy $\ket{\Phi^+}$ state prepared using either implementation. Pauli-frame tracking can be used to account for other Bell state inputs.
Under an ideal operation, the fusion is successful for click patterns in $\cC_0 = \{ (1100), (0011) \}$ or $\cC_Z = \{ (1010), (0101) \}$, and fails elsewise. 
The corresponding output, with probabilities summed over each class of click patterns, is
\be
    \ket{q}_\tout = \begin{cases}
        \ket{q}, & \cC \in \cC_0,~p_0=1/4 \\
        Z \ket{q}, & \cC \in \cC_Z,~p_Z=1/4 \\
        \textrm{non-qubit}, & \textrm{elsewise},~p_\textrm{fail}=1/2 \\
    \end{cases}.
\ee

Here we contrast the performance of the teleportation in terms of fidelity of the output qubit and probability of success when using Bell states generated in the the direct and ghost implementations. 
To make the calculation analytically tractable, we take the limit of only interferometric loss for the Bell-state generation. We take the fusion to be performed with beamsplitters of equal efficiency to those used in the Bell-state generation, $\eta_{\rm BS}$, and we take the input qubit to be perfect. 

\subsubsection{Direct implementation}
Here the input Bell state is uniformly amplitude damped, $\rho_{\Phi^+} = \cE^{\otimes 4}_{\eta_e}(\ketbra{\Phi^+}{\Phi^+})$, so has no multiphoton contributions. Thus, the type-II fusion is loss tolerant in that it will only succeed if the fused modes were in the desired two-photon subspace.
Accordingly, the output qubit modes inherit the amplitude damping of the Bell state such that successfully heralded output states are
\be
    \dir{\rho}_q = \begin{cases}
        \cE^{\otimes 2}_{\eta_e}(\ketbra{q}{q}), & \cC \in \cC_0 \\
        \cE^{\otimes 2}_{\eta_e}(Z \ketbra{q}{q} Z), & \cC \in \cC_Z 
    \end{cases}
\ee
with probabilities diminished by $\eta_{\rm BS}$ and $\eta_e \eta_{\rm BS}$ from loss on the fused qubit and Bell modes, respectively. Thus, the fidelity of the output qubit is 
\be
    \dir{\cF}_q = \eta_e = \frac{\eta_\uparrow}{2 - \eta_\downarrow} 
\ee
and the combined probability of success is
\be
    \dir{p}_\textrm{succ} = \dir{p}_0 + \dir{p}_Z 
    = \frac{1}{2} \eta_{\rm BS}^2 \eta_e.
\ee
Here we simply track the potential $Z$ correction, we do not account for any loss that may be present if it was to be implemented.

\subsubsection{Ghost implementation}
In the ghost implementation, multiphoton contributions of the Bell state together with fusion loss can induce false-positive fusion successes. Even with no signal mode loss during the Bell-state generation ($\g_0 = \g_1 = 0$), the presence of loss on the upper idler modes, $\eta_2 = \eta_\uparrow$, results in an infinite tower of such false-positives and corresponding increasingly suppressed multiphoton terms in the output state (see \appref{app:teleportation_derivations}). 
However, with only interferometric loss, $\eta_\uparrow = \eta_T$, in the hybrid-frequency implementation one can reasonably take $\eta_T \ra 1$ and obtain concise, interpretable expressions.
Namely, in these limits, one finds the output state for $\cC \in \cC_0$ to be 
\be\label{eq:ghostTeleportedQubit}
    \gho{\rho}_q = \frac{P_q \ketbra{q}{q} + P_{\rm mm} \rho_{\rm mm} + P_{11} \kbm{11}{11} }{ P_q + P_{\rm mm} + P_{11} },
\ee
which is simply a mixture of 
the target qubit state $\ketbra{q}{q}$, 
the maximally mixed dual-rail-qubit state $\rho_{\rm mm} = (\kbm{10}{10} + \kbm{01}{01})/2$,
and the multiphoton contribution $\kbm{11}{11}$ with weights
\subalign{
    P_q &= \eta_\downarrow^2,
    \\
    P_{\rm mm} &= 8 \eta_\downarrow q_\downarrow \g_{\rm BS},
    \\
    P_{11} &= 2 q_\downarrow (\eta_\downarrow + 4 q_\downarrow \g_{\rm BS}),
}
where $\eta_\downarrow = \eta_{\rm BS}^2$ here. 
For $\cC \in \cC_Z$ one gets this same state up to a known Pauli-$Z$ on the qubit branch. 
The corresponding fidelity of the output qubit and overall success probability of the fusion are 
\subalign{
    \gho{\cF}_q &= \frac{ P_q + P_{\rm mm}/2 }{ P_q + P_{\rm mm} + P_{11} }, \\
    \gho{p}_\textrm{succ} &= \frac{\eta_{\rm BS}^2}{2} 
    \frac{ P_q + P_{\rm mm} + P_{11} }{ P_q + 4 \eta_\downarrow q_\downarrow + 2 q_\downarrow^2 }.
}

\subsubsection{Comparison}
\begin{figure}[ht!]
    \subfloat[(a)]{\includegraphics[width=0.95\linewidth]{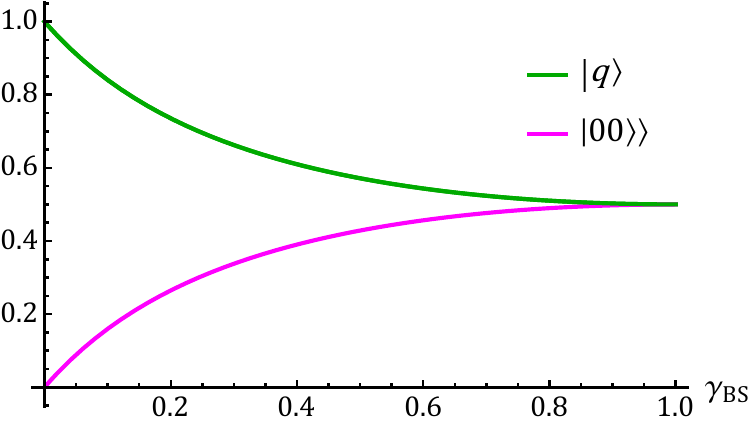}} \\
    \subfloat[(b)]{\includegraphics[width=0.95\linewidth]{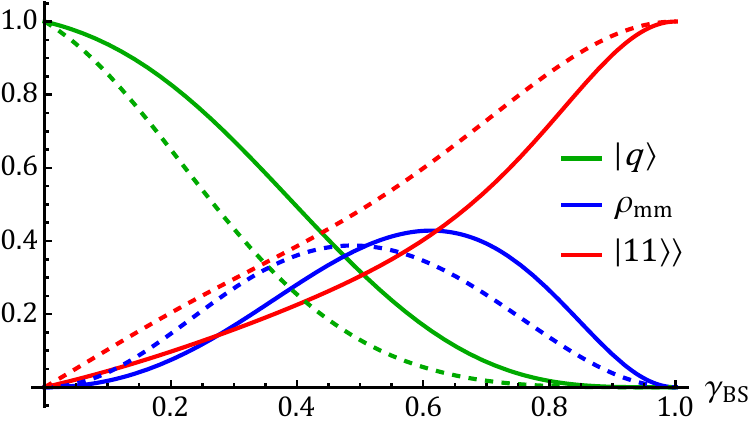}}
    \caption{Output-state populations for single-qubit teleportation versus $\g_{\rm BS}$ using (a) direct- and (b) ghost-generated Bell states. Here we assume a hybrid-frequency implementation with only frequency-beamsplitter loss. The populations in (b) depend on $\lamSq$ for which we fix two values: $\lamSq = 0.1$ (solid curves) and $\lamSq=0.25$ (dashed curves).}
    \label{fig:teleportation_populations} 
\end{figure}

Because of the simple structure of the output qubit states in the $\g_\uparrow = 0$ limit, in \figref{fig:teleportation_populations} we plot the populations of all the states that appear in each implementation. We see that the teleported qubit state has very different structure in the two implementations. In the direct case one only gets amplitude damping leading to a vacuum contribution error. Meanwhile, in the ghost case one gets a single-dual-rail-qubit depolarizing channel term, $\rho_{\rm mm}$, as well as an extra photon term, but no vacuum contribution. 

Even though the Bell-state generation yield is equal across the two implementations in the considered limit, the teleportation yields, defined as $Y_\textrm{tele} = p_\textrm{succ} \times \cF_q$, are different.
Moreover, in the parameter regimes considered here, the higher-fidelity ghost region is also a higher-yield region for teleportation, so the fidelity improvement is not paid for by a reduced teleportation yield in this primitive; see \appref{app:teleportation_derivations}.

\subsection{Larger resource generation via type-I fusion}\label{sec:typeI_fusion}
A type-I fusion measurement probabilistically combines two resource states of $N$ and $M$ qubits into a resource state of $N+M-1$ qubits
\cite{browne2005resource}. For $N,M \geq 2$, this produces a larger output resource state; the smallest nontrivial example is fusing two Bell states into a $3$-GHZ state, as shown in \figref{fig:teleportation_and_TI_fusion}(b). 
Ideally, the single-click patterns $\cC=(10)$ and $\cC=(01)$ herald $\ket{G_3}$ and $Z_1\ket{G_3}$,  
respectively, each with probability $1/4$; all zero- or two-click outcomes fail and leave the undetected modes outside the qubit subspace.

The success probability and fidelity of the fused output depend on both the noisy Bell inputs and loss in the fusion circuit. While the lossy Bell states admit analytic expressions, propagating their multiphoton components through a lossy type-I fusion is cumbersome. Fusion loss can also create false positives: events that would ideally register as two photons in one detector, and hence fail, may lose a photon and masquerade as single-click successes. Multiphoton components in the input Bell states introduce additional such branches. We therefore treat this example numerically, consistently including detector, propagation, and beamsplitter loss for both direct and ghost Bell resources.
The simulation is performed in two stages: Bell-state generation followed by type-I fusion. Both stages are simulated using Perceval~\cite{heurtel2023perceval}, with the numerical workflow described in \appref{app:fusion_simu_details}. We use comparable nominal parameters to those used in \secref{sec:HBSG}, setting $\eta_p=0.98$, $\eta_T=1$, and $\eta_d=0.98$ in the model of \eqref{eq:BellEtas}. For the fusion stage, we include additional propagation loss with $\eta_{p,f}=0.999$ and take the beamsplitter and detector efficiencies to be the same as in the Bell-state-generation stage.

Importantly, the incoherent character of the multi-mode amplitude damping channel implies that the noisy Bell states do not have coherences between sectors of distinct total photon number. This allows us to decompose them as mixtures over sectors with fixed total photon number, $S$:
\be\label{eq:simulated_bells}
    \rho^{(l)}_\textrm{Bell} = \sum_{S=0,1,2,{>2}} p_S \rho_S
\ee
with $l=d,g$. The populations and state components satisfy $\sum_S p_S = 1$ and $\Tr[\rho_S] = 1$.
The differences between direct and ghost implementations are manifest in the relations between the populations in \eqref{eq:simulated_bells}. In the low-loss limit, one finds 
\begin{subequations}\label{eq:Bell_error_hierarchy}
    \begin{align}
        \dir{\cF}_\textrm{Bell} &= p_2 \bra{\Phi^+} \dir{\rho}_2 \ket{\Phi^+} > p_1 > p_{>2}, \\
        \gho{\cF}_\textrm{Bell} &= p_2 \bra{\Phi^+} \gho{\rho}_2 \ket{\Phi^+} > p_{>2} > p_1.    
    \end{align}
\end{subequations}
With the Bell states at hand, we simulate the type-I fusion. The output state can be similarly decomposed into total-photon-number sectors:
\be\label{eq:simulated_ghzs}
    \rho^{(l)}_\textrm{3-GHZ} = \sum_{S=0,1,2,3,{>3}} p_S \rho_S.
\ee

\begin{figure}[ht!]
    \subfloat[(a)]{\begin{minipage}{\linewidth}
        \centering
        \hspace{7.5mm}\includegraphics[width=0.6\linewidth]{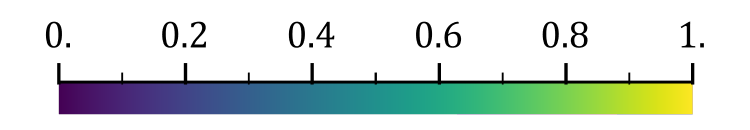}\\[-0.5ex]
        \includegraphics[width=0.9\linewidth]{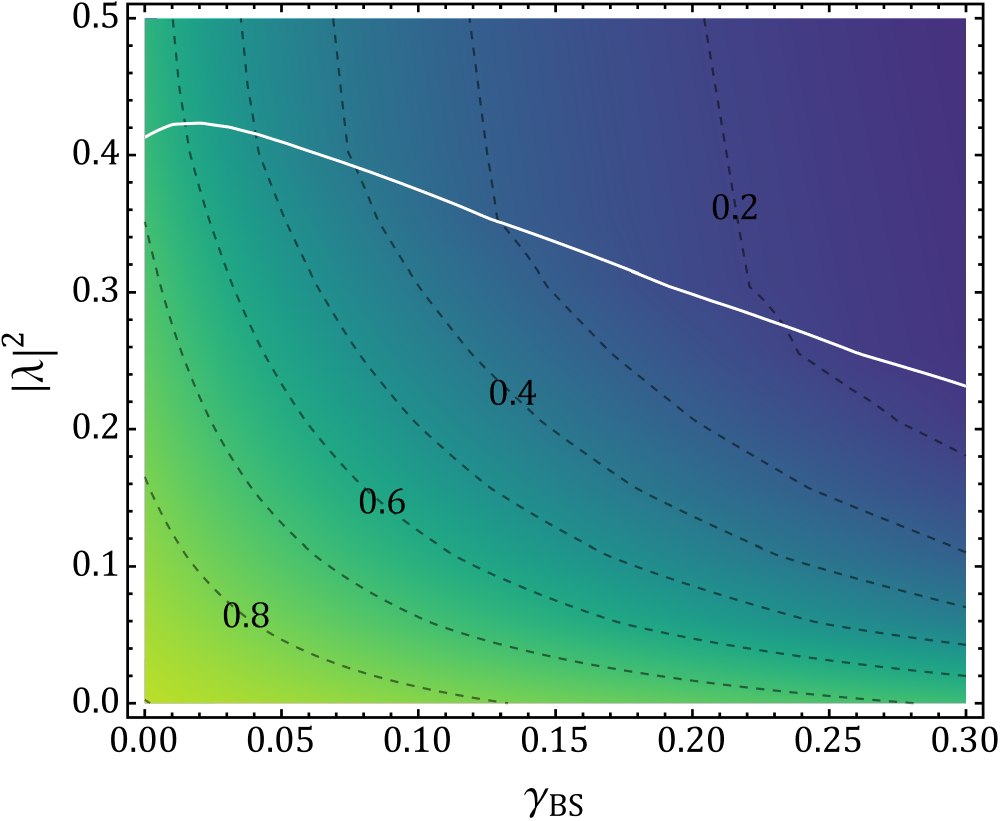}
    \end{minipage}}
    \\
    \subfloat[(b) $\lamSq = 0.108$]{ \includegraphics[width=0.9\linewidth]{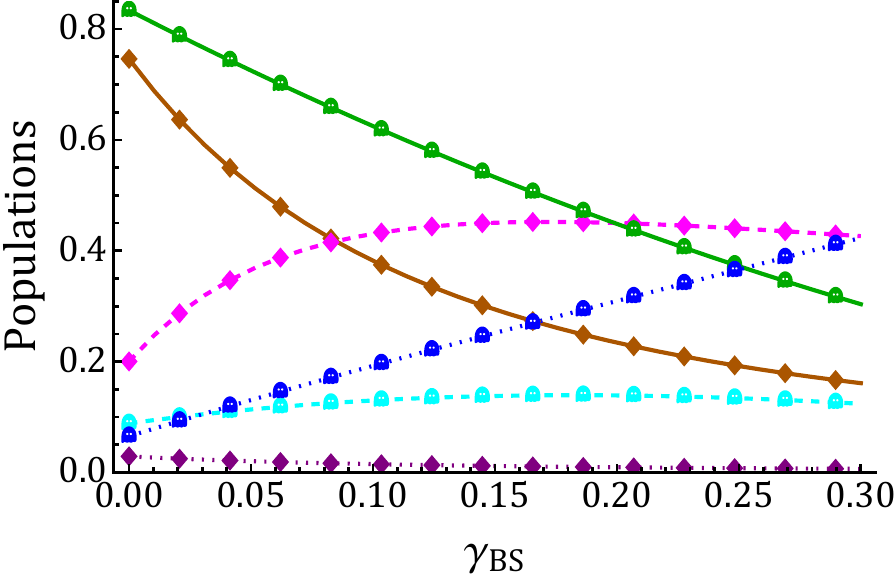}}
    \\
    \subfloat[(c) $\lamSq = 0.255$]{ \includegraphics[width=0.9\linewidth]{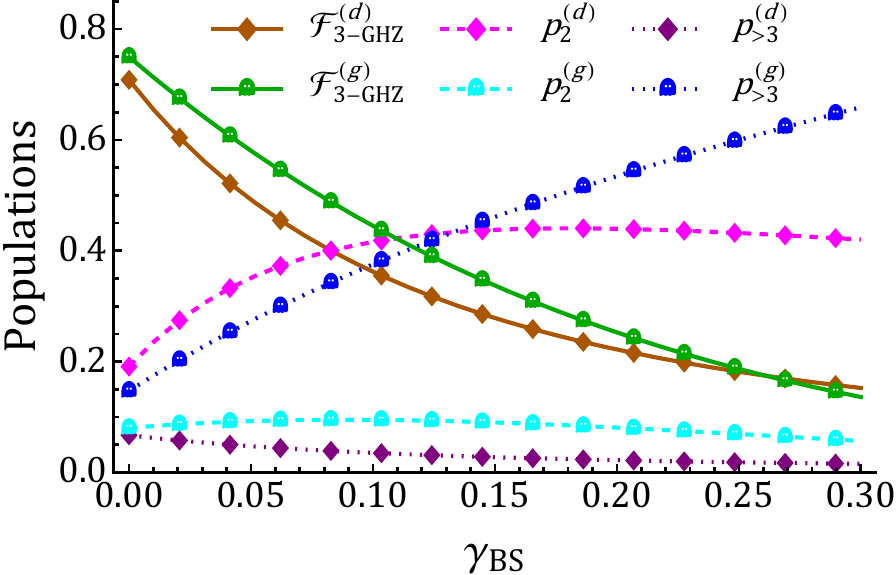}}
    \caption{
    Type-I fusion of two noisy Bell states into a 3-GHZ state. 
    (a) Maximum output fidelity $\max(\dir{\cF}_\textrm{3-GHZ}, \gho{\cF}_\textrm{3-GHZ})$ versus $\g_{\rm BS}$ and $\lamSq$; the white curve marks equal fidelities.
    (b,c) Output fidelity and dominant error-sector populations versus $\g_{\rm BS}$ in each implementation for $\lamSq$ of $0.108$ and $0.255$. Here $p_2$ and $p_{>3}$ are the two-photon and $>3$-photon populations of the 3-GHZ output; see \eqref{eq:simulated_ghzs}. Lines connect all simulated data points; markers are shown on every other point. 
    }
    \label{fig:fusion_results} 
\end{figure}

Figure~\ref{fig:fusion_results}(a) shows the larger of the direct and ghost fidelities for the fusion-generated 3-GHZ state, $\max(\gho{\cF}_\textrm{3-GHZ}, \dir{\cF}_\textrm{3-GHZ})$, as a function of $\g_{\rm BS}$ and $\lamSq$. The solid white curve marks the locus where the implementations have equal fidelity. Consistent with the Bell-state results of \secref{sec:HBSG}, ghost-generated Bell states can deliver higher-fidelity 3-GHZ states over a substantial low-squeezing, low-to-moderate-loss region. This is not automatic: fusion loss can convert multiphoton components of a ghost-generated Bell state into false-positive detections, which could eliminate the fidelity advantage in principle. The results in \figref{fig:fusion_results}(a) show that, in the parameter regimes considered, these multiphoton-induced false positives are sufficiently suppressed for the ghost implementation to remain favorable.

The fusion largely preserves the hierarchical error structure of the input state shown in \eqref{eq:Bell_error_hierarchy}, so the output state inherits it. In the direct implementation, the dominant Bell-state error lies in the single-photon sector, $p_1$, corresponding to one missing photon, which is primarily mapped to the two-photon sector, $p_2$, of the 3-GHZ output under fusion. In the ghost implementation, the dominant Bell-state error is instead an extra-photon contribution captured by $p_{>2}$, which remains a higher-photon-number contribution, $p_{>3}$, after fusion. 
Thus, while the combined effect of multiphoton errors plus loss does not prevent the ghost implementation from being advantageous in some parameter regimes, multiphoton errors remain the dominant error. This behavior is illustrated in \figref{fig:fusion_results}(b) and (c), where $\cF_\textrm{3-GHZ}$, $p_2$, and $p_{>3}$ are plotted as functions of $\g_{\rm BS}$ for two representative values of $\lamSq$ for each implementation.

\section{Discussion and outlook}\label{sec:conclusion}
We have analyzed ghost-unitary implementations of heralded photonic resource-state generation in which the transpose of a target linear-optical transformation is applied to the heralding idler modes rather than the delivered signal modes. The examples studied here show that the transpose trick, enabled by the entanglement structure of MMSV states, is not merely a formal equivalence but a practical noise-engineering tool. When the dominant loss is associated with a transformation that can be shifted to the heralding subsystem, loss that would directly degrade the output resource is instead converted into false-positive heralding events requiring both higher-pair emission and photon loss. In the low-squeezing regime, this can strongly suppress conditional infidelity.

The resource-generation examples clarify both the scope and the tradeoffs of this mechanism. For HOM-state, Bell-state, and $n$-GHZ-state generation, the ghost implementation can trade raw heralding probability for higher conditional state fidelity. In the limit of only interferometric loss, this tradeoff leaves the yield unchanged for the examples considered, making the ghost implementation favorable when high conditional fidelity is the limiting resource. 
Absent such intrinsic filtering, achieving comparable state fidelities from unmitigated heralded sources would otherwise require downstream entanglement purification protocols, which are themselves probabilistic and typically demand significant auxiliary hardware and feed-forward control \cite{pan2003experimental}.
For the $n$-GHZ generation circuits, we also proved that the direct implementation produces identically damped outputs in the setting considered: the noisy heralded state is equivalent to the ideal GHZ state followed by uniform amplitude damping. This extends the known $n=2,3$ cases to the full family.

We further examined how the distinct error structures of the two implementations propagate into downstream photonic QIP primitives. In the direct implementation, the dominant errors are missing-photon components arising from amplitude damping. In the ghost implementation, the dominant errors are instead multiphoton components associated with higher-pair emission and false heralding. For the teleportation and type-I-fusion examples considered here, these multiphoton errors do cascade into later operations, but they do not necessarily eliminate the ghost advantage. Rather, performance depends on how the primitive maps input error structure into output infidelity and success probability.

Ghost implementations are more sensitive to unequal squeezing strengths and relative phases, since they rely on interference between source-emission histories rather than only between already-heralded photons. Moreover, the strongest advantage arises when the dominant loss comes with the linear-optical components and can thus be shifted to the heralding subsystem. Residual non-commuting loss, partial distinguishability, and more general detector errors (e.g., dark counts or efficiency imbalance) are model dependent and require separate analysis.
Even when loss cannot be fully shifted, the transpose identity may still be useful by allowing a target unitary $U = V W$ to be partitioned across signal and idler modes as in \eqref{eq:subunitaryTransposeTrick}, thereby reducing the depth of the delivered-mode circuit. 
Since active frequency-domain transformations can be comparatively lossy, small frequency-bin demonstrations provide a natural near-term testbed for these ideas \cite{lu2023frequency,munoz2026modeling,congia2026fully}. More broadly, these results motivate further study of ghost-unitary constructions in photonic quantum information processing.

\acknowledgments
We thank Mohan Sarovar for useful discussions.
We acknowledge support from the Sandia National Laboratories Laboratory Directed Research and Development program (EPIQ project). Sandia National Laboratories is a multimission laboratory managed and operated by National Technology and Engineering Solutions of Sandia LLC, a wholly owned subsidiary of Honeywell International Inc.~for the U.S.~Department of Energy’s National Nuclear Security Administration under contract DE-NA0003525.

\textbf{AI Acknowledgment}: 
AI tools were used to assist with prose refinement, organization, notation checks, typo detection, and algebraic cross-checks. AI assistance also helped refine human-generated derivations in the Bargmann-representation and $n$-GHZ appendices by suggesting standard identities and clearer presentation strategies. The central physical ideas, calculations, proof strategies, and conclusions were developed and verified by the authors. All AI-generated content was reviewed and edited by the authors.

    
\bibliographystyle{unsrt} 
\bibliography{ghostUnitaryBib}

\clearpage
\onecolumngrid

\appendix
\addtocontents{toc}{\protect\setcounter{tocdepth}{2}} 
\printAppendixTOC 

\section{General formalism}\label{app:general_formalism}
\subsection{Generalized transpose trick}\label{app:gen_transpose_trick} 
Each RSG scheme we consider necessitates a heralding measurement of all the idler modes of the initial MMSV state of \eqref{eq:zeta_lambda}, $\ket{\zeta_{\lamvec}} = \otimes_{j=0}^{M-1} \ket{\textrm{TMSV}(\lambda_j)}$, where we begin without assuming equal $\lambda_j$. 
Here we consider the ghost implementation of some target transposed unitary, $U^T$, on the idler modes.  This also captures the idler-mode heralding in the direct implementations with $U = I$.
Here we denote the $j^{\rm th}$ signal and idler mode creation operators as $\hc{a}_j$ and $\hc{b}_j$, respectively.

\subsubsection{Loss properties} 
As suffices for all RSG examples in the main text (for both the direct and ghost 
implementations), we focus on cases in which all idler-mode loss can be commuted to the left onto the MMSV state and all signal-mode loss can be commuted to the right onto the signal-mode measurement and output state. 
This simplification is enabled by the compatibility of our lumped component-efficiency models, such as homogeneous idler detector efficiencies, with two standard properties of amplitude-damping channels: single-mode loss is multiplicative and uniform loss commutes with linear optics. (These are standard quantum optics properties that can easily be shown; e.g., see App.~C of \refref{randles2026limits}.)
Generic RSG implementations may contain residual loss within the circuit that cannot be commuted to the beginning or end. In such cases, backpropagating the detection POVM is less straightforward: the lossy implementation must be treated as a quantum channel, and the detector POVM is mapped by the corresponding adjoint channel to a generally mixed, non-projective POVM.

\subsubsection{Heralded signal-mode state} 
The unnormalized signal-mode output state upon idler click-pattern $\ivec$ (before any signal mode loss) is 
\be
    \tilde{\rho}_s = \Tr_i\br{U_i^T \cE_i\!\p{\ketbra{\zeta_{\lamvec}}{\zeta_{\lamvec}}} U_i^* \kbm{\ivec}{\ivec}}
    = \Tr_i\br{ \hc{\cE}_i\!\p{\ketbra{u^*}{u^*} } \ketbra{\zeta_{\lamvec}}{\zeta_{\lamvec}}  },
\ee
where $\cE_i$ is the idler-mode loss channel with Kraus operators $K_{\lvec} = \otimes_{j=0}^{M-1} K_{l_j}$ with the single-mode Kraus operators given in \eqref{eq:adChannelKrausOp}. 
Here
\be
    \ket{u} = U \ketm{\ivec}
    = f(\{\hc{b}_j\}) \vac,
\ee
which serves to define the function $f = f_{\ivec}$ of all $M$ idler-mode creation operators $\hc{b}_j$,
and $\ket{u^*}$ is the state obtained by complex conjugating the amplitudes in $\ket{u}$:
\be
    \ket{u^*} = U^* \ketm{\ivec}
    = \bar{f}(\{\hc{b}_j\}) \vac,
\ee
where we use an overbar to denote coefficient-wise conjugation. 

Expanding out the Kraus operators in the adjoint loss channel it follows that
\be
    \hc{\cE}_i\!\p{\ketbra{u^*}{u^*} }
    = \bar{f}(\{\sqrt{\eta_j} \hc{b}_j\}) 
    \br{ \Motimes_{k=0}^{M-1} \tilde{T}_k(\g_k) } 
    \hc{\bar{f}}(\{\sqrt{\eta_j} \hc{b}_j\}),
\ee
where the unnormalized thermal state terms, $\tilde{T}(\g) = \g^{\hc{b} b }$, come from the creation operators terms in the adjoint Kraus operators. The rescaling in $\bar{f}$ comes from $\eta_j^{\hc{b}_j b_j/2}$ terms and using the identity 
\be\label{eq:rescalingIdentity}
    c^{\hc{b} b } g(\hc{b}) = g(c \hc{b}) c^{\hc{b} b },
    \qquad c \in \mathbb{C},
\ee
which follows by induction on $\hc{b}$-monomials plus linearity. 
The initial MMSV state can be decomposed similar to \eqref{eq:zetaMMSV}, yet with unequal squeezing, as
\be
    \ket{\zeta_{\lamvec}} = \cN_{\lamvec}^{1/2} \sum_{\nvec} D_{\lamvec} \ketm{\nvec, \nvec}
\ee
with 
$\cN_{\lamvec} = \prod_{j=0}^{M-1} \p{1-|\lambda_j|^2}$
and 
$D_{\lamvec} = \prod_{j=0}^{M-1} \lambda_j^{\hc{b}_j b_j}$. 
It follows that 
\be
    \tilde{\rho}_s = \cN_{\lamvec}
    \Tr_i\!\bigg[ 
    (I_s \otimes E_i^T)
    \sum_{\nvec, \vec{m}}
    \kbm{\nvec, \nvec}{\vec{m}, \vec{m}}
    \bigg]
\ee
with $E_i^T = D_{\lamvec^{\!*}} \hc{\cE}_i\!\p{\ketbra{u^*}{u^*} } D_{\lamvec}$, where from \eqref{eq:rescalingIdentity} we see that the $D$ operators act simply to rescale the arguments of both $\bar{f}$ and each $\tilde{T}$, 
\be
   E_i^T = \bar{f}(\{r_j \hc{b}_j\}) 
    \br{ \Motimes_{k=0}^{M-1} \tilde{T}_k(q_k) } 
    \hc{\bar{f}}(\{r_j \hc{b}_j\}),
\ee
with $r_j = \sqrt{\eta_j} \lambda_j^*$
and $q_k = \g_k |\lambda_k|^2$.
As $\sum_{\nvec} \ketm{\nvec, \nvec}$ is maximally entangled in each fixed-photon-number sector, we can apply the transpose trick to map the linear transformation from the idler to signal modes and evaluate the partial trace
\begin{empheq}[box=\eqbox]{equation}\label{eq:generalizedTransposeTrick}
    \tilde{\rho}_s = \cN_{\lamvec} E_s
    = \cN_{\lamvec}
    f(\{r_j \hc{a}_j\}) 
    \br{ \Motimes_{k=0}^{M-1} \tilde{T}_k(q_k) }_s
    \hc{f}(\{ r_j \hc{a}_j \})
\end{empheq}
with $\hc{a}_j$ the signal-mode creation operators. We emphasize that the thermal states are now on the signal modes. 
This formula generalizes the transpose trick, showing that nonuniform loss and squeezing can naturally be incorporated along with a detection POVM on half the bipartition.

\subsection{Bargmann representation}\label{app:BargmannRep}
The Bargmann representation \cite{bargmann1961hilbert} is a powerful phase-space description that enables a compact summary of general results (\appref{app:general_Bargmann_formulae})and streamlines the $n$-qubit GHZ state calculations (\appref{app:n_ghz}). It maps pure states to holomorphic functions of complex variables and density operators to holomorphic kernels in independent ket- and bra-side variables. Operators, channels, and measurement maps are then represented by differential operators acting on these functions or kernels. For a single ket-side mode, the creation and annihilation operators are represented as
\subalign{
    a &\mapsto \partial_z, \\
    \hc{a} &\mapsto z.
}
This mapping preserves the bosonic commutation relations, e.g., $\br{a,\hc{a}}=1$ maps to $\br{\partial_z,z}=1$ on holomorphic test functions. The corresponding independent bra-side variables of an operator kernel, the dual creation and annihilation operators, are represented as
$a \mapsto \tilde{z}$,
$\hc{a} \mapsto \partial_{\tilde{z}}$.
See \refref{vourdas2006analytic} for a topical review and \refref{schleich2015quantum} for a broader treatment of quantum optics in phase space.

In the Bargmann representation, traces may be evaluated either as Gaussian coherent-state contractions or, equivalently, as differential contractions at the origin~\cite{bargmann1961hilbert,vourdas2006analytic,schleich2015quantum}. For a single-mode Bargmann kernel
\be\label{eq:nm_op_to_Bargmann}
    \sum_{n,m} A_{nm} \kbm{n}{m} \mapsto 
    A(z,\tilde{z}) = \sum_{n,m} A_{nm}
    \frac{z^n\tilde{z}^m}{\sqrt{n!m!}}
\ee
one has
\be
    \Tr(A) = \sum_n A_{nn}
    = \int_{\mathbb C}\frac{d^2z}{\pi} e^{-|z|^2} A(z,z^*)
    = \left.
        e^{\partial_z\partial_{\tilde{z}}} A(z,\tilde{z})
    \right|_{z,\tilde{z}=0}.
\ee
In the Gaussian integral form, the bra-side variable is evaluated on the coherent-state diagonal, $\tilde{z}=z^*$. In the differential form, $z$ and $\tilde{z}$ remain independent variables and the trace is implemented by the operator $e^{\partial_z\partial_{\tilde{z}}}$. The multimode generalization follows by replacing $\partial_z\partial_{\tilde{z}} \ra \sum_j \partial_{z_j}\partial_{\tilde{z}_j}$, or equivalently by integrating over all traced coherent-state variables.
More generally, we define the contraction operator
\be
    \cK_z(\g)[A] := \left.
        \exp\!\p{
            \g\partial_z\partial_{\tilde{z}}
        }
        A(z,\tilde{z})
    \right|_{z,\tilde{z}=0}.
\ee
For factorized kernels $A(z,\tilde{z}) \ra P(z)Q(\tilde{z})$, this gives
\be
    \cK_z(\g)\br{
        P(z)Q(\tilde{z})
    }
    = \sum_{n=0}^{\infty} \frac{\g^n}{n!} \left.
        \frac{\partial^n P}{\partial z^n} \frac{\partial^n Q}{\partial \tilde{z}^n}
    \right|_{z,\tilde{z} = 0}
    = \sum_{n=0}^{\infty} \frac{\g^n}{n!} P^{(n)}(0) Q^{(n)}(0).
\ee
The trace contraction corresponds to $\g=1$, while $0\le\g<1$ appears naturally in lossy PNRD contractions.

\subsection{Generic direct and ghost Bargmann formulae}\label{app:general_Bargmann_formulae}
Consider a generic heralded RSG scheme for which there is both idler-side and signal-side heralding with respective representative click patterns $\ivec$ and $\sigvec$. (If, as in the HOM example, no signal mode heralding is necessary, then $\sigvec$ does not exist and corresponding steps can be omitted.)
We can directly calculate the function of creation operators, $f_{\ivec}$, characterizing the circuit by propagating the desired single photon input $\ketm{\ivec} \equiv f_\tin \vac$ through the direct circuit implementation:
$f_\tin \ra f_{\ivec} = U f_\tin \hc{U}$.
The ghost implementation equivalent entails backpropagating the POVM element $\kbm{\ivec}{\ivec}$ through the transposed circuit.  

\subsubsection{Direct implementation}
With the circuit on the signal modes, the idler heralding is specified by \eqref{eq:generalizedTransposeTrick} with $f \ra f_\tin$:
\be
    \dir{\tilde{\rho}}_{s,0} 
    \mapsto \cN_{\lamvec}
    f_\tin(\{r_j z_j\}) 
    \cT_s
    f_\tin(\{r_j^* \tilde{z}_j \})
    \qquad
    \textrm{with}
    \qquad
    \cT_s = \br{ \prod_{k=0}^{M-1} \tilde{T}_k(q_k) }_s
\ee
with $r_j, q_k$ dictated by the idler mode efficiencies
and $\tilde{T}_k(q_k) \mapsto e^{q_k z_k \tilde{z}_k }$ in the Bargmann representation.
The probability of this idler-side heralding is $\dir{p}_{\ivec} = \Tr( \dir{\tilde{\rho}}_{s,0})$ and the normalized output is $\dir{\rho}_{s,0} = \dir{\tilde{\rho}}_{s,0} / \dir{p}_{\ivec}$.
As $f_\tin$ is simply the monomial corresponding to the input Fock pattern, the $r_j$ rescaling factors factor out as an overall scalar:
\be
    f_\tin(\{r_j z_j\}) = R_{\ivec}\, f_\tin(\{z_j\})
    \qquad
    \textrm{with}
    \qquad
    R_{\ivec} = \prod_{j\in I_{\ivec}} r_j,
\ee
where $I_{\ivec}$ is the set of occupied modes in the heralded input pattern. Thus, the $r_j$ factors affect the idler-heralding probability but, after normalization, do not deform the signal input state. This is the Bargmann expression of the familiar fact that perfect local idler heralding prepares definite signal Fock states independent of the relative squeezing amplitudes and phases.

For the circuits we consider, all loss can be commuted to the end of the signal mode circuit, so we organize the rest of the calculation as follows:
\begin{enumerate}
    \item pass $\dir{\rho}_{s,0}$ through the ideal circuit,
    \item perform lossy PNRD on the bottom signal modes focusing on the detection pattern $\sigvec$,
    and
    \item pass the itinerant output state through any loss channels on the top signal modes.
\end{enumerate} 
The first step maps the unscaled input polynomial $f_\tin(\{z_j\})$ to $f_{\ivec}(\{z_j\})$, while the thermal Gaussian background transforms as 
$\cT_s \ra \cT_U = U \cT_s \hc{U}$:
\be
    \dir{\rho}_{s,0} \ra \dir{\rho}_{s,1}
    = \frac{ \cN_{\lamvec} |R_{\ivec}|^2 }{\dir{p}_{\ivec}}
    f_{\ivec}(\{z_j\}) 
    \cT_U
    f_{\ivec}(\{\tilde{z}_j \}).
\ee
The second step can be understood by recalling the POVM element for detecting $p$ photons with a PNRD with efficiency $\eta_d$,
\be\label{eq:heralding_POVM_to_operator}
    \hat{\Pi}_p(\eta_d) = \sum_{n_\tin=p}^\infty \binom{n_\tin}{p} \eta_d^p \g_d^{n_\tin-p} \kbm{n_\tin}{n_\tin},
\ee
and then translating it into a Bargmann differential operator:
\be
    \Tr_m\!\br{ (\cdot) \hat{\Pi}_p(\eta_d) }
    \mapsto 
    \frac{1}{p!} \p{\eta_d \partial_{z_m} \partial_{\tilde{z}_m}}^p 
    \cK_{z_m}(\g_d)[\cdot] 
    \equiv \cD_p^{(m)}(\eta_d)[\cdot].
\ee 
Here the slot denotes the Bargmann density operator kernel, $m$ labels the measured mode, and the evaluation at $z_m,\tilde{z}_m=0$ implicit in the contraction operator is to be applied at the end.
With the slot implicit, the full signal-mode POVM element using the notation of Table~\ref{tab:bsLossOnlyComparison} is thus
\be
    \hat{\Pi}_{\sigvec} = \Motimes_{j=0}^{M_{s_h}-1} \hat{\Pi}_{\sigma_j} 
    \mapsto \cD_{\sigvec} = \prod_{j=0}^{M_{s_h}-1} \cD_{\sigma_j}^{(N_\tout+j)}
    = \prod_{j=N_\tout}^{M_{s_h}+N_\tout-1}
        \frac{1}{\sigma_{j-N_\tout}!} \p{\eta_j \partial_{z_j} \partial_{\tilde{z}_j}}^{\sigma_{j-N_\tout}} \cK_{z_j}(\g_j)
\ee
with $\eta_j$ the signal-mode efficiencies here.  
The output state after the signal-side heralding is thus
\be
    \dir{\tilde{\rho}}_{s,2} = \Tr_{s_h}\!\p{\dir{\rho}_{s,1} \hat{\Pi}_{\sigvec}}
    \mapsto
    \cD_{\sigvec}\br{ \dir{\rho}_{s,1} }
\ee
with heralding probability $\dir{p}_{\sigvec} = \Tr( \dir{\tilde{\rho}}_{s,2} )$ such that the normalized output is 
$\dir{\rho}_{s,2} = \dir{\tilde{\rho}}_{s,2} / \dir{p}_{\sigvec}$.
Lastly, this state is passed through any top-signal-mode loss channels,
$\dir{\rho}_{s} = \Motimes_{j=0}^{N_\tout-1} \cE^{(j)}_{\eta_j}\!\p{\dir{\rho}_{s,2}}$.
Combining these results, the Bargmann representation of the overall heralded output state can be expressed as
\be\label{eq:dir_rho_GHZ}
    \dir{\rho}_{s} = \frac{\cN_{\lamvec} |R_{\ivec}|^2}{\dir{p}_{\sigvec} \dir{p}_{\ivec}}
    \Motimes_{j=0}^{N_\tout-1} \cE^{(j)}_{\eta_j}\!\p{ 
        \cD_{\sigvec}\br{ 
            f_{\ivec}(\{z_j\}) 
            \cT_U
            f_{\ivec}(\{\tilde{z}_j \})
        }
    }.
\ee

\subsubsection{Ghost implementation}
The ghost calculation is largely analogous to the direct one. The operational difference is that the circuit is applied on the idler modes, so step 1 is included in the idler heralding as specified by \eqref{eq:generalizedTransposeTrick} with $f \ra f_{\ivec}$:
\be
    \gho{\tilde{\rho}}_{s,1} 
    \mapsto \cN_{\lamvec}
    f_{\ivec}(\{r_j z_j\}) 
    \cT_s
    f_{\ivec}(\{r_j^* \tilde{z}_j \}),
\ee
where the expressions are as defined above but, crucially, one must heed the different underlying lumped efficiencies.
Otherwise, the following steps, 2 and 3, are equivalent, thus, the final state is  
\be\label{eq:gho_rho_GHZ}
    \gho{\rho}_{s} = \frac{\cN_{\lamvec}}{\gho{p}_{\sigvec} \gho{p}_{\ivec}}
    \Motimes_{j=0}^{N_\tout-1} \cE^{(j)}_{\eta_j}\!\p{ 
        \cD_{\sigvec}\br{ 
            f_{\ivec}(\{r_j z_j\}) 
            \cT_s
            f_{\ivec}(\{r_j^* \tilde{z}_j \})
        }
    }
\ee
with the heralding probabilities 
$\gho{p}_{\ivec} = \Tr( \gho{\tilde{\rho}}_{s,1} )$
and
$\gho{p}_{\sigvec} = \Tr( \cD_{\sigvec}\br{ \gho{\rho}_{s,1} } )$.

\subsection{Fock-basis matrix elements of lossy TMSV sources}\label{app:LossyTMSVs}
The general formalism of \appref{app:general_Bargmann_formulae} expresses heralded resource-state generation in terms of lossy thermal kernels and Bargmann contractions.
We now specialize those expressions to a single lossy TMSV source and extract a small number of low-photon-number Fock-basis matrix elements.
This matrix-element viewpoint is more explicit than the full Bargmann representation, and is useful for calculations such as the HOM-state generation example in \appref{app:HOMStateFull}, as well as for checking low-order Bell- and GHZ-state expressions, where only a few idler matrix elements are needed.
The identities below should therefore be viewed as explicit Fock-basis matrix elements of the kernels introduced above, rather than as a separate model.

We consider a lossy TMSV state 
subject to signal efficiency $\eta_s=1-\g_s$ and idler efficiency $\eta_i=1-\g_i$,
$\rho_\textrm{TMSV}(\lambda;\eta_s,\eta_i)
= \cE_{\eta_s}\otimes\cE_{\eta_i}\!\p{
\ketbra{\textrm{TMSV}(\lambda)}{\textrm{TMSV}(\lambda)}}$,
as in \eqref{eq:rhoADdTMSV}.
We define the partial idler matrix elements
\be\label{eq:varrhoDef}
    \varrho_{jm}(\lambda;\eta_s,\eta_i)
    := {}_i\bram{j}
    \rho_\textrm{TMSV}(\lambda;\eta_s,\eta_i)
    \ketm{m}_i,
\ee
which can be obtained either by direct application of the amplitude-damping Kraus operators or, equivalently, by extracting the corresponding Bargmann-kernel coefficients.
To make this connection explicit, let $\kappa_\rho(z,\tilde z;w,\tilde w)$ denote the Bargmann kernel of $\rho_\textrm{TMSV}$, where $z,\tilde z$ are the signal ket- and bra-side variables and $w,\tilde w$ are the corresponding idler variables.
With the normalization convention of \eqref{eq:nm_op_to_Bargmann}, the signal-mode Bargmann kernel of \eqref{eq:varrhoDef} is
\be\label{eq:Bargmann_to_varrho}
    \kappa_{\varrho_{jm}}(z,\tilde z)
    =
    \frac{1}{\sqrt{j!m!}}
    \left.
        \partial_w^j
        \partial_{\tilde w}^m
        \kappa_\rho(z,\tilde z;w,\tilde w)
    \right|_{w,\tilde w=0}.
\ee
Thus, the quantities $\varrho_{jm}$ used below are the low-photon-number idler-sector matrix elements of the lossy TMSV state, or equivalently the corresponding idler coefficients of its Bargmann kernel.

For completeness, we sketch the Kraus-expansion calculation from which these elements may be obtained.
Using the amplitude-damping Kraus operators of \eqref{eq:adChannelKrausOp}, the lossy TMSV state can be written as
\be\label{eq:rhoTMSVKrausExpansion}
    \rho_\textrm{TMSV}(\lambda;\eta_s,\eta_i)
    =
    \sum_{\ell_s,\ell_i=0}^\infty
    \ketbra{\tau_{\ell_s,\ell_i}}{\tau_{\ell_s,\ell_i}},
\ee
where
\be\label{eq:tau_lsi}
    \ket{\tau_{\ell_s,\ell_i}}
    =
    \sqrt{1-\lamSq}
    \sum_{n=\max(\ell_s,\ell_i)}^\infty
    \lambda^n
    \sqrt{\binom{n}{\ell_s}\binom{n}{\ell_i}}
    \g_s^{\ell_s/2}\g_i^{\ell_i/2}
    \eta_s^{(n-\ell_s)/2}\eta_i^{(n-\ell_i)/2}
    \ketm{n-\ell_s}_s\ketm{n-\ell_i}_i .
\ee
Substituting \eqref{eq:tau_lsi} into \eqref{eq:varrhoDef} gives finite sums for any fixed idler indices $j,m$.
Rather than list the resulting general expression, we record only the low-order matrix elements needed below.
With 
\be\label{eq:appB_shorthand}
    q=\g_i\lamSq,\quad
    g=\g_s q,\quad
    h=1-g, 
    \quad \textrm{and} \quad Q=\eta_s q/h,
\ee
the vacuum, single photon, and two photon partial idler matrix elements are
\begin{subequations}\label{eq:varrhojj}
    \begin{align}
        \varrho_{00} &= \frac{1-\lamSq}{h} \sum_{n=0}^\infty Q^n \kbm{n}{n},
        \label{eq:varrho00} \\
        \varrho_{11} &= \frac{(1-\lamSq)\eta_i}{\g_i h^2}
        \sum_{n=0}^\infty (n+g) Q^n \kbm{n}{n},
        \label{eq:varrho11} \\
        \varrho_{22} &= \frac{(1-\lamSq)\eta_i^2}{2\g_i^2 h^3}
        \sum_{n=0}^\infty \br{n(n-1)+2g(2n+g)} Q^n \kbm{n}{n}.
        \label{eq:varrho22}
    \end{align}
\end{subequations}
Note that $\varrho_{jj}$ is the unnormalized signal state heralded by the detection of $j$ idler photons.
Meanwhile, the coherences between the one- and two-photon sectors and the vacuum sector are
\begin{subequations}\label{eq:varrhoij}
    \begin{align}
        \varrho_{10} &= \varrho_{01}^\dagger
        = \frac{(1-\lamSq)\lambda\sqrt{\eta_s\eta_i}}{h^2}
        \sum_{n=0}^\infty \sqrt{n+1}\,Q^n \kbm{n+1}{n}, 
        \label{eq:varrho10} \\
        \varrho_{20} &= \varrho_{02}^\dagger
        = \frac{(1-\lamSq)(\lambda\sqrt{\eta_s\eta_i})^2}{\sqrt{2}h^3}
        \sum_{n=0}^\infty \sqrt{(n+1)(n+2)} Q^n \kbm{n+2}{n}. 
        \label{eq:varrho20}
    \end{align}
\end{subequations}
Here the Fock states in Eqs.~(\ref{eq:varrhojj}) and (\ref{eq:varrhoij}) refer to the signal mode. 
The factors of $\g_i^{-1}$ and $\g_i^{-2}$ in Eqs.~(\ref{eq:varrho11}) and (\ref{eq:varrho22}) arise only from factoring the sums using $q=\g_i\lamSq$; the limits $\g_i\to0$ are finite and are obtained by taking the corresponding continuous limits.

The traces of these unnormalized matrix elements are useful consistency checks and arise in later calculations. They depend only on the idler loss, as expected after tracing over the signal mode:
\begin{subequations}\label{eq:varrhoTraces}
    \begin{align}
        \Tr\br{\varrho_{00}} &= \frac{1-\lamSq}{1-q} 
        =: p_0(\lambda), \label{eq:appB_direct_p0} \\
        \Tr\br{\varrho_{11}} &= \frac{(1-\lamSq)\eta_i\lamSq}{(1-q)^2}
        =: p_1(\lambda), \label{eq:appB_direct_p1} \\
        \Tr\br{\varrho_{22}} &= \frac{(1-\lamSq)(\eta_i\lamSq)^2}{(1-q)^3}
        =: p_2(\lambda), \label{eq:appB_direct_p2} \\
        \Tr\br{\varrho_{10}} &= \Tr\br{\varrho_{01}}
        = \Tr\br{\varrho_{20}} = \Tr\br{\varrho_{02}}
        = 0.
    \end{align}
\end{subequations}
In particular, $\Tr\br{\varrho_{jj}}$ is the probability of observing $j$ photons in the lossy idler mode of a single TMSV source.
In \appref{app:HOMStateFull}, $\varrho_{11}$ describes the directly heralded single-mode input state, while $\varrho_{00}$, $\varrho_{22}$, and $\varrho_{20}=\varrho_{02}^\dagger$ appear in the ghost projection onto the HOM-state sector.
The diagonals $\varrho_{00}$, $\varrho_{11}$ and the coherences $\varrho_{10}=\hc{\varrho}_{01}$ similarly appear in low-order Bell- and GHZ-state contractions, though the $n$-GHZ appendix below is organized using the Bargmann representation.

\section{HOM-state generation details}\label{app:HOMStateFull}
Here we provide the details of the HOM-state generation calculation summarized in \secref{sec:HOM}.
To recap, the target state is
$\ket{\psi_\textrm{HOM}} = \frac{1}{\sqrt{2}}\p{\ketm{20}+\ketm{02}}$ and
we use the symmetric beamsplitter convention, 
$B 
= \tfrac{1}{\sqrt{2}} 
\left( \begin{smallmatrix} 1 & i \\ i & 1 \end{smallmatrix} \right) 
= B^T$,
so that
$a_0^\dagger a_1^\dagger
\overset{B}{\mapsto}
\frac{i}{2}\p{a_0^{\dagger 2}+a_1^{\dagger 2}}$,
which directly prepares $\ket{\psi_\textrm{HOM}}$ from $\ketm{11}$, up to an irrelevant global phase.
As in the main text, we use the generalized transpose-trick perspective of \appref{app:gen_transpose_trick} to commute and lump losses so that both the direct and ghost implementations can be described using lossy TMSV sources followed by ideal linear optics and ideal detection.
The effective input state is
\be\label{eq:appHOM_rhoin}
    \rho_\tin = \rho_\textrm{TMSV}(\lambda_0;\eta_s,\eta_i)
    \otimes \rho_\textrm{TMSV}(\lambda_1;\eta_s,\eta_i),
\ee
with implementation-dependent lumped efficiencies given in \eqref{eq:HOM_etas}.
The formulae below allow unequal squeezing parameters $\lambda_0$ and $\lambda_1$, but use common lumped efficiencies for the two sources.
We use the shorthand of \eqref{eq:appB_shorthand} source by source:
\be\label{eq:appHOM_shorthand}
    q_j=\g_i|\lambda_j|^2,\quad
    g_j=\g_s q_j,\quad
    h_j=1-g_j,
    \quad \textrm{and} \quad
    Q_j=\eta_s q_j/h_j,
    \qquad j=0,1.
\ee
For equal squeezing, $\lambda_0=\lambda_1=\lambda$, we omit the source index and write $q,g,h,Q$. (One can naturally extend the analysis to source-dependent losses, i.e., $\eta_s \ra \eta_{s,j}$, $\eta_i \ra \eta_{i,j}$, etc., though the expressions get more notationally cumbersome.) 

For either implementation, a $(1,1)$ idler click pattern heralds success, the corresponding ideal PNRD POVM element is
$\hat{\Pi}_i^{(11)}=\kbm{11}{11}$.
The unnormalized heralded signal state is therefore
\be\label{eq:appHOM_general_tilderho}
    \tilde\rho_\textrm{HOM} \equiv
    p_\textrm{HOM}\rho_\textrm{HOM} =
    \Tr_i\!\br{
        U_\textrm{HOM} \rho_\tin \hc{U}_\textrm{HOM}
        \hat{\Pi}_i^{(11)}
    }
    \qquad \textrm{with} \qquad
    U_\textrm{HOM}
    = \begin{cases}
        B_s, & \textrm{direct}, \\
        B_i^T, & \textrm{ghost}
    \end{cases}
\ee
and its fidelity is
\be\label{eq:appHOM_fidelity_def}
    \cF_\textrm{HOM}
    =
    \bra{\psi_\textrm{HOM}}
    \rho_\textrm{HOM}
    \ket{\psi_\textrm{HOM}}
    =
    \frac{
        \bra{\psi_\textrm{HOM}}
        \tilde\rho_\textrm{HOM}
        \ket{\psi_\textrm{HOM}}
    }{
        p_\textrm{HOM}
    } .
\ee
In the next two subsections we evaluate \eqref{eq:appHOM_general_tilderho} for the direct and ghost beamsplitter placements. The results used in the main text follow using the common effective efficiencies $(\eta_s,\eta_i)$ specified in \eqref{eq:HOM_etas}.

\subsection{Direct implementation}\label{app:HOM_direct}
In the direct implementation, each TMSV source is first locally heralded by detecting one photon in its idler mode.
Using the notation of \appref{app:LossyTMSVs}, the unnormalized signal state heralded from source $j$ is
\be\label{eq:appHOM_direct_rho1_def}
    \rho_1(\lambda_j) = \frac{1}{p_1(\lambda_j)}
    \varrho_{11}(\lambda_j;\eta_s,\eta_i),
\ee
so, conditioned on the two single-photon heralds, the state incident on the signal beamsplitter is $\rho_1(\lambda_0)\otimes\rho_1(\lambda_1)$.
The fidelity calculation is simplified by back-propagating the target state through the beamsplitter,
$\hc{B}_s \ket{\psi_\textrm{HOM}} = -i\ketm{11}$,
so only the $\kbm{1}{1}$ component of each heralded single-mode state contributes to the HOM-state fidelity.
Thus, the direct HOM fidelity factorizes as 
\be\label{eq:appHOM_direct_fidelity_factorized}
    \dir{\cF}_\textrm{HOM} = \cF_1(\lambda_0)\cF_1(\lambda_1)
\ee
in terms of the single-photon fidelities that follow from 
Eqs.~(\ref{eq:varrho11}) and (\ref{eq:appB_direct_p1}):
\be\label{eq:appHOM_F1}
    \cF_1(\lambda_j) := \bram{1}\rho_1(\lambda_j)\ketm{1}
    = \eta_s \frac{(1-q_j)^2(1+g_j)}{h_j^3}.
\ee
The corresponding direct HOM success probability is simply the probability of the two local idler heralds,
\be\label{eq:appHOM_direct_prob}
    \dir{p}_\textrm{HOM} = p_1(\lambda_0)p_1(\lambda_1).
\ee
Though the fidelity calculation is eased via the back-propagation method, one can nevertheless compute the heralded output state, finding
\be\label{eq:dir_rhoHOM}
    \dir{\rho}_\textrm{HOM} = \frac{(1-q_0) (1-q_1)}{4} \cE_{\eta_s}^{\otimes 2}\!\br{
    \p{a_0^{\dagger 2} + a_1^{\dagger 2}} 
    (q_0 q_1)^{(\hc{a}_0 a_0 + \hc{a}_1 a_1)/2}
    \p{\frac{q_0}{q_1}}^{i(\hc{a}_1 a_0 - \hc{a}_0 a_1)/2}
    \p{a_0^2 + a_1^2}
    }.
\ee
This expression contains the desired two-photon addition term, together with correlated thermal factors that reduce to a beamsplitter-invariant thermal background when $q_0=q_1$.

\subsection{Ghost implementation}\label{app:HOM_ghost}
In the ghost implementation, the beamsplitter acts on the idler modes before the $(1,1)$ detection.
Using the cyclic nature of the trace, the unnormalized output of \eqref{eq:appHOM_general_tilderho} can be written in terms of the back-propagated projection POVM as
\be\label{eq:appHOM_ghost_backprop}
    \gho{\tilde\rho}_\textrm{HOM} =
    \Tr_i\!\br{ \rho_\tin \hat{\Pi}_i^{(\psi)} },
    \qquad
    \hat{\Pi}_i^{(\psi)} = B_i^* \kbm{11}{11} B_i^T
    = \ketbra{\psi_\textrm{HOM}}{\psi_\textrm{HOM}}_i.
\ee
Thus, the ghost implementation projects the idler modes onto the same HOM state that one wishes to herald on the signal modes. 
Using the partial matrix elements of \appref{app:LossyTMSVs}, the unnormalized signal state is
\be\label{eq:appHOM_ghost_varrho}
    \gho{\tilde\rho}_\textrm{HOM} =
    \frac{1}{2}\br{
         \varrho_{22}(\lambda_0)\varrho_{00}(\lambda_1)
        + \varrho_{00}(\lambda_0)\varrho_{22}(\lambda_1)
        +\varrho_{20}(\lambda_0)\varrho_{02}(\lambda_1)
        +\varrho_{02}(\lambda_0)\varrho_{20}(\lambda_1)
    },
\ee
where all $\varrho_{jm}(\lambda_k)$ have common effective efficiencies $(\eta_s,\eta_i)$.
Equivalently, because signal loss commutes with the idler projection, this state can be written in the compact operator form
\be\label{eq:appHOM_ghost_compact_tilde}
    \gho{\tilde\rho}_\textrm{HOM}
    =
    \frac{\eta_i^2(1-|\lambda_0|^2)(1-|\lambda_1|^2)}{4}
    \cE_{\eta_s}^{\otimes 2}\!\br{
        \p{
            \lambda_0^{*2}a_0^{\dagger 2}
            + \lambda_1^{*2}a_1^{\dagger 2}
        }
        q_0^{\hat n_0} q_1^{\hat n_1}
        \p{
            \lambda_0^2a_0^2
            + \lambda_1^2a_1^2
        }
    },
\ee
with $\hat n_j=\hc{a}_j a_j$.
This form makes the mechanism transparent: idler loss produces thermal tails $q_j^{\hat n_j}$, while the idler HOM projection coherently adds two photons to either signal mode.

Taking the trace gives the ghost heralding probability
\be\label{eq:appHOM_ghost_prob}
    \gho{p}_\textrm{HOM}
    =
    \frac{
        \eta_i^2 (1-|\lambda_0|^2) (1-|\lambda_1|^2)
    }{
        2(1-q_0)(1-q_1)
    }
    \br{
        \p{\frac{|\lambda_0|^2}{1-q_0}}^2
        + \p{\frac{|\lambda_1|^2}{1-q_1}}^2
    } .
\ee
The fidelity with $\ket{\psi_\textrm{HOM}}$ is obtained by extracting the corresponding matrix elements of \eqref{eq:appHOM_ghost_varrho}, giving
\be\label{eq:appHOM_ghost_fidelity}
\begin{aligned}
    \gho{\cF}_\textrm{HOM}
    &=
    \frac{(\eta_s\eta_i)^2}{4\gho{p}_\textrm{HOM}}
    \frac{1-|\lambda_0|^2}{h_0^3}
    \frac{1-|\lambda_1|^2}{h_1^3}
    \bigg[
        |\lambda_0|^4
        \frac{1+4g_0+g_0^2}{h_0^2}
        h_1^2
        +
        |\lambda_1|^4
        \frac{1+4g_1+g_1^2}{h_1^2}
        h_0^2
        \\
        &\hspace{3.8cm}
        +
        \p{\lambda_0^*\lambda_1}^{2}
        +
        \p{\lambda_0\lambda_1^*}^{2}
        +
        2(\g_s\g_i)^2|\lambda_0|^4|\lambda_1|^4
    \bigg].
\end{aligned}
\ee
The first two terms arise from the two incoherent alternatives in which both idler photons are associated with source $0$ or source $1$, while the phase-sensitive middle terms arise from the coherence between these alternatives.
The last term is a false-positive contribution involving both signal and idler loss, and vanishes when either side is lossless.

\subsection{Unified structure for equal squeezing}\label{app:homYield}
For equal squeezing, $\lambda_0=\lambda_1=\lambda$, the direct and ghost calculations reduce to the same algebraic expressions for any fixed effective efficiencies $(\eta_s,\eta_i)$.
This equality is the HOM instance of the generalized transpose-trick symmetry: when the two sources are identical, moving the beamsplitter from the signal side to the idler side only changes which physical losses are absorbed into $\eta_s$ and $\eta_i$.
Using the shorthand of \eqref{eq:appB_shorthand}, the common heralding probability and conditional fidelity are
\be\label{eq:appHOM_equal_pAndF}
    p_\textrm{HOM} = \br{ \frac{\eta_i(1-\lamSq)\lamSq}{(1-q)^2} }^2,
    \qquad
    \cF_\textrm{HOM} = \br{ \eta_s \frac{(1-q)^2(1+g)}{h^3} }^2.
\ee
At the state level, the equal-squeezing (unified model) result also admits a useful photon-number-sector decomposition. Let $T_j(Q)=(1-Q)Q^{\hat n_j}$ be a normalized thermal state on signal mode $j$, and define $\cR=Q/q=\eta_s/h$. For fixed effective efficiencies, the normalized conditional HOM state can be written as
\be\label{eq:appHOM_convex_mixture}
    \rho_\textrm{HOM}
    =
    w_2\rho_2(Q)
    +
    w_1\rho_1(Q)
    +
    w_0\rho_0(Q),
\ee
where
\subalign{\label{eq:appHOM_convex_states}
    \rho_2(Q) &= \frac{(1-Q)^2}{4}
    \p{a_0^{\dagger 2}+a_1^{\dagger 2}} T_0(Q)T_1(Q)  \p{a_0^2+a_1^2},
    \\
    \rho_1(Q) &= \frac{1-Q}{2}
    \br{
        \hc{a}_0T_0(Q)a_0T_1(Q)
        +
        T_0(Q)\hc{a}_1T_1(Q)a_1
    },
    \\
    \rho_0(Q) &= T_0(Q)T_1(Q),
}
and
\be\label{eq:appHOM_convex_weights}
    w_2 = \cR^2,
    \qquad
    w_1 = 2\cR(1-\cR),
    \qquad
    w_0 = (1-\cR)^2.
\ee
Here $\rho_2$, $\rho_1$, and $\rho_0$ are normalized states in the two-, one-, and zero-photon sectors, respectively, and $w_0+w_1+w_2=1$.
This decomposition makes explicit that signal loss converts the ideal two-photon HOM component into lower-photon-number sectors, while idler loss is encoded in the thermal parameter $Q$.

From \eqref{eq:appHOM_equal_pAndF} one finds the yield
\be\label{eq:appHOM_equal_Y}
    Y_\textrm{HOM} =
    p_\textrm{HOM} \times \cF_\textrm{HOM}
    = \br{ \frac{\eta_s\eta_i(1-\lamSq)\lamSq(1+g)}{h^3} }^2,
\ee
where the direct and ghost implementations differ after substituting the implementation-dependent efficiencies of \eqref{eq:HOM_etas}.
In the beamsplitter-loss-only limit, $\eta_p=\eta_d=1$, the direct implementation has $(\eta_s,\eta_i)=(\eta_{\rm BS},1)$, while the ghost implementation has $(\eta_s,\eta_i)=(1,\eta_{\rm BS})$.
Equation (\ref{eq:appHOM_equal_Y}) gives the same yield in both cases,
\be\label{eq:appHOM_equal_Y_BSloss}
    Y_\textrm{HOM} =  \br{ \eta_{\rm BS}(1-\lamSq)\lamSq }^2.
\ee
Thus, in this idealized loss model, the ghost implementation trades raw heralding probability for conditional-state fidelity without changing the yield.
With additional propagation or detector loss, this exact yield equality need not hold because the two implementations assign the lumped efficiencies differently.
In particular, $Y_\textrm{HOM}$ depends not only on the product $\eta_s\eta_i$, but also on $g=\g_s\g_i\lamSq$ and $h=1-g$.
Moreover, in the convention of \eqref{eq:HOM_etas}, propagation loss $\eta_p$ affects both the signal and idler modes in the ghost implementation, while it affects only the signal modes in the direct implementation.
For the nominal parameters considered in the main text with $\eta_p,\eta_d<1$, the direct implementation therefore tends to have a somewhat larger yield, though the difference is modest in the regimes of interest.

\subsection{Unequal squeezing}\label{app:HOM_unequal_squeezing}
The formulae above allow the two sources to have unequal squeezing parameters, $\lambda_0\neq\lambda_1$.
In the direct implementation, this imbalance changes the two local heralding probabilities $p_1(\lambda_j)$ and the thermal parameters $q_j$, but it does not coherently rebalance the two-photon HOM amplitude.
The full direct output state in \eqref{eq:dir_rhoHOM} makes this structure explicit: the desired two-photon addition operator, $a_0^{\dagger 2}+a_1^{\dagger 2}$, is unchanged, while unequal $q_0$ and $q_1$ generate the nontrivial thermal mixing factor
$\p{q_0 / q_1}^{i(\hc{a}_1 a_0-\hc{a}_0 a_1)/2}$.
When $q_0=q_1$, this factor is the identity and the thermal background is beamsplitter invariant.
When $q_0\neq q_1$, the thermal background becomes a correlated two-mode Gaussian, but the target-state fidelity still reduces to the factorized result \eqref{eq:appHOM_direct_fidelity_factorized}.
Thus, conditioned on the two local idler heralds, unequal squeezing primarily changes the rate and thermal background of the prepared state rather than introducing a coherent imbalance between $\ketm{20}$ and $\ketm{02}$.

The ghost implementation behaves differently because the idler measurement coherently combines two source-emission histories. This is clearest in the absence of loss.
Then \eqref{eq:appHOM_ghost_compact_tilde} reduces to a pure state
\be\label{eq:appHOM_ghost_unequal_pm}
    \ket{\gho{\psi}_\textrm{HOM}}
    \propto \lambda_0^{*2}\ketm{20} + \lambda_1^{*2}\ketm{02}
    \propto \p{\lambda_0^{*2}+\lambda_1^{*2}}\ket{\psi_+}
    + \p{\lambda_0^{*2}-\lambda_1^{*2}}\ket{\psi_-},
\ee
where we defined $\ket{\psi_\pm}=(\ketm{20}\pm\ketm{02})/\sqrt{2}$.
The target HOM state is $\ket{\psi_\textrm{HOM}}=\ket{\psi_+}$, so the corresponding no-loss fidelity is
\be\label{eq:appHOM_ghost_unequal_fidelity_noloss}
    \gho{\cF}_\textrm{HOM}|_{\eta_s=\eta_i=1} =
    \frac{|\lambda_0^2+\lambda_1^2|^2}{ 2\p{|\lambda_0|^4+|\lambda_1|^4} }.
\ee
Thus, unequal magnitudes or phases of the squeezing parameters produce a coherent phase-flipped component $\ket{\psi_-}$.
Physically, the successful idler detection erases which-path information when the two histories, ``source $0$ emitted two pairs'' and ``source $1$ emitted two pairs,'' are indistinguishable in amplitude and phase. With loss included, the same phase-sensitive terms appear explicitly in the general ghost fidelity \eqref{eq:appHOM_ghost_fidelity}.

\subsection{Related single-photon examples}\label{app:SPSs}
The same methods can be applied to simple one-photon analogues of the HOM calculation.
We consider the interferometric-loss-only setting used in Table~\ref{tab:bsLossOnlyComparison}: the relevant component has efficiency $\eta = 1-\g$, with $q=\g\lamSq$, and $P_n=(1-\lamSq)|\lambda|^{2n}$.
First, consider a single TMSV source used to herald one photon through a lossy single-mode operation, such as a lossy phase shifter or identity channel.
This example is not useful as an isolated phase operation, since the phase could be absorbed into the source, but it gives the simplest illustration of the direct/ghost tradeoff.
In the direct implementation, the loss acts on the signal mode after a perfect one-photon herald, giving
\be\label{eq:appSPS_direct}
    \dir{p}_1 = P_1,
    \qquad \dir{\cF}_1 = \eta.
\ee
In the ghost implementation, the same loss is shifted to the heralding idler mode, giving
\be\label{eq:appSPS_ghost}
    \gho{p}_1 = \frac{\eta P_1}{(1-q)^2},
    \qquad \gho{\cF}_1 = (1-q)^2 .
\ee
The yield is therefore common to both implementations,
\be\label{eq:appSPS_yield}
    Y_1 = p_1\cF_1
    = \eta P_1
    = \eta(1-\lamSq)\lamSq .
\ee
The ghost implementation has higher conditional fidelity when
\be\label{eq:appSPS_fidelity_condition}
    \gho{\cF}_1 \ge \dir{\cF}_1
    \quad\Longleftrightarrow\quad
    \lamSq \le \frac{1}{1+\sqrt{\eta}},
\ee
which is the same condition as for HOM-state generation; see \eqref{eq:HOM_bsLoss_curve}. 

As a slightly less trivial example, consider preparing a single dual-rail qubit using the same two-source beamsplitter geometry as HOM, but heralding on a one-photon idler pattern.
With the symmetric beamsplitter convention, the two single-click patterns $(0,1)$ and $(1,0)$
prepare the two $Y$-basis states (up to irrelevant global phases)
\be\label{eq:appPlusY_states}
    \ket{\pm y} = \frac{\ketm{01} \pm  i\ketm{10}}{\sqrt{2}},
\ee
respectively. Either outcome can be used if the corresponding Pauli-frame phase is tracked.

Fixing $(0,1)$ as the accepted pattern, the direct implementation gives
\be\label{eq:appPlusY_direct}
    \dir{p}_{+y} = P_0 P_1,
    \qquad
    \dir{\rho}_{+y} = \cE^{\otimes 2}_{\eta}\!\p{\ketbra{+y}{+y}},
    \qquad
    \dir{\cF}_{+y} = \eta.
\ee
The ghost calculation is analogous to the HOM case, but with one photon added coherently rather than two, giving
\be\label{eq:appPlusY_ghost_state}
    \gho{p}_{+y} = \frac{\eta P_0P_1}{(1-q)^3},
    \qquad
    \gho{\rho}_{+y} = \frac{1-q}{2} (\hc{a}_1+i \hc{a}_0) T_0(q)T_1(q) (a_1 - i a_0),
    \qquad
    \gho{\cF}_{+y}  = (1-q)^3.
\ee
The common yield for one accepted pattern is
\be\label{eq:appPlusY_yield}
    Y_{+y}  = p_{+y}\cF_{+y}
    = \eta P_0 P_1
    = \eta(1-\lamSq)^2\lamSq.
\ee
Accepting both $(0,1)$ and $(1,0)$ doubles the heralding probabilities and yields, while the conditional fidelities are unchanged assuming one can apply or track the appropriate conditional phase correction.
For this $\ket{+y}$ example, the ghost implementation has higher conditional fidelity when
\be\label{eq:appPlusY_fidelity_condition}
    \gho{\cF}_{+y} \ge \dir{\cF}_{+y}
    \quad\Longleftrightarrow\quad
    \lamSq
    \le
    \frac{1-\eta^{1/3}}{1-\eta},
\ee
which is stricter than the condition of \eqref{eq:appSPS_fidelity_condition} for the single-photon and HOM-state examples, e.g., there one needs $\lamSq < 1/2$ for the ghost implementation to win for all $\eta$, yet here one needs $\lamSq < 1/3$.
These one-photon examples mirror the HOM behavior in a simpler setting: in the ghost implementation, loss in the resource-generation operation is converted into suppressed false-positive heralds, improving the conditional fidelity at sufficiently low squeezing while preserving the yield in this idealized loss model.

\section{Heralded Bell-state generation (HBSG) results}\label{app:FullHBSGResults}
This appendix collects details for the HBSG circuit used in \secref{sec:HBSG}.
We focus on the particular scheme of \figref{fig:Bell_circuits} for four reasons.
First, among the HBSG schemes compared in \refref{randles2026limits}, this scheme has the largest heralding probability.
Second, its loss structure is compatible with the lumped-loss model used in the main text, so source, propagation, detector, and interferometric losses can be incorporated into the four effective efficiencies $\eta_0,\ldots,\eta_3$.
Third, the circuit is comparatively robust to multiphoton errors among the four-photon schemes considered in \refref{randles2026limits}. 
Finally, for frequency-bin encoded Bell-state generation, the circuit topology admits a hybrid spatial-frequency implementation that performs well in the loss-limited regimes relevant to near-term experiments.

The circuit may be viewed as the $n=2$ member of the GHZ-generation family treated in \appref{app:n_ghz}, with two Bell-specific practical modifications. First, the final dual-rail output beamsplitters can be omitted as these beamsplitters map dual-rail Bell states to dual-rail Bell states.
Second, beyond the strict GHZ click-pattern family, one may accept additional signal-heralding patterns that generate permuted Bell states $\ket{\chi^\pm} \propto \ketm{1100} \pm \ketm{0011}$, after known mode relabeling or Pauli-frame updates.  
Notably, these two modifications work together: with the output beamsplitters omitted, the additional accepted patterns give the permuted Bell states, $\ket{\chi^\pm}$, rather than the corresponding bunched-state superpositions $\ketm{2000}+\ketm{0200} \pm (\ketm{0020}+\ketm{0002})$.
Additionally, the circuit of \figref{fig:Bell_circuits}(a) has a heralding signal mode permutation relative to the GHZ prescription, which amounts to a frame update.
We therefore treat the Bell case explicitly, while using the $n=2$ GHZ formulae as important consistency checks.

\subsection{Direct implementation: 4 versus 8 TMSV states}\label{app:HBSG_4vs8}
Given the shown target click-pattern on the idler modes, $\ivec = (0000 1111)$, using $M=8$ TMSV states may appear quite wasteful as we are using half of them to generate the vacuum, $\ketm{0}$, in given modes. 
Meanwhile, one can trivially inject the vacuum into a given mode by not placing a source there. 
So why go through the extra work of making probabilistic vacuum state generators that may include multiphoton contributions?
The reasoning is threefold:
\begin{enumerate}
    \item Doing so allows us to leverage the input-photon configuration flexibility to increase the overall generation probability. \label{item:input_flexibility}
    \item As we will demonstrate in this section, the fidelity results are typically not very different in the two cases, and often the input flexibility can be used to more than make up for any fidelity degradation while keeping the overall successful probability of the protocol at least as large.
    \item Lastly, it simplifies the analysis by allowing us to construct a unified model for the direct and ghost implementations. 
\end{enumerate}
This comparison concerns optical performance only; it does not include engineering costs associated with additional sources, detectors, or feed-forward hardware.

To the first point, we note that in the absence of idler-side loss, the total probabilities of heralding a Bell state in the two cases are:
\be\label{eq:4vs8-prob-comp}
    \dir{p}_{\rm Bell}(M) = \frac{3}{16} \br{ \eta_\downarrow (2-\eta_\downarrow)}^2  p_\iota^{(M)}(\lamSq)
    \qquad 
    \textrm{with}
    \qquad 
    p_\iota^{(M)}(\lamSq) = \begin{cases}
        \br{ 2 P_0(\lamSq) P_1(\lamSq) }^4, & M=8 \\
        \br{ P_1(\lamSq) }^4, & M=4
    \end{cases}
\ee
and $P_0(\lamSq) = 1-\lamSq$, $P_1(\lamSq) = (1-\lamSq) \lamSq$ the ideal probabilities of heralding $0,1$ photons from a TMSV source. 
The $3/16$ prefactor reflects the three accepted Bell-equivalent signal-output classes, rather than the two appearing in the strict $n=2$ GHZ count.
A quick calculation reveals 
\be
    \max\br{p_\iota^{(8)}(\lamSq)} = p_\iota^{(8)}(1/3) = \p{\frac{8}{27}}^4 
    > 
    \max\br{p_\iota^{(4)}(\lamSq)} = p_\iota^{(4)}(1/2) = \p{\frac{1}{4}}^4 
\ee
(note that $8/27 = 0.\overbar{296} > 1/4 = 0.25$).
Moreover, this same comparison is relevant for $n$-GHZ generation, where 
$p_\iota^{(4n)}(\lamSq) > p_\iota^{(2n)}(\lamSq)$ for $\lamSq < 1/2$. Thus, for each scheme ($n \geq 2$) the additional input-photon configuration flexibility of the $M=4n$ direct and ghost implementations more than makes up for the degradation of having to probabilistically generate vacuum.

\begin{figure*}[t]
  \centering
  \begin{minipage}[t]{0.51\textwidth}
    \centering
    \subfloat[(a)]{%
      \label{fig:hbs-4v8-fidelity}
      \includegraphics[width=\linewidth]{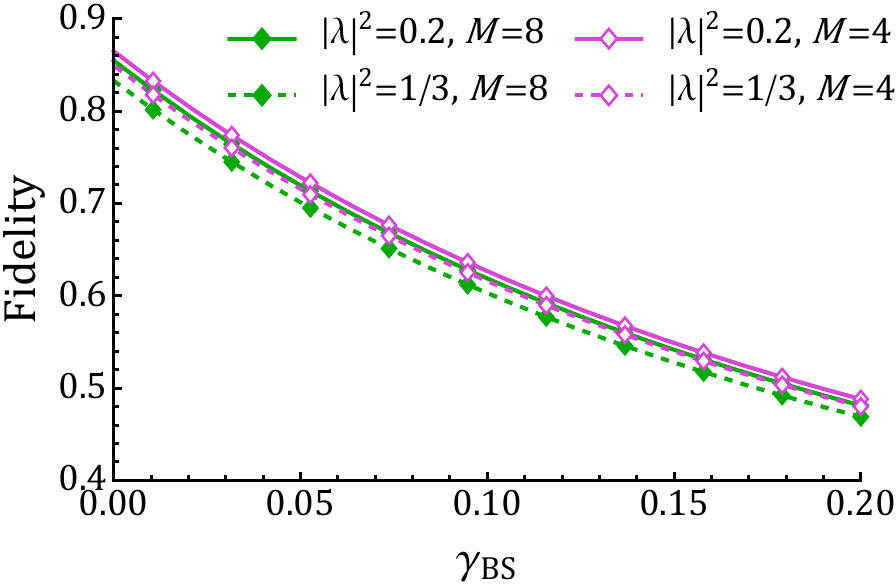}%
    }
    \vspace{5mm}
    \subfloat[(b)]{%
      \label{fig:hbs-4v8-yield}
      \includegraphics[width=\linewidth]{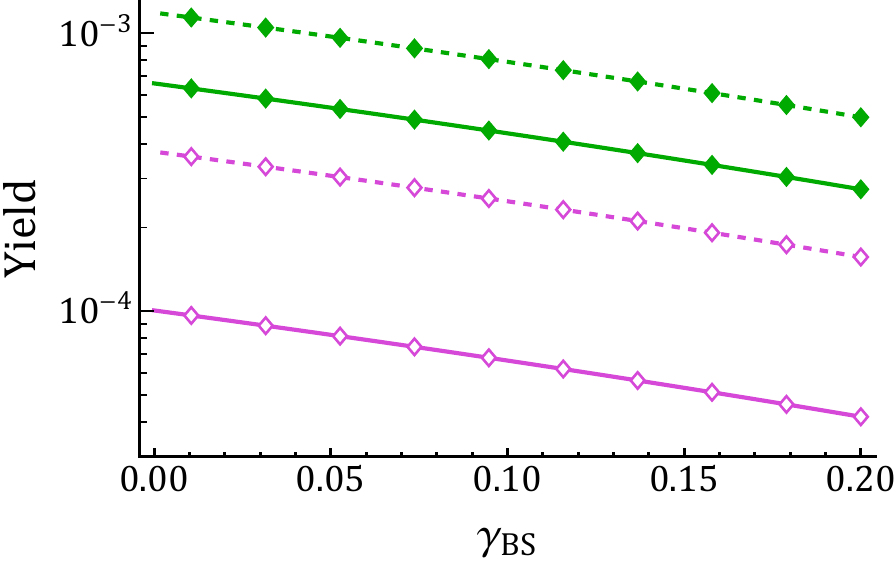}%
    }
  \end{minipage}
  \hspace{0.05\textwidth}
  \begin{minipage}[t]{0.42\textwidth}
    \centering
    \subfloat[(c)]{%
      \label{fig:hbs-4v8-idler-ratio}
      \begin{minipage}[t]{\linewidth}
        \centering
        \makebox[\linewidth][r]{%
          \includegraphics[width=0.5\linewidth]{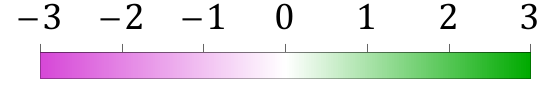}%
        }
        \vspace{0.1em}
        \includegraphics[width=0.99\linewidth]{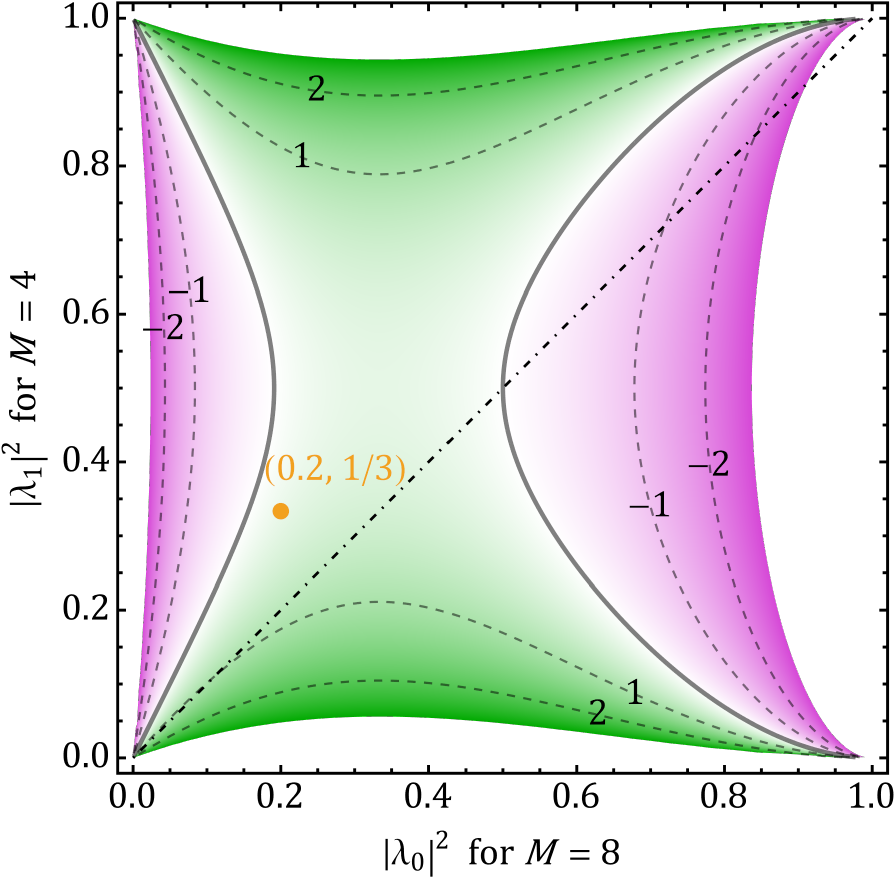}
        \\
        \subfloat[(d)]{%
          \label{fig:hbs8-sim-vs-closed-form}
          \includegraphics[width=\linewidth]{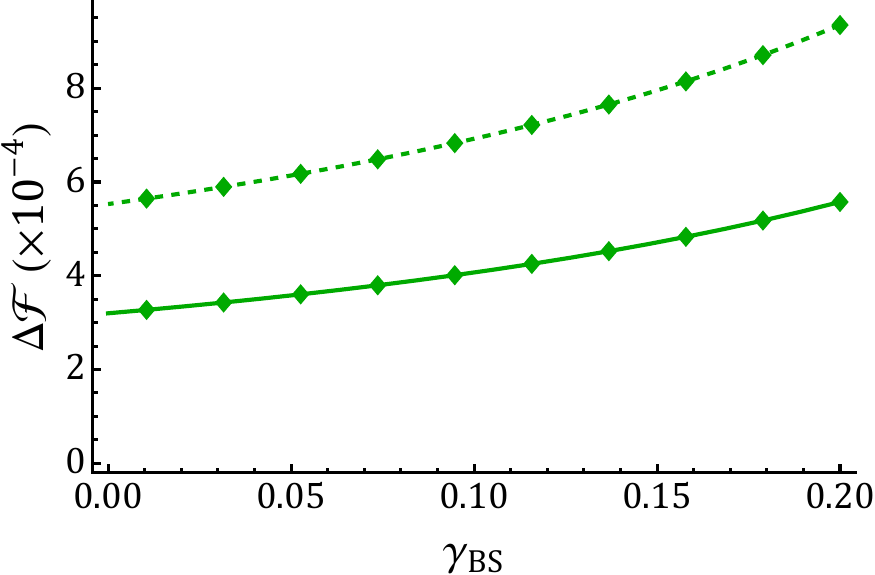}%
    }
      \end{minipage}%
    }
  \end{minipage}
  \caption{
    (a-c) Comparison of simulated direct HBSG using $M=8$ and $M=4$
    TMSV-source implementations.
    (a) Bell state fidelity versus beamsplitter loss, $\g_{\rm BS}$.
    (b) Total yield versus $\g_{\rm BS}$.
    (c) Lossless idler-pattern probability comparison, plotting values of
    $R_i$ of \eqref{eq:log-ratio-4vs8} with modulus $<3$ (other values are excluded, shown in white).
    The marked point corresponds to
    $(|\lambda_0|^2,|\lambda_1|^2)=(0.2,1/3)$, the parameter pair used in
    panels (a) and (b).
    (d) Finite-cutoff validation of the closed-form representative-pattern fidelity, $\cF_{\rm rep}$, for the $M=8$ direct implementation.
    We plot $\Delta\cF=\cF_{\rm Perceval}-\cF_{\rm rep}$ versus $\g_{\rm BS}$ for $\lamSq=0.2$ and $\lamSq=1/3$ (see \appref{app:HBSG_rep_contraction}, Check 3).
    For the reported numerical results we set $\eta_p = 0.98$, $\eta_T=1$, and $\eta_d = 0.98$ for the model in \eqref{eq:BellEtas}, and use the maximum-photon-number parameter $S_\textrm{max} = 6$; see \eqref{eq:signal_input_state_bell_direct}. 
    For (a,b,d), we take $20$ uniformly spaced points $\g_{\rm BS} \in [0,0.2]$; markers are shown on every other point for visual clarity.  
  }
  \label{fig:hbsg-4vs8-direct-comparison}
\end{figure*}

To the second point, having additional TMSV sources also gives more opportunity for something to go wrong in the idler heralding. 
Thus, we want to compare the $M=8$ versus $M=4$ direct HBSG implementations beyond the interferometric-loss-only limit. 
As we do not have general analytic formulae including heralding loss in the $M=4$ case,
here we perform the comparison numerically using Perceval~\cite{heurtel2023perceval} for the simulations as described in \appref{app:HBSG_simulation}.
With idler-side loss we indeed find that the $M=4$ TMSV case does deliver a higher fidelity than when using $M=8$ TMSV states for fixed $\lamSq$. However, this difference is marginal for small squeezing and low idler loss as seen in \figref{fig:hbsg-4vs8-direct-comparison}(a).  
Moreover, one can lower the squeezing for the $M=8$ case, $|\lambda_0|^2$, relative to the $M=4$ case, $|\lambda_1|^2$, to obtain a higher fidelity 
and, as the $M=8$ case can deliver an appreciably higher yield, 
it turns out that you can do better in terms of both fidelity and yield.
This is seen in \figref{fig:hbsg-4vs8-direct-comparison}(a,b) by comparing the $M=8$, $|\lambda_0|^2 = 0.2$ curve (solid green) to the $M=4$, $|\lambda_1|^2 = 1/3$ curve (dashed pink). For these values, one finds that by lowering the squeezing for the $M=8$ case it delivers a both a (slightly) better fidelity and a higher yield. 
To motivate the parameter choices used in this comparison, it is useful to again
consider the limit of no idler loss. In this limit the conditional Bell-state fidelity is the same for each $M$, so the overall yields are proportional to the probability of the idler-side heralding given in \eqref{eq:4vs8-prob-comp}:
$Y^{(d,M)}_{\rm Bell}(\lamSq) = \frac{3}{16} \p{ \eta_\uparrow \eta_\downarrow }^2  p_\iota^{(M)}(\lamSq)$.
Thus, in \figref{fig:hbsg-4vs8-direct-comparison}(c) we plot
\be\label{eq:log-ratio-4vs8}
    R_i = \log_{10}\!\br{
        \frac{Y^{(d,8)}_{\rm Bell}(|\lambda_0|^2)}
        {Y^{(d,4)}_{\rm Bell}(|\lambda_1|^2)}
    }
    = \log_{10}\!\br{
        \frac{p_\iota^{(8)}(|\lambda_0|^2)}
         {p_\iota^{(4)}(|\lambda_1|^2)}
    },
\ee
so $R_i>0$ indicates a larger lossless heralding probability for
$M=8$, $R_i<0$ indicates a larger probability for $M=4$, and $R_i=0$ gives the equal-probability (and equal-yield) boundary.

\subsection{Unified-model analysis: representative-pattern contraction and closed forms}\label{app:HBSG_rep_contraction}
We first describe the calculation for the representative click patterns shown in \figref{fig:Bell_circuits},
$\sigvec=(1100)$ on the signal heralding modes and $\ivec=(00001111)$ on the idler modes.
The target state for this representative pattern is
$\ket{\Phi^+}  = \frac{1}{\sqrt{2}}\p{\ketm{1010}+\ketm{0101}}$.
As in the main text, for equal squeezing, both the ghost and direct implementations can be mapped into a unified model in terms of the four lumped efficiencies $\eta_0,\eta_1,\eta_2,\eta_3$ that label the top signal, bottom signal, top idler, and bottom idler modes of \figref{fig:Bell_circuits}(b), respectively, with $\g_j=1-\eta_j$. 
The implementation differences are accounted for by changing how the lumped efficiencies are composed of physical-component efficiencies as in \eqref{eq:BellEtas}.

The calculation can be organized directly in the Bargmann representation of \appref{app:general_Bargmann_formulae}: the interferometer maps the input creation variables to linear forms, the accepted signal and idler click patterns select coefficients of the resulting kernel, and the Bell-state fidelity is obtained by contracting the four remaining output variables with $\ket{\Phi^+}$.
For the $n=2$ Bell case this contraction is somewhat unwieldy but tractable: expanding the same Bargmann kernel in the Fock basis produces coefficient sums of the type collected in \appref{app:LossyTMSVs}, all of which can be evaluated in closed form.
We use this coefficient-level evaluation to display compact closed forms for the representative pattern; it is equivalent to the Bargmann contraction and reduces to the corresponding $n=2$ GHZ results under the direct and ghost interferometric-loss-only specializations discussed below.

After projecting the signal heralding modes onto $\sigvec$, the unnormalized state of the remaining signal and idler modes is
\be\label{eq:appBell_sig_projection}
    \tilde\rho_{\sigvec} =
    {}_{s_h}\!\bram{\sigvec} \rho_\tin \ketm{\sigvec}_{s_h}.
\ee
The idler click pattern is evaluated by backpropagating through the transposed interferometer,
\be\label{eq:appBell_idler_povm}
    \hat\Pi^{(i)}_{\ivec} = U_i^* \kbm{\ivec}{\ivec} U_i^T,
\ee
so that the representative unnormalized Bell-state output is
\be\label{eq:appBell_rep_output}
    \tilde\rho_{\rm rep} = \Tr_i\!\br{ \tilde\rho_{\sigvec}\hat\Pi^{(i)}_{\ivec} }.
\ee
From which our goal is to compute
\be\label{eq:appBell_rep_p_f}
    p_{\rm rep} = \Tr\br{\tilde\rho_{\rm rep}},
    \qquad
    \tilde \cF_{\rm rep} = \bra{\Phi^+}\tilde\rho_{\rm rep}\ket{\Phi^+},
    \qquad
    \cF_{\rm rep} = \frac{\tilde \cF_{\rm rep}}{p_{\rm rep}}.
\ee

Although \eqref{eq:appBell_rep_output} is compact, it hides two sources of algebraic complexity.
First, $\rho_\tin$ is a tensor product of $M=8$ lossy TMSV states [see \eqref{eq:rho0_HOM}], and the coefficient representation of each factor follows from the double-sum Kraus expansion in \eqref{eq:rhoTMSVKrausExpansion}.
Second, the back-propagated idler POVM $\hat\Pi^{(i)}_{\ivec}$ is not a single Fock projector in the source basis: for the representative Bell circuit, $U_i^*\ketm{\ivec}$ expands into many (144) Fock-basis terms.
Operationally, we group said terms according to their measured signal-heralding patterns (of which there are 46), form the corresponding outer products on the four unmeasured signal modes, replace each single-mode factor by the appropriate lossy-TMSV partial matrix element $\varrho_{jk}$, and use Mathematica for algebra assistance.
The probability, $p_{\rm rep}$, follows by tracing these factors.
The unnormalized  Bell overlap, $\tilde \cF_{\rm rep}$, requires the four Fock-basis components of $\ketbra{\Phi^+}{\Phi^+}$, namely two population terms and two coherence terms, and therefore reduces to scalar overlaps $\varrho_{jk}^{(ab)}:=\bram{a}\varrho_{jk}\ketm{b}$.
These scalar overlaps reduce the infinite sums in the lossy-TMSV coefficients to closed-form geometric sums.

In computing the quantities of \eqref{eq:appBell_rep_p_f}, the following local shorthand keeps the closed forms readable:
\be\label{eq:appBell_shorthand_arbitrary_eta}
\begin{gathered}
    q_1=\g_1\lamSq,\qquad
    q_2=\g_2\lamSq, \\
    g_{13}=\g_3 q_1=1-h_{13},\qquad
    g_{02}=\g_0 q_2=1-h_{02}, \\
    x=\frac{\eta_3 q_1}{h_{13}},
    \qquad
    y=\frac{\eta_2\lamSq}{1-q_2},
    \qquad
    \omega_2=\frac{\g_2}{\eta_2}.
\end{gathered}
\ee
Here $q_1$ and $q_2$ are the single-mode thermal parameters associated with loss on the bottom signal and top idler modes, respectively, while $g_{13}$ and $g_{02}$ encode the two signal-idler loss products that appear after conditioning.
With this notation, the representative-pattern heralding probability is
\be\label{eq:appBell_p_rep_arbitrary_eta}
    p_{\rm rep}
    =
    \frac{
        (1-\lamSq)^4(1-q_1)^6(x+y)^2
    }{
        32 h_{13}^6 (1-q_2)^4q_1^2
    }
    \left[
        2y^2g_{13}^2
        +
        4xyg_{13}(1+g_{13})
        +
        x^2(1+4g_{13}+2g_{13}^2)
    \right].
\ee
The corresponding unnormalized Bell-state overlap is
\be\label{eq:appBell_f_rep_arbitrary_eta}
    \tilde \cF_{\rm rep}
    =
    \frac{
        \eta_0^2 q_2^2 (1-\lamSq)^4(1-q_1)^6
    }{
        64\,\omega_2^4 h_{13}^6 q_1^2 h_{02}^4
    }
    \cP_F ,
\ee
where
\be\label{eq:appBell_PF_arbitrary_eta}
\begin{aligned}
    \cP_F
    &= 4g_{13}^2g_{02}^2 
    + 4(\omega_2x)g_{13}(1+2g_{13})g_{02}
    + (\omega_2x)^2
    \br{
        (1+2g_{13})^2
        + 2\bigl(4g_{13}(1+g_{13})-1\bigr)g_{02}h_{02}
        + h_{02}^4
    } \\
    &\quad
    + 2(\omega_2x)^3(1+6g_{13}+4g_{13}^2)h_{02}
    + 2(\omega_2x)^4(1+4g_{13}+2g_{13}^2)h_{02}^2
\end{aligned}
\ee
is a polynomial organized in powers of the combination $\omega_2 x$.
The factor $x=\eta_3q_1/h_{13}$ plays the same algebraic role as the effective thermal ratio $Q=\eta_s q/h$ in \appref{app:LossyTMSVs}: it is the ratio that remains after summing over the loss events associated with the conditioned modes.
The representative-pattern fidelity is then $\cF_{\rm rep}=\tilde \cF_{\rm rep}/p_{\rm rep}$.
The apparent singularities in \eqref{eq:appBell_p_rep_arbitrary_eta} and \eqref{eq:appBell_f_rep_arbitrary_eta} are removable; they are artifacts of the compact parametrization used.

These expressions provide several useful consistency checks for the symbolic calculation: 
\\
\emph{Check 1.}
In the no-loss limit, one obtains the expected representative-pattern result
\be\label{eq:appBell_no_loss_check}
    p_{\rm rep}
    =
    \tilde \cF_{\rm rep}
    =
    \frac{1}{32}(1-\lamSq)^4|\lambda|^4,
    \qquad
    \cF_{\rm rep}=1.
\ee
\\
\emph{Check 2.}
The direct and ghost interferometric-loss-only limits reproduce the $n=2$ cases of the $n$-GHZ formulae derived in \appref{app:n_ghz}.
For the direct implementation, we take the perfect-idler limit $\eta_2=\eta_3=1$ and set $\eta_0=\eta_\uparrow$, $\eta_1=\eta_\downarrow$ giving
\be\label{eq:app_directBell_GHZ_n2_check}
    \dir{\cF}_{\rm rep} = \eta_e^2,
    \qquad
    \eta_e = \frac{\eta_\uparrow}{2-\eta_\downarrow}.
\ee
For the ghost implementation, we take the perfect-signal limit $\eta_0=\eta_1=1$ and set $\eta_2=\eta_\uparrow$, $\eta_3=\eta_\downarrow$ giving
\be\label{eq:app_ghostBell_GHZ_n2_check}
    \gho{\cF}_{\rm rep} = \frac{(1-q_\uparrow)^6}{1+4\Upsilon+2\Upsilon^2},
    \qquad
    \Upsilon = \frac{\eta_\uparrow q_\downarrow}{\eta_\downarrow(1-q_\uparrow)},
    \qquad
    q_{\uparrow,\downarrow}=(1-\eta_{\uparrow,\downarrow})\lamSq.
\ee
These are precisely the $n=2$ direct and ghost cases of the $n$-GHZ expressions. 
\\
\emph{Check 3.}
Finally, as a numerical consistency check, \figref{fig:hbsg-4vs8-direct-comparison}(d) compares the closed-form representative-pattern fidelity,  $\cF_{\rm rep}=\tilde \cF_{\rm rep}/p_{\rm rep}$, with a finite-cutoff Perceval calculation for the $M=8$ direct implementation.
The two agree at the $10^{-3}$--$10^{-4}$ level over the plotted range.
The residual $\cF_{\rm Perceval}-\cF_{\rm closed}$ is consistently positive and grows mildly with $\lamSq$ and $\g_{\rm BS}$, which is consistent with truncation removing small higher-photon-number contributions that can otherwise produce false-positive heralds and reduce the exact conditional fidelity.

Equations (\ref{eq:appBell_p_rep_arbitrary_eta}) and (\ref{eq:appBell_f_rep_arbitrary_eta}) are for a single representative pair of accepted click patterns.
For equal squeezing, the accepted idler patterns are equiprobable, and the accepted signal patterns are related by permutations, local phases, and Pauli-frame updates as noted in the main text and described in \appref{app:GHZPermutations}.
Thus, pattern multiplicities multiply the total heralding probability but do not change the representative conditional fidelity after the corresponding frame corrections.
In the Bell case we accept $N_\iota=16$ idler patterns and $N_\sigma=6$ signal patterns, giving the total probability factor $N_\iota N_\sigma=96$ for the Bell-equivalent output classes considered here.

\section{GHZ-state generation on any number of qubits}\label{app:n_ghz}
Here we derive analytical representations of the $n$-qubit GHZ states generated using the heralded schemes of \refref{gimeno2016towards} in the presence of loss. In the interferometric-loss-only limit, these expressions lead to closed-form heralding probabilities and output-state fidelities for both the direct and ghost implementations.
The calculations are most naturally organized in the Bargmann representation of \appref{app:BargmannRep}, which makes the circuit polynomial, heralding contractions, and loss-induced thermal factors explicit.
Up to Bell-specific accepted-pattern freedoms, the results of \appref{app:FullHBSGResults} correspond to the $n=2$ member of this family.

\subsection{Circuit specification}
For each $n$, the direct signal mode circuit can be specified by three layers of symmetric 50:50 beamsplitters. Denoting such a beamsplitter on modes $m_0$ and $m_1$ as the pair $(m_0, m_1)$ we have 
\begin{enumerate}
    \item Transverse layer (T): 
    $(j, j+2n),~j=0,1, \cdots, 2n-1$,
    \item Ladder layer (L):
    $(2j, 2j+1),~j=0,1, \cdots, 2n-1$,
    \item Heralding layer (H): 
    $(2n+2j+1, 2n + (2j+2 \mod{2n})),~j=0, \cdots, n-1$.
\end{enumerate}
These layers fully specify the circuit; an illustration can be found in Fig.~18 of \refref{randles2026limits}. 
There are $N_\iota = 2^{2n}$ valid idler mode heralding patterns, $\ivec$, each involving detecting $2n$ single photons, one per mode linked by a beamsplitter in the T layer in the ghost implementation (in the direct implementation this corresponds to one input single photon per T beamsplitter). 
There are $N_\sigma = 2^n$ valid signal mode heralding patterns, $\sigvec$, each involving detecting $n$ single photons one per mode linked by a beamsplitter in the H layer.
For concreteness, we focus on representative patterns $\ivec = (0^{2n}, 1^{2n})$ and $\sigvec = (01)^n$; the symmetry argument in \appref{app:GHZPermutations} shows that the same probability and fidelity results apply to the other valid click patterns.

\subsection{Circuit transformation and polynomial factorization}
To leverage the results of \appref{app:general_Bargmann_formulae}, we need to calculate $f_{\ivec}$ for which we consider the ideal direct implementation with input
\be
    \ketm{\ivec} = \ketm{0}^{\otimes 2n} \otimes \ketm{1}^{\otimes 2n}
    = f_\tin \vac
    \qquad \textrm{with} \qquad
    f_\tin = \Motimes_{j=2n}^{4n-1} \hc{a}_j 
    \mapsto \prod_{j=2n}^{4n-1} z_j.
\ee
It will turn out to be useful to separate the Bargmann variables into 4 classes based on mode parity and whether the mode is in the top or bottom $2n$:
\be
    \hc{a}_j \mapsto z_j 
    \ra \begin{cases}
        z_{2k} &= e_k \\
        z_{2k+1} &= o_k \\
        z_{2n+2k} &= u_k \\
        z_{2n+2k+1} &= v_k 
    \end{cases}
\ee
with $k = 0, \cdots, n-1$.
Each layer of the circuit acts to transform the creation operators in $f$ and thus transform this Bargmann polynomial. Namely,
\subalign{
    f_\tin = \prod_{k=0}^{n-1} u_k v_k 
    &\overset{T}{\ra}
    \prod_{k=0}^{n-1} \frac{ u_k + i e_k }{\sqrt{2}} \frac{ v_k + i o_k }{\sqrt{2}} 
    \\
    &\overset{L}{\ra}
    \prod_{k=0}^{n-1} \frac{ u_k + i v_k + i e_k + i^2 o_k}{2} \frac{ v_k + i u_k + i o_k + i^2 e_k }{2}
    \\
    &\overset{H}{\ra}
    \prod_{k=0}^{n-1}
    \br{ 
        \frac{1}{2\sqrt{2}} \p{ u_k + i v_{k-1} + i v_k + i^2 u_{k+1} }
        + \frac{1}{2} \p{ i e_k + i^2 o_k }
    }
    \\ &\qquad\times 
    \br{ 
        \frac{1}{2\sqrt{2}} \p{ v_k + i u_{k+1} + i u_k + i^2 v_{k-1} }
        + \frac{1}{2} \p{ i o_k + i^2 e_k }
    } \nonumber 
    \\
    &= \prod_{k=0}^{n-1} \p{\frac{i}{8} F_k}
    \equiv f_\tout = f_{\ivec}, \label{eq:fGHZ_out}
}
where in the last step we observed that each multiplicand can be nicely factored as
\begin{subequations}\label{eq:fGHZ_factor}
   \begin{align}
        F_k &= x_k^2 - y_k^2, \\
        x_k &= u_k + i v_{k-1} + i \sqrt{2} e_k, \\
        y_k &= u_{k+1} - i v_k + \sqrt{2} o_k.
   \end{align} 
\end{subequations}
All $k$-indices in these expressions are understood modulo $n$.
With this simple factorized form we now proceed to analyze the direct and ghost implementations.

\subsection{Interferometric-loss-only limit}

\subsubsection{Direct implementation} 
Our task now is to evaluate \eqref{eq:dir_rho_GHZ} in the limit of only signal mode loss which contains the interferometric-loss-only limit as a special case. Thus, taking $\eta_j \ra 1$ for the idler heralding such that $r_j \ra \lambda_j^*$ and $q_j \ra 0$, we have $\cT_s \ra 1$ (the Bargmann vacuum), which is invariant under unitaries, so $\cT_U \ra 1$.
The heralded output is 
\be
    \dir{\rho}_{s} = \frac{\cN_{\lamvec} |R_{\ivec}|^2}{\dir{p}_{\sigvec} \dir{p}_{\ivec}}
    \Motimes_{j=0}^{2n-1} \cE^{(j)}_{\eta_j}\!\p{ 
        \dir{\rho}_*
    }
\ee
with
\be\label{eq:dir_rho_star}
    \dir{\rho}_* 
    \equiv \cD_{\sigvec}\br{ 
            f_{\ivec}(\{z_j\}) 
            f_{\ivec}(\{\tilde{z}_j \})
        }
\ee
the unnormalized heralded signal-mode state before any direct loss on the output modes.

The signal-mode heralding is captured by the operator $\cD_{\sigvec}$ of \eqref{eq:heralding_POVM_to_operator}. For our nominal $\sigvec = (01)^n$,
\be\label{eq:cD_sigvec}
    \cD_{\sigvec}
    = \prod_{k=0}^{n-1} \big[ 
        \eta_{v_k} \partial_{v_k} \partial_{\tilde{v}_k}
        \cK_{u_k}(\g_{u_k}) \cK_{v_k}(\g_{v_k})
    \big],
\ee
where $\eta_{v_k} = 1 - \g_{v_k}$ is the efficiency for bottom signal mode corresponding to $v_k = z_{2n+2k+1}$, and similarly for $u_k = z_{2n+2k}$.
This heralding operator contains a combination of click terms, $\partial_z$, and contractions, $\cK_z$, which commute with one another. 
Focusing first on the ket-side click operator,
$\partial_{\sigvec} \equiv \prod_k \partial_{v_k}$,
we see that any contributing herald mode term must provide at least one $v_k$ for each $k$ from $f_{\ivec}(\{z_j\})$. Similarly, the corresponding bra-side operator
$\tilde{\partial}_{\sigvec} \equiv \prod_k \partial_{\tilde{v}_k}$
requires at least one $\tilde{v}_k$ for each $k$ from $f_{\ivec}(\{\tilde{z}_j\})$. Since these two requirements act on independent sets of variables, we can analyze the $v$ and $\tilde{v}$ contributions separately.
From Eqs.~(\ref{eq:fGHZ_out}) and (\ref{eq:fGHZ_factor}) we recall that
$f_{\ivec}(\{z_j\}) \propto \prod_k F_k = \prod_k (x_k^2 - y_k^2)$.
Any given $v_k$ can be provided by exactly two neighboring factors:
\begin{enumerate}
    \item $F_k$ via $y_k = u_{k+1} - i v_k + \sqrt{2} o_k$, and
    \item $F_{k+1}$ via $x_{k+1} = u_{k+1} + i v_k + i \sqrt{2} e_{k+1}$.
\end{enumerate}
Thus, after expanding the product of differences in $f_{\ivec}$, each monomial chooses from each factor $F_k$ either the $x_k^2$ contribution or the $-y_k^2$ contribution. Equivalently, each $F_k$ contributes either $v_{k-1}$ through $x_k$, or $v_k$ through $y_k$.
This choice can be represented as an assignment of $n$ edges on the alternating $2n$-cycle with alternating vertex labels
\be
\{ F_0, v_0, F_1, v_1, \cdots, F_{n-1}, v_{n-1} \},
\ee
where choosing $x_k^2$ corresponds to assigning the edge $F_k \leftrightarrow v_{k-1}$, while choosing $-y_k^2$ corresponds to assigning the edge $F_k \leftrightarrow v_k$. The vertex labels are defined modulo $n$. Since each monomial chooses exactly one contribution from each $F_k$, it selects exactly one edge incident to each $F_k$, and hence exactly $n$ edges total.

\begin{figure}[ht!]
    \subfloat[(a)]{ \includegraphics[width=0.9\linewidth]{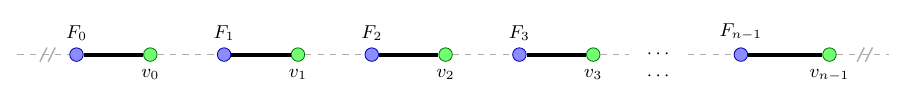}}
    \\
    \subfloat[(b)]{ \includegraphics[width=0.9\linewidth]{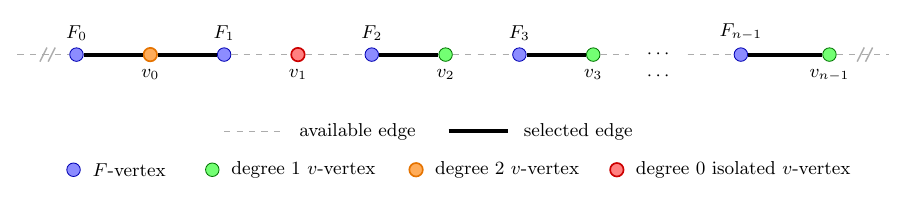}}
    \caption{
    Cycle-graph representation of terms in $\partial_{\sigvec}f_{\ivec}$.
    (a) A surviving term: the assignment $F_k\leftrightarrow v_k \forall k$, corresponding to $\prod_k(-y_k^2)$, is a perfect matching, so every $v_k$ is supplied exactly once. The other surviving term is the complementary perfect matching $F_k\leftrightarrow v_{k-1}$, corresponding to $\prod_k x_k^2$.
    (b) An annihilated crossing term: if, relative to (a), $F_1$ supplies $v_0$ instead of $v_1$, then $v_0$ has degree two while $v_1$ is isolated. The corresponding monomial is $x_1^2 \prod_{k\neq 1}(-y_k^2)$. Such terms lie in $\ker(\partial_{\sigvec})$ and do not contribute.
    }
    \label{fig:n_GHZ_cycle_graph} 
\end{figure}

The click operator $\partial_{\sigvec} = \prod_k \partial_{v_k}$ annihilates any monomial that fails to contain at least one factor of every $v_j$. In the graph representation, this means that no $v_j$ may be isolated. However, there are exactly $n$ selected edges and exactly $n$ vertices of the form $v_j$, so the condition that every $v_j$ is incident to at least one selected edge forces every $v_j$ to be incident to exactly one selected edge. Hence the surviving monomials are precisely those whose edge assignments are perfect matchings of this alternating cycle.
There are exactly two such assignments:
\subalign{
    F_k &\leftrightarrow v_k \quad \forall~k, \\
    F_k &\leftrightarrow v_{k-1} \quad \forall~k,
}
which correspond to either choosing the $-y_k^2$ term from every factor, or the $x_k^2$ term from every factor; see \figref{fig:n_GHZ_cycle_graph}(a).
Any crossing term containing both $x$- and $y$-type choices must change choice somewhere around the cycle. Such a change necessarily leaves at least one $v_j$ isolated, while another $v_\ell$ is supplied twice; see \figref{fig:n_GHZ_cycle_graph}(b). Therefore every crossing term fails to provide at least one required $v_j$, and hence lies in $\ker(\partial_{\sigvec})$:
\be\label{eq:click_nullspace}
    \partial_{\sigvec} f_{\ivec}(\{z_j\}) =
    \partial_{\sigvec}\p{ 
        \prod_k x_k^2
        + \prod_k (-y_k^2)
        + \text{crossing terms}
    }
    = \partial_{\sigvec}\p{
        \prod_k x_k^2 + \prod_k (-y_k^2)
    }.
\ee
The same argument applies independently to the bra-side variables: every crossing term in $f_{\ivec}(\{\tilde{z}_j\})$ fails to provide at least one $\tilde{v}_j$ and hence lies in $\ker(\tilde{\partial}_{\sigvec})$. 
Moreover, a different choice of $\sigvec$ always corresponds to the joint replacement $v_k \leftrightarrow u_{k+1}$ (understood modulo $n$) in the click pattern and operator. These variables occur in pairs, $(v_k, u_{k+1})$, in both $y_k$ and $x_{k+1}$, hence this argument holds for other $\sigvec$ with the vertex labels $v_k$ promoted to $(v_k, u_{k+1})$.

From \eqref{eq:dir_rho_star}, the signal-side heralding operator $\cD_{\sigvec}$ is applied to the unscaled circuit polynomial $f_{\ivec}(\{z_j\})$ and its bra-side counterpart. 
By the matching argument above, after the ket- and bra-side click derivatives act, only the two perfect-matching branches survive. Following \eqref{eq:click_nullspace}, these surviving branches are generated by $x_{k+1}$ and $y_k$, which are the local factors associated with the $e$- and $o$-branches, respectively.
Likewise for the bra-side conjugates $\tilde{x}_{k+1}$ and $\tilde{y}_k$.
The remaining contraction operators act locally on the pair $(v_k, u_{k+1})$. For nonuniform signal-heralding losses, defining 
\be
    \Sigma_k = \g_{u_{k+1}}+\g_{v_k},
    \qquad
    \Delta_k = \g_{u_{k+1}}-\g_{v_k},
\ee
one finds the corresponding contractions
\subalign{
    \cK_{u_{k+1}}(\g_{u_{k+1}}) \cK_{v_k}(\g_{v_k})
    \br{ x_{k+1}\tilde x_{k+1} }
    &= \Sigma_k + 2 e_{k+1}\tilde e_{k+1}, 
    \\
    \cK_{u_{k+1}}(\g_{u_{k+1}}) \cK_{v_k}(\g_{v_k})
    \br{ y_k\tilde y_k }
    &=
    \Sigma_k + 2 o_k\tilde o_k, 
    \\
    \cK_{u_{k+1}}(\g_{u_{k+1}}) \cK_{v_k}(\g_{v_k})
    \br{ x_{k+1}\tilde y_k }
    &= \Delta_k + 2i e_{k+1}\tilde o_k, 
    \\
    \cK_{u_{k+1}}(\g_{u_{k+1}}) \cK_{v_k}(\g_{v_k})
    \br{ y_k\tilde x_{k+1} }
    &=
    \Delta_k - 2i o_k\tilde e_{k+1}.
}
These identities show why the identically damped structure is not generic. If $\Delta_k \neq 0$, the coherence terms acquire lower-degree contributions that are not produced by ordinary amplitude damping of the GHZ state.

For the result quoted in the main text, we now specialize to the symmetric lumped-loss setting. In the four-efficiency notation used above, the top/output signal modes have efficiency $\eta_0 = \eta_\uparrow$, while the bottom/heralding signal modes have efficiency $\eta_1 = \eta_\downarrow$. Thus, in the Bargmann variables used here,
$\g_{u_k}=\g_{v_k}\equiv \g_\downarrow$ for all $k$. 
Then $\Sigma_k = 2\g_\downarrow$ and $\Delta_k=0$ for all $k$. Multiplying the local identities over $k$ gives
\be\label{eq:evaluated_dirRho_star}
    \dir{\rho}_* =
    \p{\frac{\eta_\downarrow}{2^3}}^n \br{
        \prod_{k=0}^{n-1} \p{e_k\tilde e_k+\g_\downarrow}
        + \prod_{k=0}^{n-1} \p{o_k\tilde o_k+\g_\downarrow}
        + \prod_{k=0}^{n-1} (i e_k \tilde o_k)
        + \prod_{k=0}^{n-1} (-i o_k \tilde e_k)
    }
\ee
with the signal-side heralding probability $p_{\sigvec} = 2 \br{\eta_\downarrow (2-\eta_\downarrow)/8}^n$.
This is precisely the Bargmann kernel of an amplitude-damped $n$-qubit GHZ state. Indeed from \eqref{eq:nGHZ}, with $e^{i \phi_n}  = i^n$ in this case,  
\be\label{eq:target_GHZ_Bargmann}
    \ket{G_n} \mapsto G_n(\vec e,\vec o)
    = \frac{1}{\sqrt{2}}
    \p{
        \prod_{k=0}^{n-1} o_k
        + i^n \prod_{k=0}^{n-1} e_k
    },
\ee
where up to normalization single-mode amplitude damping with efficiency $\eta$ maps 
$e_k \tilde e_k \mapsto e_k \tilde e_k + \g/\eta$
and likewise $ o_k \tilde o_k \mapsto o_k \tilde o_k + \g/\eta$, 
while leaving the full coherences unchanged (lower photon number sectors do not retain any coherences).  
Thus, the kernel of \eqref{eq:evaluated_dirRho_star} is proportional to
\be
    \cE_{\eta_*}^{\otimes 2n}
    \p{\ketbra{G_n}{G_n}}
    \qquad \textrm{with} \qquad
    \eta_*=\frac{1}{1+\g_\downarrow} = \frac{1}{2-\eta_\downarrow}. 
\ee
Thus, the lossy signal-side heralding alone induces effective uniform amplitude damping.
Including the final uniform loss on the top/output modes, with efficiency $\eta_0 = \eta_\uparrow$ and composing the amplitude-damping channels yields the normalized direct output state 
\be
    \dir{\rho}_{G_n} =
    \cE_{\eta_e}^{\otimes 2n} \p{\ketbra{G_n}{G_n}}
    \qquad \textrm{with} \qquad
    \eta_e = \eta_\uparrow \eta_*
    = \frac{\eta_\uparrow}{2-\eta_\downarrow},
\ee
which is the identically damped form used in the main text; \eqref{eq:direct_GHZ_identically_damped}.

An informative intermediate case is when each heralding-layer beamsplitter has equal loss on its two output modes, then $\g_{u_{k+1}}=\g_{v_k}$ and hence $\Delta_k=0$, even if this common loss varies with $k$. The lower-degree coherence terms then vanish, and the output is still independently amplitude damped at the optical-mode level, but with mode-dependent efficiencies
$\eta_{*,k} = 1/\p{2-\eta_{H,k}}$,
where $\eta_{H,k}$ is the efficiency of the $k^{\rm th}$ heralding-layer beamsplitter on modes $(v_k, u_{k+1})$. 
The identically damped result quoted in the main text is the special case $\eta_{H,k}=\eta_\downarrow$ for all $k$.

\subsubsection{Ghost implementation}\label{app:nGHZ_ghost}
We now evaluate \eqref{eq:gho_rho_GHZ} for the ghost implementation in the interferometric-loss-only limit. In this case the signal modes are taken to be lossless during the GHZ generation circuit, so $\g_0=\g_1=0$, while the relevant loss occurs on the idler modes implementing the transposed RSG unitary. In the notation of the main text, we write
$\eta_2 \equiv \eta_\uparrow$ 
and 
$\eta_3 \equiv \eta_\downarrow$
with corresponding thermal parameters
$q_{\uparrow,\downarrow} = \g_{\uparrow, \downarrow} \lamSq$.
We also assume equal squeezing, $\lambda_j=\lambda$, so that the generalized transpose trick gives the mode rescalings
\be
    r_j =
    \begin{cases}
        \sqrt{\eta_\uparrow}\,\lambda^*, & j=0,\ldots,2n-1,\\
        \sqrt{\eta_\downarrow}\,\lambda^*, & j=2n,\ldots,4n-1
    \end{cases}.
\ee
The signal-side heralding is ideal in this limit. Thus, the Bargmann POVM operator reduces to the click derivatives
$\cD_{\sigvec} = \partial_{\sigvec}\tilde{\partial}_{\sigvec}$
with all measured variables evaluated at zero after differentiation. 
For the representative pattern $\sigvec=(01)^n$, there are no derivatives with respect to the $u_k$ or $\tilde u_k$ variables, so we may set them to zero immediately.
Using Eqs.~(\ref{eq:fGHZ_out}) and (\ref{eq:fGHZ_factor}), the rescaled ghost polynomial reduces to
\be
    f_{\ivec}(\{r_j z_j\})\big|_{\vec u=\vec 0}
    = \Lambda^n f_0(\vec v,\vec e,\vec o)
\ee
with
$\Lambda = i \eta_\downarrow(\lambda^*)^2 / 8$,
$\mu = \sqrt{2\eta_\uparrow / \eta_\downarrow}$,
and
\be\label{eq:f0_ghost_GHZ}
    f_0(\vec v,\vec e,\vec o) =
    \prod_{k=0}^{n-1}
    \br{
        \p{v_k+i\mu o_k}^2
        - \p{v_{k-1}+\mu e_k}^2
    }.
\ee
All indices are understood modulo $n$. The bra-side polynomial is the coefficient-wise conjugate, denoted $\tilde f_0$.

The thermal part of the Bargmann kernel separates into a top/output contribution and a bottom/heralding contribution,
\be
    \cT_s
    \mapsto
    \exp\!\br{
        q_\uparrow \sum_{k=0}^{n-1} \p{e_k\tilde e_k+o_k\tilde o_k}
        + q_\downarrow \sum_{k=0}^{n-1}  v_k\tilde v_k
    },
\ee
where the $u_k\tilde u_k$ terms have been set to zero. Hence the unnormalized state after signal-side heralding, in the notation of \appref{app:general_Bargmann_formulae}, is represented by the Bargmann kernel
\be\label{eq:ghost_GHZ_reduced_kernel}
\begin{aligned}
    \gho{\tilde\rho}_{s,2} \mapsto\;&
    \cN_{\lamvec}|\Lambda|^{2n}
    \exp\!\br{
        q_\uparrow \sum_{k=0}^{n-1}
        \p{e_k\tilde e_k+o_k\tilde o_k}
    }
    \left.
    \partial_{\sigvec}\tilde\partial_{\sigvec}
    \br{
        f_0(\vec v,\vec e,\vec o)
        \exp\!\br{ q_\downarrow \sum_{k=0}^{n-1} v_k\tilde v_k }
        \tilde f_0(\vec{\tilde v},\vec{\tilde e},\vec{\tilde o})
    }
    \right|_{\vec v,\vec{\tilde v}=\vec 0}.
\end{aligned}
\ee
The factor proportional to $q_\uparrow$ produces thermal tails on the output modes, while the factor proportional to $q_\downarrow$ accounts for terms where accepted signal-heralding clicks are supplied by thermal-background photons.

We first compute the unnormalized overlap with the target GHZ state,
$\gho{\tilde{\cF}}_{G_n} \equiv \bra{G_n}\gho{\tilde{\rho}}_{s,2}\ket{G_n}$.
Since $\ket{G_n}$ has exactly $n$ photons, only the terms in \eqref{eq:ghost_GHZ_reduced_kernel} that leave exactly $n$ photons on the output modes can contribute.
Only the vacuum components of the thermal factors contribute to this overlap: terms proportional to $q_\uparrow$ directly add output photons, while terms proportional to $q_\downarrow$ replace true heralding clicks with thermal-background clicks and leave higher-photon-number output components. Thus,
\be
    \gho{\tilde{\cF}}_{G_n}
    =
    \cN_{\lamvec}|\Lambda|^{2n}
    \bra{G_n}
    \partial_{\sigvec}\tilde\partial_{\sigvec}
    \br{
        f_0(\vec v,\vec e,\vec o)
        \tilde f_0(\vec{\tilde v},\vec{\tilde e},\vec{\tilde o})
    }
    \bigg|_{\vec v,\vec{\tilde v}=\vec 0}
    \ket{G_n}.
\ee
The same perfect-matching argument used in the direct implementation implies that, under $\partial_{\sigvec}$, all crossing terms in $f_0$ vanish. Thus, only the two perfect-matching branches contribute:
\be
    \partial_{\sigvec} f_0 =
    \partial_{\sigvec} \br{
        \prod_{k=0}^{n-1} \p{v_k+i\mu o_k}^2
        + (-1)^n \prod_{k=0}^{n-1} \p{v_k+\mu e_{k+1}}^2
    },
\ee
where we reindexed the second product so that every factor contains $v_k$.
Applying the click derivatives and setting $\vec v=\vec 0$ gives
\be
    \partial_{\sigvec}f_0\big|_{\vec v=\vec 0}
    = (2i\mu)^n \prod_{k=0}^{n-1}o_k
    + (-2\mu)^n \prod_{k=0}^{n-1}e_k 
    = \sqrt{2}\,(2\mu)^n i^n G_n(\vec e,\vec o),
\ee
which is proportional to the target GHZ Bargmann polynomial of \eqref{eq:target_GHZ_Bargmann}. 
Therefore,
\be\label{eq:ghost_GHZ_fid_num}
    \gho{\tilde{\cF}}_{G_n}
    = 2\cN_{\lamvec} \p{2\mu|\Lambda|}^{2n}
    = 2\cN_{\lamvec}
    \br{
        \frac{\eta_\uparrow \eta_\downarrow |\lambda|^4}{8}
    }^n,
\ee
where $\cN_{\lamvec}=(1-\lamSq)^{4n}$ for equal squeezing. This unnormalized fidelity numerator is precisely the yield contribution of the desired GHZ branch for the representative click patterns.

We now compute the corresponding heralding probability for the representative pair of click patterns. This is the trace of the unnormalized state in \eqref{eq:ghost_GHZ_reduced_kernel},
$\gho{p}_{\ivec}\gho{p}_{\sigvec} = \Tr\!\p{\gho{\tilde\rho}_{s,2}}$.
The trace over the output variables can be evaluated using the Bargmann contraction identity
\be
    \cK_z(\alpha)\br{ e^{q z\tilde z}A(z,\tilde z) }
    = \frac{1}{1 - \alpha q}
    \cK_z\!\p{\frac{\alpha}{1 - \alpha q}}\br{A(z,\tilde z)}
\ee
with $\alpha = 1$.
Defining $h_\uparrow \equiv 1-q_\uparrow$, the trace over the $2n$ output modes contributes a prefactor $h_\uparrow^{-2n}$ and scales the output-mode trace contractions giving
\be\label{eq:ghost_GHZ_prob_start}
\begin{aligned}
    \gho{p}_{\ivec}\gho{p}_{\sigvec}
    =
    \cN_{\lamvec}
    \p{\frac{|\Lambda|}{h_\uparrow}}^{2n}
    \left.
    \partial_{\sigvec}\tilde\partial_{\sigvec}
    \br{
        e^{q_\downarrow\sum_k v_k\tilde v_k}
        \prod_{k=0}^{n-1}
        \cK_{e_k}\!\p{h_\uparrow^{-1}}
        \cK_{o_k}\!\p{h_\uparrow^{-1}}
        \br{
            f_0\tilde f_0
        }
    }
    \right|_{\vec v,\vec{\tilde v}=\vec 0}.
\end{aligned}
\ee
The signal-heralding clicks can arise in two ways. A click in mode $v_j$ can be supplied desirably by a ``true'' photon from the polynomial $f_0$, or it can be a false positive supplied by the thermal factor $e^{q_\downarrow v_j\tilde v_j}$. There are no mixed thermal-true clicks, since the thermal factor supplies $v_j$ and $\tilde v_j$ in equal powers. Thus, each accepted heralding pattern is specified by a matching on the same $2n$-cycle graph used in the direct calculation: a selected edge indicates a true click supplied by a neighboring factor $F_k$. Meanwhile, an unmatched $v$-vertex can be supplied by a thermal-background photon, so unlike in the direct case, the matching need not be perfect.

A matching of size $c$ supplies $c$ true clicks and leaves $n-c$ $v$-vertices to be supplied by thermal-background photons. For a true-click edge, there are two possibilities:
\be
    \partial_{v_j}F_j\big|_{\vec v=\vec 0}
    =
    2i\mu o_j,
    \qquad
    \partial_{v_j}F_{j+1}\big|_{\vec v=\vec 0}
    =
    -2\mu e_{j+1}.
\ee
The bra-side derivatives give the conjugate factors. 
By contrast, an unmatched $v$-vertex is supplied by the linear term $q_\downarrow v_j\tilde v_j$ from the measured-mode thermal factor. The corresponding unmatched $F_j$-vertex contributes its residual output polynomial,
\be
    F_j\big|_{v_j,v_{j-1}=0}  = -\mu^2\p{e_j^2+o_j^2},
\ee
which will be weighted by $q_\downarrow$ due to the thermal factor.
The relevant output contractions are
\be
    \cK_{e_j}\p{1/h_\uparrow}\br{e_j^p \tilde e_j^p}
    = \cK_{o_j}\p{1/h_\uparrow}\br{o_j^p\tilde o_j^p}
    = p! h_\uparrow^{-p},
\ee
for $p=1,2$, while cross terms such as $o_j\tilde e_j$ vanish. 
Thus, the respective Bargmann norms for each true-click edge and  thermally supplied click are
\be
    w_{\rm edge} = 4\mu^2 h_\uparrow^{-1},
    \qquad
    w_{\rm th} = 4q_\downarrow\mu^4 h_\uparrow^{-2}.
\ee
Therefore, a matching with $c$ selected edges contributes
$w_{\rm edge}^c w_{\rm th}^{n-c}$. 

Summing over all $c$-edge matchings of the $2n$-cycle gives
\be
    \gho{p}_{\ivec}\gho{p}_{\sigvec} =
    \cN_{\lamvec} \p{\frac{|\Lambda|}{h_\uparrow}}^{2n}
    \sum_{c=0}^{n} M(2n,c)\, 4^n
    \p{\frac{\mu^2}{h_\uparrow}}^{2n-c} q_\downarrow^{n-c},
\ee
where
\be
    M(2n,c) = \frac{2n}{2n-c} \binom{2n-c}{c}
\ee 
is the number of $c$-edge matchings on a $2n$-cycle.
Using $h_\uparrow=1-q_\uparrow$, $|\Lambda|=\eta_\downarrow|\lambda|^2/8$, and $\mu^2=2\eta_\uparrow/\eta_\downarrow$, this can be written as
\be
    \gho{p}_{\ivec}\gho{p}_{\sigvec}  =
    2\,\cM_n(\Upsilon)
    \br{
        \frac{
            \eta_\uparrow\eta_\downarrow
            (1-\lamSq)^4 |\lambda|^4
        }{
            8(1-q_\uparrow)^3
        }
    }^n,
\ee
where
\be\label{eq:matching_polynomial}
    \cM_n(\Upsilon) = \frac{1}{2}
    \sum_{c=0}^{n} M(2n,c)(2\Upsilon)^{n-c},
    \qquad
    \Upsilon = \frac{
        \eta_\uparrow q_\downarrow
    }{
        \eta_\downarrow(1-q_\uparrow)
    }.
\ee
Multiplying by the number of valid idler and signal heralding patterns, $N_\iota N_\sigma = 2^{3n}$, gives the total ghost-generation probability
\be
    \gho{p}_{G_n} = 2\cM_n(\Upsilon)
    \br{
        \frac{
            \eta_\uparrow \eta_\downarrow (1-\lamSq)^4 |\lambda|^4
        }{ (1-q_\uparrow)^3 }
    }^n,
\ee
which is the expression quoted in \eqref{eq:ghost_GHZ_prob} of the main text.
The matching polynomial of \eqref{eq:matching_polynomial} admits a compact closed form in terms of the Chebyshev polynomial of the first kind:
\be
    \cM_n(\Upsilon) = T_n(1+\Upsilon)
    = \frac{1}{2}
    \br{ \p{1+\Upsilon+\sqrt{\Upsilon(\Upsilon+2)}}^n
    + \p{1+\Upsilon-\sqrt{\Upsilon(\Upsilon+2)}}^n }.
\ee
This follows from the standard matching generating function of the cycle graph $C_{2n}$. Consequently, for $\Upsilon\ge0$, the large-$n$ behavior is controlled by the larger root, giving 
\be
    \lim_{n\to\infty} \cM_n(\Upsilon)^{1/n}
    = 1+\Upsilon+\sqrt{\Upsilon(\Upsilon+2)}.
\ee

Combining the unnormalized fidelity numerator with the representative-patterns heralding probability gives the normalized fidelity
\be
    \gho{\cF}_{G_n} =
    \frac{
        \bra{G_n}\gho{\tilde\rho}_{s,2}\ket{G_n}
    }{
        \gho{p}_{\ivec}\gho{p}_{\sigvec}
    }
    =
    \frac{
        (1-q_\uparrow)^{3n}
    }{
        \cM_n(\Upsilon)
    }.
\ee
This is the ghost-implementation fidelity quoted in the main text; see \eqref{eq:ghost_GHZ_fid}. 
The total yield follows immediately by multiplying the total heralding probability by the normalized fidelity.
In the interferometric-loss-only, equal-squeezing setting, it is equal across both implementations:
\be
    Y_{G_n}^{(\textrm{tot})}
    = \gho{p}_{G_n}\gho{\cF}_{G_n}
    = \dir{p}_{G_n}\dir{\cF}_{G_n}
    = 2(\eta_\uparrow\eta_\downarrow)^n (P_0 P_1)^{2n}
\ee
with $P_0 P_1 = (1-\lamSq)^2 \lamSq$.
The matching expansion also gives a useful qualitative picture of the ghost output state. A matching of size $c$ corresponds to $c$ true signal-heralding clicks and $n-c$ clicks supplied by thermal-background photons on the measured modes. Before accounting for the output thermal tail, such a term contains $2n-c$ photons on the output modes: the desired GHZ component arises from the perfect matchings with $c=n$, while matchings with $c<n$ produce higher-photon-number sectors. The remaining thermal tail,
$\exp\!\br{ q_\uparrow \sum_k\p{e_k\tilde e_k+o_k\tilde o_k} }$,
then further dresses each sector with output-mode thermal photons. Thus, in contrast to the direct implementation, whose false-positive heralds appear as missing-photon components, the ghost implementation produces multiphoton errors. The extreme $c=0$ contribution already contains $2n$ output photons before this additional thermal dressing.

\subsection{Coherent errors from unequal squeezing}\label{app:nGHZ_unequal_lambda} 
We now isolate the coherent errors induced by unequal squeezing in the ghost implementation. To do so, we consider the ideal-loss limit, so that the thermal factors are absent, $q_\uparrow=q_\downarrow=0$, and the heralded output is pure. Unlike in the direct implementation, the rescaling factors $r_j=\lambda_j^*$ appear inside the nontrivial circuit polynomial $f_{\ivec}$ and therefore unequal squeezing can deform the heralded state itself.

For the representative click patterns $\ivec = (0^{2n}, 1^{2n})$ and $\sigvec=(01)^n$, the relevant polynomial after setting the unclicked heralding variables $u_k$ to zero can be written as
\be
    f_0 = \prod_{k=0}^{n-1}\p{x_k^2-y_k^2},
    \qquad
    x_k = \lambda_{2n+2k+1}^* v_k + i\sqrt{2}\lambda_{2k+1}^* o_k,
    \qquad
    y_k = \lambda_{2n+2k-1}^* v_{k-1} + i\sqrt{2}\lambda_{2k}^* e_k.
\ee
As the squeezing values are different, we leave them in $f_0$ as they do not factor out globally. The same perfect-matching argument used above implies that all crossing terms lie in $\ker(\partial_{\sigvec})$. Hence only the two perfect-matching branches contribute,
$\partial_{\sigvec} f_0  =
\partial_{\sigvec} \br{ \prod_k x_k^2 + \prod_k \p{-y_k^2} }$, so 
\be
    \partial_{\sigvec}f_0\big|_{\vec v=\vec 0}
    \propto
    \p{ \prod_{k=0}^{n-1} \lambda_{2n+2k+1}^* }
    \br{
        \prod_{k=0}^{n-1} \p{ \lambda_{2k+1}^* o_k }
        + \prod_{k=0}^{n-1} \p{i \lambda_{2k}^* e_k }
    }.
\ee
The prefactor involving the bottom/heralding-mode squeezing amplitudes is common to both GHZ branches. It therefore affects the heralding probability for this click pattern, but not the normalized output-state fidelity.
The normalized heralded state is thus
\be
    \ket{\tilde G_n}  = \frac{
        c_o^*\ketm{01}^{\otimes n}
        + i^n c_e^*\ketm{10}^{\otimes n}
    }{ \sqrt{|c_o|^2+|c_e|^2} },
    \qquad \textrm{with} \qquad
    c_o = \prod_{k=0}^{n-1}\lambda_{2k+1},
    \qquad
    c_e = \prod_{k=0}^{n-1}\lambda_{2k},
\ee
whereas the target state for this representative click pattern is 
$\ket{G_n} = (\ketm{01}^{\otimes n} + i^n\ketm{10}^{\otimes n} ) / \sqrt{2}$.
Thus, unequal squeezing produces a coherent imbalance between the two GHZ branches, with fidelity
\be
    \gho{\cF}_{G_n}(\lamvec) = |\bra{G_n} \tilde G_n\rangle|^2
    = \frac{ |c_o+c_e|^2 }{ 2\p{|c_o|^2+|c_e|^2} }
    = \frac{1}{2} +
    \frac{ \Re\p{c_e c_o^*} }{ |c_e|^2+|c_o|^2 }.
\ee
This is analogous to the coherent NOON imbalance discussed for HOM-state generation in \appref{app:HOM_unequal_squeezing}.
For other valid signal-heralding patterns, the common bottom-mode prefactor changes by replacing the corresponding selected $\lambda_{2n+j}$ factors, and the target GHZ phase may change by a known click-pattern-dependent phase. However, after tracking this phase, the normalized branch imbalance is still governed by the even- and odd-rail products $c_e$ and $c_o$ above.

\subsection{Invariance under valid click-pattern permutations}\label{app:GHZPermutations}
In the main derivations we evaluate a representative pair of heralding patterns,
$\ivec=(0^{2n}1^{2n})$ and $\sigvec=(01)^n$. Here we briefly justify why this suffices in the symmetric loss model considered in the main text. The key observation is that the symmetric beamsplitter is invariant under input or output swaps up to known phases:
\be
    PB = BP = iZBZ
    \qquad \textrm{with} \qquad
     B = \scalebox{1.25}{\raisebox{-0.4\totalheight}{\beamsplitter{0.6}}}
    = \frac{1}{\sqrt2}
    \begin{pmatrix}
        1&i\\ i&1
    \end{pmatrix},
    \qquad
    P = \begin{pmatrix}
        0&1\\ 1&0
    \end{pmatrix},
    \qquad
    Z = \begin{pmatrix}
        1&0\\ 0&-1
    \end{pmatrix}.  
\ee
Thus, exchanging the two modes associated with one of the relevant beamsplitters is equivalent to inserting local phase shifts, up to an overall phase. Photon-number detection probabilities are unchanged by these phases, while the unmeasured output state acquires only known Pauli-frame phases.

For the signal-side heralding patterns of the $n$-GHZ circuit, each valid $\sigvec$ chooses one detector from each pair $(v_k,u_{k+1})$ associated with the heralding layer. Relative to the representative pattern $(01)^n$, any other valid pattern is obtained by applying a product of swaps $v_k\leftrightarrow u_{k+1}$. In the Bargmann derivation this replacement leaves the same $2n$-cycle incidence structure: each selected derivative still chooses one variable from each heralding-layer pair, so the same perfect-matching and matching-polynomial arguments apply. The only change is a known phase in the surviving GHZ branches.
The same reasoning applies to the idler-side patterns. Each valid $\ivec$ selects one mode from each transverse-layer beamsplitter pair. Under equal squeezing and symmetric loss within the mode classes, all such choices are related to the representative pattern by local swaps at the transverse beamsplitters and hence have the same heralding probability and conditional fidelity, up to known output phases.

Consequently, for the $n$-GHZ family in the symmetric loss and equal-squeezing setting considered here, all valid $(\ivec,\sigvec)$ pairs are equivalent up to Pauli-frame phases. It is therefore sufficient to compute a single representative pattern and multiply the corresponding probability by
$N_\iota N_\sigma =2^{2n}2^n=2^{3n}$.
The $n=2$ Bell-state circuit has additional signal-side patterns that may also be accepted because of the extra symmetry of the four-mode Hadamard sub-interferometer; see \appref{app:FullHBSGResults}.

\section{Teleportation, type-I fusion, and numerical simulation details}\label{app:QISTasks}
\subsection{Qubit teleportation}\label{app:teleportation_derivations}
Here we give the details behind the single-qubit teleportation comparison of \secref{sec:qubit_tele}.
We consider an ideal input qubit 
$\ket{q} = \alpha\ket{0}+\beta\ket{1} = \alpha\ketm{01}+\beta\ketm{10}$ 
teleported through a noisy $\ket{\Phi^+}$ Bell state using a type-II fusion measurement, as in \figref{fig:teleportation_and_TI_fusion}(a).
The successful fusion click patterns are grouped into two Pauli-frame classes,
$\cC_0=\{(1100),(0011)\}$ and $\cC_Z=\{(1010),(0101)\}$.
For $\cC\in\cC_0$ the ideal output is $\ket{q}$, while for $\cC\in\cC_Z$ it is $Z\ket{q}$.
In what follows, probabilities are summed over both successful classes, and the $Z$ correction is tracked in the Pauli frame.

We focus on the interferometric-loss-only setting used in the main text.
The type-II fusion beamsplitters are taken to have the same efficiency, $\eta_{\rm BS}$, as the beamsplitters used to generate the Bell state, while the input qubit and fusion detectors are otherwise ideal.
For the direct implementation, the generated Bell state is equivalent to an ideal Bell state followed by uniform amplitude damping with effective efficiency
$\eta_e = \frac{\eta_\uparrow}{2-\eta_\downarrow}$.
Consequently, conditioned on successful fusion, the output states are
\be\label{eq:appTele_direct_output}
    \dir{\rho}_q =
    \begin{cases}
        \cE^{\otimes 2}_{\eta_e}\!\p{\ketbra{q}{q}}, & \cC\in\cC_0,\\
        \cE^{\otimes 2}_{\eta_e}\!\p{Z\ketbra{q}{q}Z}, & \cC\in\cC_Z.
    \end{cases}
\ee
The teleported-qubit fidelity and total fusion success probability are therefore
\be\label{eq:appTele_direct_F_p}
    \dir{\cF}_q = \eta_e,
    \qquad
    \dir{p}_{\rm succ} = \frac{1}{2}\eta_{\rm BS}^2\eta_e.
\ee
Here the factor $1/2$ is the ideal type-II fusion success probability, while the two factors of $\eta_{\rm BS}$ come from loss on the two fused modes.

The ghost implementation has a different error structure. Even when the delivered signal modes of the Bell-state generator are lossless, multiphoton components in the ghost-generated Bell state can combine with fusion loss to produce false-positive fusion successes.
Accordingly, it is instructive bookkeeping to separate the noisy Bell state according to the number of photons in the Bell modes entering the fusion measurement.
After commuting the fusion loss into the fused qubit and Bell modes, only terms with exactly two photons on the fused modes can contribute to a successful type-II fusion event.
Thus, the one-photon sector of the Bell state combines with the one-photon component of the input qubit, while the two-photon sector of the Bell state combines with the vacuum component produced by fusion loss on the input qubit.
All other photon-number sectors are rejected by the ideal PNRD fusion patterns.

For general $\eta_\uparrow<1$, these false positives form an infinite tower of increasingly suppressed multiphoton contributions that along with fusion loss can get mapped into the one- and two-photon sectors of the Bell state. The resulting calculation is not particularly, illuminated so we focus on the hybrid-frequency implementation where to leading order one can reasonably take $\eta_\uparrow=\eta_T\to1$, $\eta_\downarrow=\eta_{\rm BS}^2$, and $q_\downarrow=(1-\eta_\downarrow)\lamSq$.
For $\cC\in\cC_0$, the normalized ghost output state can then be written as
\be\label{eq:appTele_ghost_output}
    \gho{\rho}_q = \frac{
        P_q\ketbra{q}{q}
        + P_{\rm mm}\rho_{\rm mm}
        + P_{11}\kbm{11}{11}
    }{
        W_{\rm tele}
    },
    \qquad
    W_{\rm tele}=P_q+P_{\rm mm}+P_{11}
\ee
with
\subalign{\label{eq:appTele_ghost_weights}
    P_q = \eta_\downarrow^2,
    \qquad
    P_{\rm mm} = 8\eta_\downarrow q_\downarrow\g_{\rm BS},
    \qquad
    P_{11} = 2q_\downarrow\p{\eta_\downarrow+4q_\downarrow\g_{\rm BS}}.
}
For $\cC\in\cC_Z$, the qubit branch is replaced by $Z\ketbra{q}{q}Z$, while the $\rho_{\rm mm}$ and $\kbm{11}{11}$ branches are unchanged.
The three terms in \eqref{eq:appTele_ghost_output} have simple interpretations.
The $P_q$ branch is the desired teleported qubit, 
$P_{\rm mm}$ is a false-positive contribution that remains in the single-photon dual-rail subspace but is maximally mixed, $\rho_{\rm mm} = \frac{1}{2}\p{\kbm{10}{10}+\kbm{01}{01}}$,
and $P_{11}$ is a two-photon output outside the dual-rail-qubit subspace.
After applying the appropriate Pauli-frame correction, the fidelity is therefore
\be\label{eq:appTele_ghost_fidelity}
    \gho{\cF}_q  =
    \frac{P_q+P_{\rm mm}/2}{W_{\rm tele}}.
\ee
The corresponding success probability follows from the trace of the unnormalized output state.
In this limit the ghost-generated Bell-state normalization entering the fusion calculation is
\be
    D_{\rm Bell} =
    P_q+4\eta_\downarrow q_\downarrow+2q_\downarrow^2,
    \qquad \textrm{so} \qquad
    \gho{p}_{\rm succ}  =
    \frac{\eta_{\rm BS}^2}{2} \frac{W_{\rm tele}}{D_{\rm Bell}}.
\ee
The prefactor $\eta_{\rm BS}^2/2$ is the lossy type-II-fusion success factor: $1/2$ is the ideal fusion probability and $\eta_{\rm BS}^2$ accounts for loss on the two fused modes.

\begin{figure}[ht!]
    \begin{minipage}{\linewidth}
        \centering
        \hspace{7mm}\includegraphics[width=0.32\linewidth]{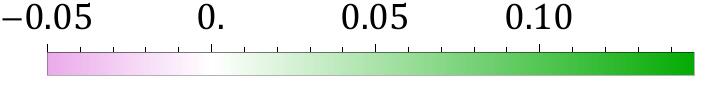}\\[-0.5ex]
        \includegraphics[width=0.45\linewidth]{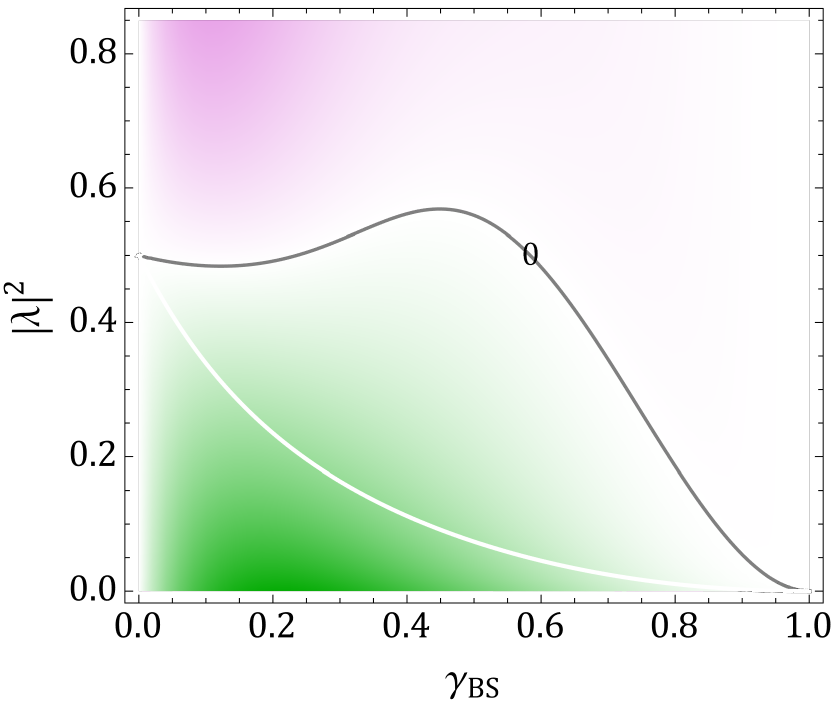}
    \end{minipage}%
    \caption{Teleportation yield comparison. Density plot of $\gho{Y}_{\rm tele}-\dir{Y}_{\rm tele}$ versus $\g_{\rm BS}$ and $\lamSq$. The equal-yield locus is shown in gray, while the equal-fidelity locus is shown as the white curve.}
    \label{fig:teleportation_yield_comp}
\end{figure}

In the same $\eta_\uparrow\to1$ limit, the direct and ghost teleportation yields, $Y_{\rm tele}=p_{\rm succ}\times\cF_q$, are
\subalign{\label{eq:appTele_yields}
    \dir{Y}_{\rm tele} &=
    \frac{1}{2}\eta_{\rm BS}^2\eta_e^2
    = \frac{1}{2}\p{\frac{\eta_{\rm BS}}{2-\eta_\downarrow}}^2,
    \\
    \gho{Y}_{\rm tele} &=
    \frac{1}{2}\eta_{\rm BS}^2
    \frac{
        P_q+P_{\rm mm}/2
    }{
        D_{\rm Bell}
    }
    = \frac{1}{2}\eta_{\rm BS}^2
    \frac{
        \eta_\downarrow^2
        + 4\eta_\downarrow q_\downarrow\g_{\rm BS}
    }{
        \eta_\downarrow^2
        + 4\eta_\downarrow q_\downarrow
        + 2q_\downarrow^2
    }.
}
Although the Bell-state-generation yield is equal for the direct and ghost implementations in this interferometric-loss-only setting, the downstream teleportation yields need not be equal.
For the parameter range shown in \figref{fig:teleportation_yield_comp}, the region in which the ghost implementation gives a higher teleported-qubit fidelity is also a region of higher teleportation yield.
Thus, for this primitive, the conditional-fidelity advantage is not offset by a reduced downstream yield.

\subsection{Numerical simulation details for Bell-state generation and type-I fusion}\label{app:fusion_simu_details}
The type-I fusion results of \secref{sec:typeI_fusion} require simulated noisy Bell-state inputs generated by the direct and ghost heralded Bell-state generation (HBSG) implementations. Naive simulations of the full HBSG schemes of \figref{fig:Bell_circuits} would require a $M=8$ MMSV input on $2M=16$ signal and idler modes. Accounting for loss via the beamsplitter loss model doubles the number of modes to $32$. 
Meanwhile, one needs to include enough total-photon-number sectors, $S$ in \eqref{eq:zetaMMSV}, to reliably capture multiphoton errors, yet the Hilbert-space dimension grows combinatorially with the photon-number cutoff. Therefore, simulating the entire simulation at once is impractical and we are obliged to exploit the structure of the schemes to decompose the calculation into smaller composable and tractable simulations, as summarized in \figref{fig:simulation_sketches}.

\subsubsection{Simulation of the Bell-state generation}\label{app:HBSG_simulation}
For the direct implementation depicted in Fig.~\ref{fig:Bell_circuits}(a), we can exploit the known click pattern on the eight detectors on the idler modes, $\ivec = (00001111)$, and the lack of linear optical components acting on these modes, to precompute the corresponding mixed state onto which the signal modes get projected as a consequence of the imperfect heralding. This action immediately cuts the required number of modes in half, and provides a controllable way of including multi-photon errors coming from imperfect heralding.

\begin{figure*}[t!]
    \centering{\includegraphics[width=0.99\linewidth]{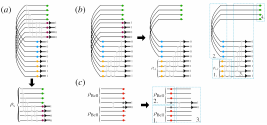}}
    \caption{
    Simulation workflows for the direct and ghost HBSG implementations as well as for type-I fusion. (a) Direct HBSG: the idler click pattern projects the MMSV into an eight mode state, $\rho_s$. (b) Ghost HBSG: the signal-side heralding click pattern projects out part of the MMSV into the state $\rho_{i,\downarrow}$; the remaining circuit is decomposed into four smaller simulations as shown on the right. (c) Type-I fusion of two noisy Bell-states: asymmetric loss is applied independently to each Bell state, then the resulting mixed-state input is passed through the ideal fusion circuit with heralding detection. Such splittings are illustrated via the dashed light-blue boxes. 
    }
    \label{fig:simulation_sketches} 
\end{figure*}

Taking all idler modes to experience the same level of loss, characterized by the efficiency $\eta_d$, and assuming equal squeezing, the resulting input state on the signal modes is constructed as
\be\label{eq:signal_input_state_bell_direct}
    \tilde{\rho}_s = (1-|\lambda|^2)^8 \sum_{S=0}^{S_\textrm{max}} |\lambda|^{2 S} \sum_{\vec{l}\in l_S} 
    (1-\eta_d)^{S-4} \eta_d^4 {\textstyle{\p{\prod_{j=4}^{7}l_j}}} \kbm{\lvec\,}{\lvec\,}.
\ee
Here $S_\textrm{max}$ is a cutoff parameter that should be $\infty$ to include all multi-photon contributions, but in practice must be truncated to enable the simulations (we use $S_\textrm{max}=6$ for the reported results),
$\lvec = (l_0,l_1,...,l_7)$ is an occupation vector, 
and $l_S = \{ \vec{l}: S_{\vec{l}} = S, l_4>0,l_5>0,l_6>0,l_7>0 \}$ is the subset of the occupation vectors with total photon number $S$ that have nonzero photon occupations in the last four modes. These are the states which upon losing a total of $S-4$ photons can result in the expected heralding pattern $(00001111)$. In the simulation we normalize $\tilde{\rho}_s \ra \rho_s$ before propagation through
the signal circuit and track $\Tr\br{\tilde{\rho}_s}$ as the corresponding
heralding probability factor.
Once $\rho_s$ has been constructed, we perform strong linear optical simulation of the smaller circuit shown in \figref{fig:simulation_sketches}(a), which only requires $M=8$ physical modes and $M=8$ loss modes, rendering the simulation of the direct implementation tractable.
The same workflow applies to the $M=4$ direct implementation of \appref{app:HBSG_4vs8}, with the occupation vector restricted to the four TMSV-seeded input modes. The remaining four inputs are ideal vacuum modes and therefore do not introduce idler-heralding false positives.
 
Focusing on the simulation of the ghost implementation in \figref{fig:Bell_circuits}(b), the naive simulation of the full circuit suffers from the same draw backs as the direct implementation. Thus, we again decompose the simulation into tractable parts as illustrated in \figref{fig:simulation_sketches}(b).
First, we write the input MMSV as 
\be
    \ket{\zeta} = \ket{\textrm{TMSV}(\lambda)}^{\otimes 8} = \ket{\textrm{TMSV}(\lambda)}^{\otimes 4}_{\uparrow} \otimes \ket{\textrm{TMSV}(\lambda)}_{\downarrow}^{\otimes 4},
\ee
explicitly separating the inputs for the top ($\uparrow$) and bottom ($\downarrow$) four signal and idler modes. The heralding signal click pattern $\sigvec = (1100)$ projects $\ket{\textrm{TMSV}(\lambda)}_{\downarrow}^{\otimes 4}$ into a mixed idler state, $\rho_{i,\downarrow}$, of all the occupation vectors allowed by the click pattern and losses with efficiency $\eta_1$ on the heralding signal mode. The efficiencies $\eta_j$ are as specified in \eqref{eq:BellEtas}. At this point we have removed $1/4$ of the initial required modes.

The remaining circuit is still quite modular, which we exploit to render the full simulation tractable. We thus split the simulation into four smaller ones:
\begin{enumerate}
    \item Evolve $\rho_{i,\downarrow}$ through the lossy $H_4^T$ interferometer on the bottom four idler modes with efficiency $\eta_3$.
    \item Evolve $\ket{\textrm{TMSV}(\lambda)}^{\otimes 4}_{\uparrow}$ through loss on the top four idler modes with efficiency $\eta_2$.
    \item Evolve the tensor product of the outputs from steps 1~and 2~through a noiseless circuit composed of nothing on the top signal modes and the transversal beamsplitters on all the idler modes followed by ideal heralding detection on the idler modes (as the corresponding loss has already been commuted through the circuit and captured in $\eta_2, \eta_3$).
    \item Evolve the four mode state from step 3~through loss on the top signal modes with efficiency $\eta_0$.
\end{enumerate}
Steps 1~and 2~are independent and can be performed in parallel, then steps 3~and 4~are to be performed sequentially. With this decomposition, no individual simulation, even with loss modes, uses more than twelve modes, making the whole procedure tractable.

\subsubsection{Simulation of the type-I fusion of two Bell states}
Regardless of the implementation used to produce the Bell state (direct or ghost), the simulation of the lossy type-I fusion is performed in the same manner. In \figref{fig:simulation_sketches}(c) we show the lossy linear optical circuit to be simulated, where the input Bell-states are defined on modes 0 to 3 and 4 to 7, respectively. This is equivalent to \figref{fig:teleportation_and_TI_fusion}(b), but with loss channels indicated.
The modes being interfered and detected (modes 3 and 4) experience loss with an effective efficiency $\eta_f = \eta_{p,f} \eta_d \eta_{\rm BS}$, the other modes only experience propagation loss with $\eta_{p,f}$.
With this setup the simulation is performed in two steps:
\begin{enumerate}
    \item We commute back the loss channels to the beginning of the circuit, and evolve each input Bell state through its respective set of loss channels.
    \item We evolve, in a noiseless manner, the tensor product of the output states of step 1.~through the simple interferometer of the type-I fusion circuit then perform ideal heralding detection.
\end{enumerate}
Notice that the two evolutions in step 1~are independent of each other and require only $8$ modes each, making them quite tractable. 

For step 2, we find the success probability and quality of the fused $3$-GHZ state to be the same for either one of the success click-patterns and thus focus solely on $(10)$. Further, we find that depending on the parameter values the tensor product state can be quite large (a Perceval SVDistribution object with over $10^6$ components). Here, to make the subsequent evolution tractable we resort to a truncation strategy. Since ${\rm Tr}[\rho_{\rm Bell} \otimes \rho_{\rm Bell}] = 1$, we define $\rho_{\rm Bell \otimes Bell}^{\rm trunc}$ to be a mixed state constructed with only the fraction of components from $\rho_{\rm Bell} \otimes \rho_{\rm Bell}$ satisfying $1 - {\rm Tr}[\rho_{\rm Bell \otimes Bell}^{\rm trunc}]<\epsilon$ and take $\epsilon = 10^{-6}$. For all our numerical simulations we found that cutoff to bring the size of the mixed state to something tractable without a considerable sacrifice on precision. 
For the simulation results shown in \figref{fig:fusion_results} of the main text, we consider an $30 \times 11$ grid of points in the plane $(\eta_{\rm BS}, \lamSq)$ uniformly spaced in each direction with $\eta_{\rm BS} \in[0.7,1.0]$ and $\lamSq \in[0.01, 0.5]$. The remaining efficiencies are fixed to the nominal values stated in \secref{sec:qip_applications}. The maximum fidelity plot of \figref{fig:fusion_results}(a) is obtained by linear interpolation over these simulated grid points.

\end{document}